\documentclass[12pt,preprint]{aastex}

\def\SgrA {Sgr~A*}

\def\p    {\phantom{0}}
\def\q    {\phantom{00}}

\def\kms  {km~s$^{-1}$}
\def\kmsperkpc {km~s$^{-1}$~kpc$^{-1}$}
\def\kmskpc    {km~s$^{-1}$~kpc$^{-1}$}
\def\masy {mas~y$^{-1}$}

\def\uas  {\ifmmode {\mu{\rm as}}\else{$\mu$as}\fi}
\def\deg  {\ifmmode {^\circ}\else {$^\circ$}\fi}
\def\porm {\ifmmode {\pm}\else {$\pm$}\fi}
\def\chisqpdf {\ifmmode {\chi^2_{\rm pdf}}\else {$\chi^2_{\rm pdf}$}\fi}
\def\chisq    {\ifmmode {\chi^2}\else {$\chi^2$}\fi}
\def\Msun {M$_\odot$}
\def\Lsun {L$_\odot$}
\def\HI   {H~{\small I}}
\def\HII  {H~{\small II}}

\def\etal {et al.~}
\def\eg   {e.g.,~}
\def\ie   {i.e.,~}

\def\d    {\ifmmode {{\rlap{.}}^\circ}\else {${\rlap{.}}^\circ$}\fi}
\def\s    {\ifmmode {{\rlap{.}}^s}\else {${\rlap{.}}^s$}\fi}
\def\as   {\ifmmode {{\rlap{.}}^{''}}\else {${\rlap{.}}^{''}$}\fi}

\newbox\grsign \setbox\grsign=\hbox{$>$} \newdimen\grdimen \grdimen=\ht\grsign
\newbox\laxbox \newbox\gaxbox
\setbox\gaxbox=\hbox{\raise.5ex\hbox{$>$}\llap
     {\lower.5ex\hbox{$\sim$}}}\ht1=\grdimen\dp1=0pt
\setbox\laxbox=\hbox{\raise.5ex\hbox{$<$}\llap
     {\lower.5ex\hbox{$\sim$}}}\ht2=\grdimen\dp2=0pt

\def\lax{\mathrel{\copy\laxbox}}

\def\pa    {\ifmmode {\psi} \else {$\psi$}\fi}
\def\rPpm  {\ifmmode {r_{\Ro,\To}} \else {$r_{Ro,\To}$}\fi}

\def\vlsr  {\ifmmode {v_{\rm LSR}}\else {$v_{\rm LSR}$}\fi}
\def\vlsrr {\ifmmode {v^r_{\rm LSR}}\else {$v^r_{\rm LSR}$}\fi}
\def\vhelio{\ifmmode {v_{Helio}}\else {$v_{Helio}$}\fi}

\def\ura   {\ifmmode {\mu_\alpha}\else {$\mu_\alpha$}\fi}
\def\udec  {\ifmmode {\mu_\delta}\else {$\mu_\delta$}\fi}

\def\ul    {\ifmmode {\mu_l}\else {$\mu_l$}\fi}
\def\ub    {\ifmmode {\mu_b}\else {$\mu_b$}\fi}

\def\uml   {\ifmmode {v_{gr}}\else {$v_{gr}$}\fi}
\def\umb   {\ifmmode {v_b}\else {$v_b$}\fi}
\def\vsrad {\ifmmode {v_{rad}}\else {$v_{rad}$}\fi}

\def\upl   {\ifmmode {v^p_{gr}}\else {$v^p_{gr}$}\fi}
\def\upb   {\ifmmode {v^p_b}\else {$v^p_b$}\fi}
\def\vprad {\ifmmode {v^p_{rad}}\else {$v^p_{rad}$}\fi}

\def\Vo    {\ifmmode {V^{Std}_\odot}\else {$V^{Std}_\odot$}\fi}
\def\Uo    {\ifmmode {U^{Std}_\odot}\else {$U^{Std}_\odot$}\fi}
\def\Wo    {\ifmmode {W^{Std}_\odot}\else {$W^{Std}_\odot$}\fi}
\def\VH    {\ifmmode {V^H_\odot}\else {$V^H_\odot$}\fi}
\def\UH    {\ifmmode {U^H_\odot}\else {$U^H_\odot$}\fi}
\def\WH    {\ifmmode {W^H_\odot}\else {$W^H_\odot$}\fi}
\def\V     {\ifmmode {V_\odot}\else {$V_\odot$}\fi}
\def\U     {\ifmmode {U_\odot}\else {$U_\odot$}\fi}
\def\W     {\ifmmode {W_\odot}\else {$W_\odot$}\fi}

\def\Vs    {\ifmmode {V_s}\else {$V_s$}\fi}
\def\Us    {\ifmmode {U_s}\else {$U_s$}\fi}
\def\Ws    {\ifmmode {W_s}\else {$W_s$}\fi}

\def\Vsbar {\ifmmode {\overline{V_s}}\else {$\overline{V_s}$}\fi}
\def\Usbar {\ifmmode {\overline{U_s}}\else {$\overline{U_s}$}\fi}
\def\Wsbar {\ifmmode {\overline{W_s}}\else {$\overline{W_s}$}\fi}

\def\aone  {\ifmmode {a_1}\else {$a_1$}\fi}
\def\atwo  {\ifmmode {a_2}\else {$a_2$}\fi}
\def\athr  {\ifmmode {a_3}\else {$a_3$}\fi}

\def\pars  {\ifmmode{\pi_s}\else{$\pi_s$}\fi}

\def\Ts    {\ifmmode{\Theta_s}\else{$\Theta_s$}\fi}
\def\Tdot  {\ifmmode{d\Theta\over dR}\else{$d\Theta\over dR$}\fi}

\def\Rp    {\ifmmode{R_p}\else{$R_p$}\fi}

\def\To    {\ifmmode{\Theta_0}\else{$\Theta_0$}\fi}
\def\Ro    {\ifmmode{R_0}\else{$R_0$}\fi}

\def\Dp    {\ifmmode{d_p}\else{$d_p$}\fi}
\def\Zsun  {\ifmmode {Z_\odot}\else {$Z_\odot$}\fi}
\def\Ytilt {\ifmmode{\psi_Y}\else{$\psi_Y$}\fi}
\def\Xtilt {\ifmmode{\psi_X}\else{$\psi_X$}\fi}

\def\ZIAU  {\ifmmode {Z_{IAU}}\else {$Z_{IAU}$}\fi}

\usepackage{listings}
\usepackage{appendix}

\slugcomment{2019 September 19}
\shorttitle{Our View of the Milky Way} 
\shortauthors{Reid \etal}

\begin{document}

\title{TRIGONOMETRIC PARALLAXES OF HIGH-MASS STAR FORMING REGIONS: \\
       OUR VIEW OF THE MILKY WAY   
       }

\author{M. J. Reid\altaffilmark{1}, K. M. Menten\altaffilmark{2}, 
        A. Brunthaler\altaffilmark{2}, X. W. Zheng\altaffilmark{3}, 
        T. M. Dame\altaffilmark{1}, Y. Xu\altaffilmark{5}, 
        J. Li\altaffilmark{5}, N. Sakai\altaffilmark{12},
        Y. Wu\altaffilmark{13,14}, K. Immer\altaffilmark{11},   
        B. Zhang\altaffilmark{6}, A. Sanna\altaffilmark{2},
        L. Moscadelli\altaffilmark{4}, K. L. J. Rygl\altaffilmark{7}, 
        A. Bartkiewicz\altaffilmark{8}, B. Hu\altaffilmark{9},
        L. H. Quiroga-Nu\~nez\altaffilmark{10,11} \& H. J. van Langevelde\altaffilmark{11,10}
       }

\altaffiltext{1}{Center for Astrophysics~$\vert$~Harvard \& Smithsonian, 
   60 Garden Street, Cambridge, MA 02138, USA}
\altaffiltext{2}{Max-Planck-Institut f\"ur Radioastronomie, 
   Auf dem H\"ugel 69, 53121 Bonn, Germany}
\altaffiltext{3}{Department of Astronomy, Nanjing University
   Nanjing 210093, China} 
\altaffiltext{4}{Arcetri Observatory, Firenze, Italy}
\altaffiltext{5}{Purple Mountain Observatory, Chinese Academy of
   Sciences, Nanjing 210008, China}
\altaffiltext{6}{Shanghai Astronomical Observatory, 80 Nandan Rd., Shanghai, China}
\altaffiltext{7}{Italian ALMA Regional Centre, INAF-Istituto di Radioastronomia, 
   Via P. Gobetti 101, 40129 Bologna, Italy}
\altaffiltext{8}{Centre for Astronomy, Faculty of Physics, Astronomy and Informatics, Nicolaus Copernicus University, Grudziadzka 5, 87-100 Torun, Poland}

\altaffiltext{9}{Purple Mountain Observatory, Chinese Academy of
   Sciences, Nanjing 210008, China}
\altaffiltext{10}{Leiden Observatory - Leiden University, Niels Bohrweg 2,
   NL-2333CA, Leiden, The Netherlands}
\altaffiltext{11}{Joint Institute for VLBI ERIC, Postbus 2,, 7990 AA Dwingeloo,
  The Netherlands}
\altaffiltext{12}{Korea Astronomy $\&$ Space Science Institute, 776, Daedeokdae-ro,
  Yuseong-gu, Daejeon 34055, Republic of Korea}
\altaffiltext{13}{National Time Service Center, Key Laboratory of Precise
  Positioning and Timing Technology, Chinese Academy of Sciences, Xi'an 710600, China}
\altaffiltext{14}{National Astronomical Observatory of Japan, 2-21-1 Osawa,
  Mitaka, Tokyo 181-8588, Japan}

\begin{abstract}
We compile and analyze approximately 200 trigonometric parallaxes and proper motions of 
molecular masers associated with very young high-mass stars.   Most of the measurements
come from the BeSSeL Survey using the VLBA and the Japanese VERA project.
These measurements strongly suggest that the Milky Way is a four-arm spiral, with
some extra arm segments and spurs.
Fitting log-periodic spirals to the locations of the masers, allowing for ``kinks'' in
the spirals and using well-established arm tangencies in the $4^{th}$ Galactic quadrant,
allows us to significantly expand our view of the structure of the Milky Way.  
We present an updated model for its spiral structure and incorporate it into
our previously published parallax-based distance-estimation program for sources associated
with spiral arms.  Modeling
the three-dimensional space motions yields estimates of the distance to the Galactic
center, $\Ro=8.15\pm0.15$ kpc, the circular rotation speed at the Sun's position,
$\To=236\pm7$ \kms, and the nature of the rotation curve.  Our data strongly
constrain the full circular velocity of the Sun, $\To+\V=247\pm4$ \kms, and its angular
velocity, $(\To+\V)/\Ro=30.32\pm0.27$ \kmskpc.   Transforming the measured
space motions to a Galactocentric frame which rotates with the Galaxy, we find 
non-circular velocity components typically $\lax$10 \kms.  However, near the 
Galactic bar and in a portion of the Perseus arm, 
we find significantly larger non-circular motions.  Young high-mass
stars within 7 kpc of the Galactic center have a scale height of only 19 pc and, thus,
are well suited to define the Galactic plane.  We find that the orientation of the plane
is consistent with the IAU-defined plane to within $\pm$0\d1, and that the Sun 
is offset toward the north Galactic pole by $\Zsun=5.5\pm5.8$ pc.  Accounting for 
this offset places the central supermassive black hole, \SgrA, in the midplane 
of the Galaxy.  The measured motions perpendicular to the plane of the Galaxy limit 
precession of the plane to $\lax$4 \kms\ at the radius of the Sun.  Using our improved
Galactic parameters, we predict the Hulse-Taylor binary pulsar to be at a distance of
$6.54\pm0.24$ kpc, assuming its orbital decay from gravitational radiation follows 
general relativity.

\end{abstract}


\section{Introduction}

Our view of the Milky Way from its interior does not easily reveal its
properties.  The Sun is near the mid-plane of the Galaxy, resulting
in multiple structures at different distances being superposed on the sky.
Directly mapping the spiral structure of the Milky Way has proven to be 
a challenging enterprise, since distances are very large and dust extinction 
blocks most of the Galactic plane at optical wavelengths.  Thus, {\it Gaia}, 
even with a parallax accuracy of $\pm0.02$ mas, will not be able to freely 
map the Galactic plane.  However, Very Long Baseline Interferometry (VLBI) at 
radio wavelengths is unaffected by extinction and can detect molecular masers 
associated with massive young stars that best trace spiral structure in galaxies.  
Current parallax accuracy for VLBI allows distance measurements across 
most of the Milky Way.  The Bar and Spiral Structure Legacy (BeSSeL) Survey
\footnote{http://bessel.vlbi-astrometry.org}
 and the Japanese VLBI Exploration of Radio Astrometry (VERA) project
\footnote{http://veraserver.mtk.nao.ac.jp}
have now measured approximately 200 parallaxes for masers with accuracies typically 
about $\pm0.02$ mas.  Indeed, recently \citet{Sanna17} measured the parallax
of a maser of $0.049\pm0.006$ mas, placing its young, massive star 
at a distance of 20 kpc, or about 12 kpc further than the Galactic center.

VLBI parallaxes offer a unique opportunity to determine where we are in the 
Milky Way in 3-dimensions, as well as to reveal its spiral structure and 
kinematics.  Massive young stars located within 7 kpc of the 
Galactic center are distributed in a plane with a small perpendicular dispersion
($\approx20$ pc) and, thus, can be used to define the Galactic plane and locate 
the Sun relative to it.  
This allows us to robustly determine the orientation of the plane and the 
distance of the Sun perpendicular to the plane (\Zsun).  
In addition, proper motions, when coupled with parallax distances, 
give linear motions on the sky, and, when combined with line-of-sight velocities, 
provide 3-dimensional space motions.   These can be fitted to simple models of 
Galactic rotation to yield our distance from the Galactic center (\Ro), as well 
as the 3-dimensional motion of the Sun ($\U,\To+\V,\W$) in its orbit about the 
Galaxy (where the peculiar motion components are defined as \U\ toward 
the Galactic center, \V\ toward $90\deg$ longitude, and \W\ toward the
north Galactic pole, and \To\ is the circular rotation of the Galaxy at the Sun).

\citet{Reid:09b}, \citet{Honma:12} and \citet{Reid:14} have summarized
VLBI parallaxes available at the time.  Now, about twice as many parallaxes exist 
than presented in those papers, and we update our understanding of
Galactic structure and kinematics.
In Section \ref{sect:parallaxes} we collect published parallaxes,  
as well as those in preparation or submitted for publication from the BeSSeL
Survey group through 2018.  Using this large data set, we improve upon our view 
of the spiral structure of the Milky Way in Section \ref{sect:spiral_structure}, 
and we fit the space motions in order to estimate fundamental Galactic and Solar 
parameters in Section~\ref{sect:modeling}.  With these parameters, we then
calculate non-circular (peculiar) motions in Section~\ref{sect:peculiar_motions}. 
Using stars interior to the solar
orbit where the Galactic plane is very flat, we evaluate the orientation of the 
IAU-defined plane and estimate the Sun's location perpendicular to the plane in 
Section \ref{sect:GalacticPlane}.  Based on motions of massive young stars perpendicular 
to the plane, we place limits on the precession of the Galactic plane in Section 
\ref{sect:precession}.  We discuss some implications of our results in 
Section \ref{sect:discussion}, and look forward to future advances in Section 
\ref{sect:future}.

\section{Parallaxes and Proper Motions} \label{sect:parallaxes}

VLBI arrays have now been used to measure parallaxes and proper motions for about
200 maser sources associated with young, massive stars, and these are listed in  
Table \ref{table:parallaxes}.  They include results from 
the National Radio Astronomy Observatory's Very Long Baseline Array (VLBA), 
the Japanese VERA project, the European VLBI Network (EVN), and the Australian
Long Baseline Array (LBA).  For sources that have multiple parallax 
measurements (indicated with multiple references), either based on different 
masing molecules, transitions, and/or measured with different 
VLBI arrays, we present averaged results.  Some judgment was used in how to
weight the values.  Typically we used variance weighting, but for results
in tension we evaluated the robustness of each result and adjusted weights 
accordingly.  Note that, in these cases, source coordinates correspond to
a measurement of one of the masing molecules and transitions and are not averages;  
one should consult the primary references when using coordinates. 

The proper motion components, $\mu_x$ and $\mu_y$, and Local Standard of Rest (LSR) 
velocities, \vlsr, given here are meant to apply to the central star which 
excites the masers.  However, the proper motions are usually estimated from
a small number of maser spot motions, and using them to infer the motion of
the central star can involve significant uncertainty.  For example, we gave preference 
to methanol over water maser motions, since the former generally have much smaller 
motions ($\approx5$ \kms) with respect to their exciting star, compared to 
water masers which occur in outflows with expansion speeds of tens of \kms\ 
\citep[\eg][]{Sanna:10a,Sanna:10b,Moscadelli:11b}.   Some papers reporting 
proper motions give only formal measurement uncertainties, and for these we 
estimated an additional error term associated with the uncertainty in 
transferring the maser motions to that of the central star. Typically this error 
term was $\pm5$ \kms\ for methanol masers and $\pm10$ \kms\ for water masers, and
these were added in quadrature with the measurement uncertainties.  For LSR 
velocities, one often has extra information, for example from CO emission from 
associated molecular clouds.  We generally treated methanol maser and CO LSR 
velocities as equally robust estimates of the central star's velocity, and 
preferred over water maser measurements.

\section{Spiral Structure} \label{sect:spiral_structure}

\begin{figure}[htp]
\epsscale{0.85} 
\plotone{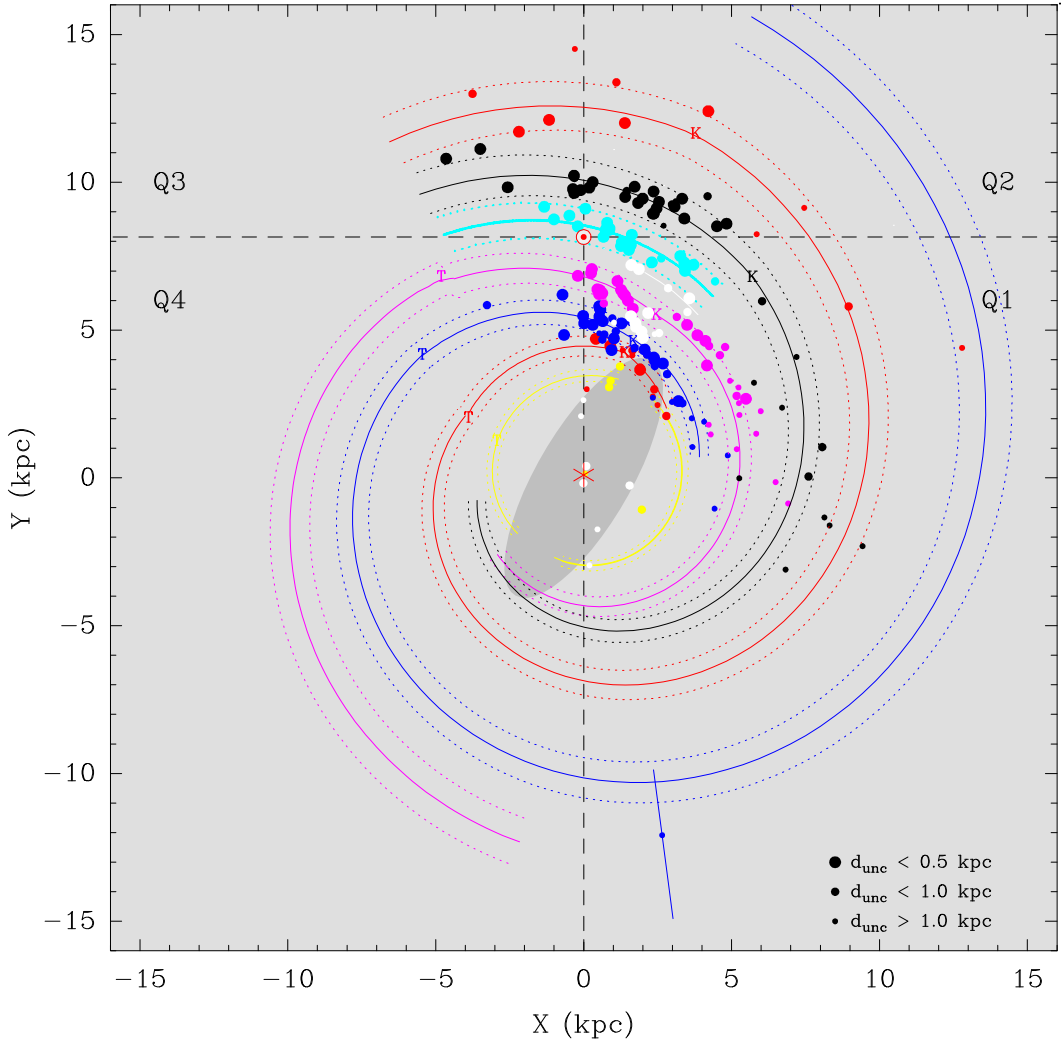}
\caption{\small
Plan view of the Milky Way from the north Galactic pole showing locations of 
high-mass star-forming regions with measured trigonometric parallaxes.
Galactic rotation is clockwise.  
The Galactic center ({\it red asterisk}) is at (0,0) and the Sun 
({\it red Sun symbol}) is at (0,8.15) kpc.
The assignment of sources to spiral arms is discussed in the text:
3-kpc arm, {\it yellow}; Norma-Outer arm, {\it red};  
Scutum-Centaurus-OSC arm, {\it blue}; Sagittarius-Carina arm, {\it purple}; and 
Local arm, {\it cyan}; Perseus arm, {\it black}.  The location of G007.47+00.05
from the parallax of \citet{Sanna17} is shown in {\it blue} with an error bar.
{\it White dots} indicate spurs or sources for which the arm assignment is unclear.
Distance uncertainties are indicated by the inverse size of the symbols, as
given in the legend at the lower right.
Galactic quadrants are divided by {\it grey dashed lines}.
The ``long'' bar is indicated with a shaded ellipse after \citet{Wegg:15}.
The {\it solid} curved lines trace the centers (and {\it dotted} lines the widths
enclosing 90\% of sources) of the fitted spiral arms (see \S3).  The locations
of arm tangencies and kinks listed in Table 2 are marked with the letters ``T''
and ``K.''
       }
\label{fig:parallaxes}
\end{figure}

\begin{figure}[htp]
\epsscale{0.85} 
\plotone{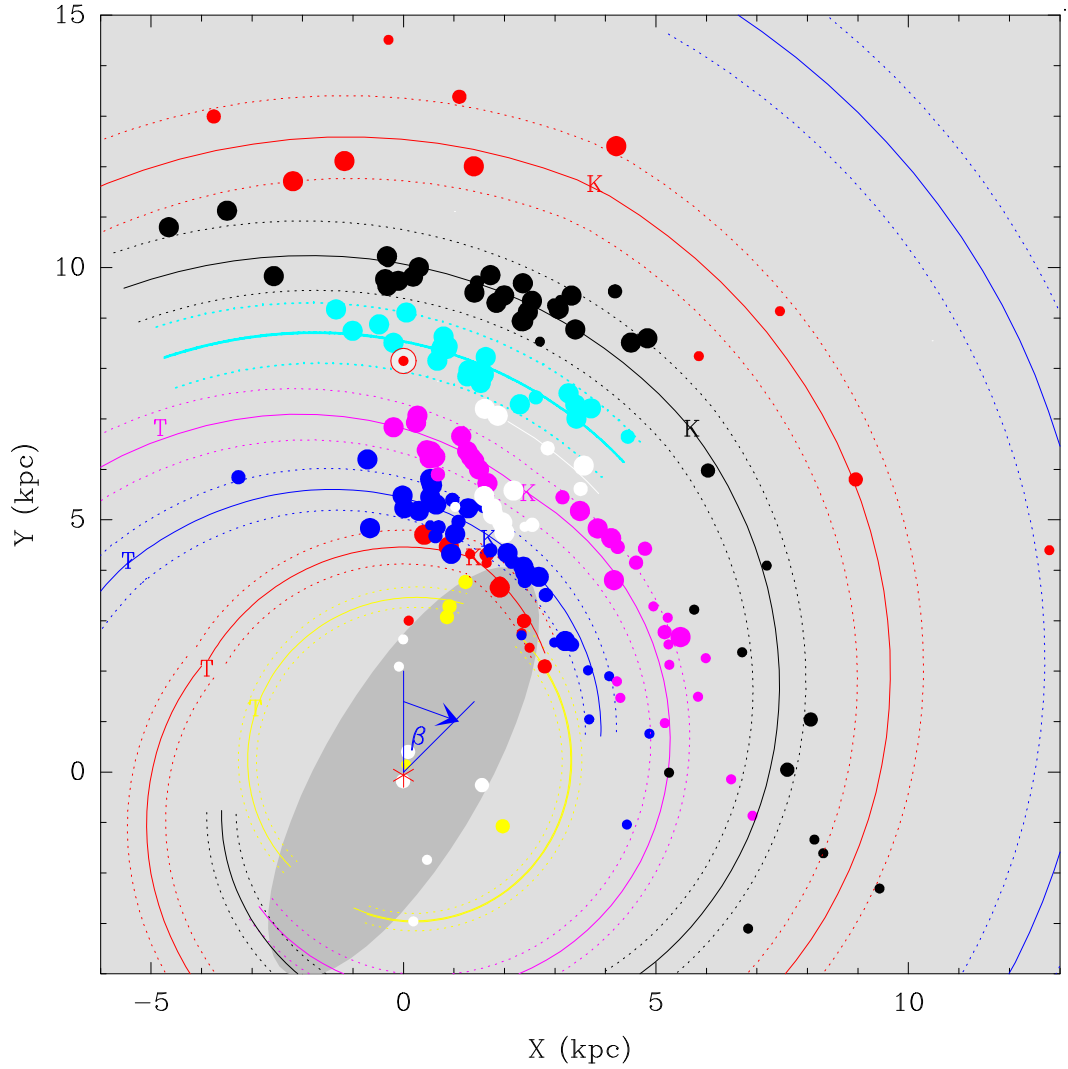}
\caption{\small
Expanded view of the Milky Way from Fig. \ref{fig:parallaxes}.
Galactic azimuth, $\beta$, is the angle formed between
a line from the center toward the Sun and from the center toward
a source as indicated by the {\it blue} arrow near the center.
       }
\label{fig:zoom}
\end{figure}

\begin{figure}[htp]
\epsscale{1.0} 
\plotone{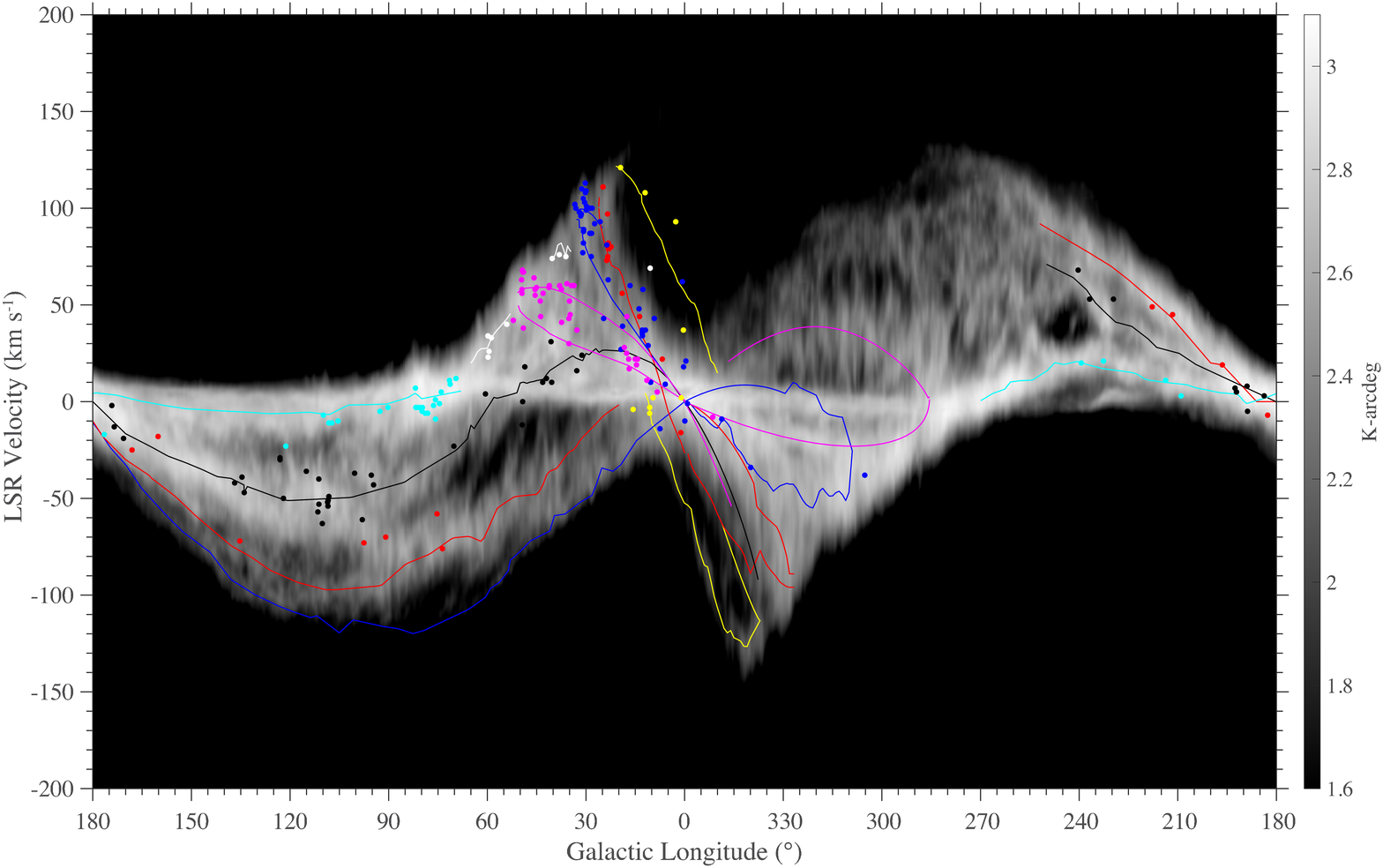}
\caption{\small
\HI\ emission in grey scale as a function of \vlsr\ and Galactic longitude.
Colored dots are sources with parallax measurements shown in Fig \ref{fig:parallaxes}
and the colored lines trace the spiral arms updated from \citet{Reid:16} and 
discussed in Section \ref{sect:armmodel}.  Note, we have connected the somewhat
irregular ($l,v$) locations of giant molecular clouds seen in CO or \HI\ emission, 
giving the slightly jagged spiral arm traces.  
       }
\label{fig:LV}
\end{figure}

The locations of the maser stars listed in Table~\ref{table:parallaxes} 
projected onto the Galactic plane are shown in Fig. \ref{fig:parallaxes} 
superposed on a schematic plot of the Milky Way as viewed from the north 
Galactic pole where rotation is clockwise.  An expanded view of the portion
of the Milky Way for which we currently have most parallax measurements is shown
in Fig. \ref{fig:zoom}.  Distance uncertainties are indicated by the size 
of the dots, with sources having smaller uncertainties emphasized with 
larger dots.  Note that for a given {\it fractional} parallax uncertainty, 
distance uncertainty increases linearly with distance.  Thus, dot size should 
not be considered as, for example, a mass or luminosity indicator. The masers are 
color coded by spiral arms, which have been assigned in part by traces 
of quasi-continuous structure seen in CO and \HI\ Galactic 
longitude-velocity plots, as well as Galactic latitude information.
In Fig. \ref{fig:LV} we overplot the parallax sources on a longitude-velocity
plot of \HI\ emission.
For the majority of cases, these arm assignments are unambiguous.  
For cases with uncertain arm identification, we used all information available, 
including the parallax and kinematic distances (from both radial and proper motions), 
using an updated version of our parallax-based distance estimator (see the Appendix A).

We fitted log-periodic spirals to the locations of the masers, expanding upon the
approach described in \citet{Reid:14}.  We now allow for a ``kink'' in an arm, 
with different pitch angles on either side of the kink.    
The basic form of the spiral model is given by 
$$\ln{(R/R_{kink})} = -(\beta - \beta_{kink}) \tan{\pa}~~,$$
where $R$ is the Galactocentric radius at a Galactocentric azimuth $\beta$
in radians (defined as 0 toward the Sun and increasing in the direction of Galactic
rotation)
for an arm with a radius $R_{kink}$ at a ``kink'' azimuth $\beta_{kink}$.  
An abrupt change in pitch angle, $\pa$, at $\beta_{kink}$ allows for a 
spiral arm to be described by segments, as found in large-scale simulations by 
\citet{DOnghia:13}, who suggest that spiral arms are formed from multiple segments 
that join together.  Kinks are also observed in spiral galaxies, for example, 
by \citet{Honig:15}, who found that arm segments have characteristic lengths
of 5 to 8 kpc, often separated by kinks or gaps.  Since, we currently have 
parallax measurements that trace arms over typically $\lax12$ kpc in length, 
allowing for only one kink is a reasonable simplification.  When assigning kink 
locations we often relied on apparent gaps in the spiral arms.  

In order to extend arm fits into the $4^{th}$ Galactic quadrant 
(awaiting parallax measurements from southern hemisphere VLBI arrays), we
constrain the fitted arms to pass near observed enhancements of CO and \HI\ 
emission at spiral arm tangencies.  Priors for these tangencies are listed in the
notes for Table~\ref{table:pitchangles}.  

As before, we fitted a straight line to 
[$x,y$]=[$\beta,\ln{(R/R_{kink})]}$ using a Markov chain Monte Carlo (MCMC) approach, 
in order to estimate the parameters by minimizing the distance perpendicular to the 
best fit line. Our model included an adjustable parameter giving the (Gaussian $1\sigma$) 
intrinsic width of a spiral arm, 
$w(R) = w(R_{kink}) + (dw/dR)(R-R_{kink})$, where $(dw/dR)= 42$ pc kpc$^{-1}$ 
was adopted from \citet{Reid:14}.  The data were variance weighted by adding in 
quadrature the effects of parallax-distance uncertainty and the component of the 
arm width perpendicular to the arm.  The arm-width parameter $w(R_{kink})$ was adjusted 
along with the other model parameters and resulted in a reduced $\chi^2_\nu$ near 
unity.  

We first weighted the data by assuming uncertainties that had a probability density 
function (PDF) with Lorentzian-like wings, which makes the fits insensitive to 
outliers (see ``conservative formulation'' of \citet{Sivia:06}), in order to identify 
and then remove sources with $>3\sigma$ residuals (see discussions of individual arms
for source names).  Then we re-fitted assuming a Gaussian PDF, and the best-fitting 
parameter values are listed in Table \ref{table:pitchangles}.
Based on this approach, we find evidence for significant kinks in the Norma, 
Sagittarius, and Outer arms, as discussed in Section \ref{sect:armmodel}.
As an example of the improved
quality of fits by adding a kink, we did three fittings for the Sagittarius-Carina arm:
1) solving for a constant pitch angle gave $\chi^2 = 34.1$ for 32 degrees of freedom (dof);
2) solving for different pitch angles about a kink fixed at an azimuth of $52^\circ$
gave $\chi^2 = 30.9$ for 31 dof; and then 3) also solving for the kink azimuth, which 
gave a value of $24^\circ$ with a $\chi^2 = 25.5$ for 30 dof.

\begin{figure}[htp]
\epsscale{0.85} 
\plotone{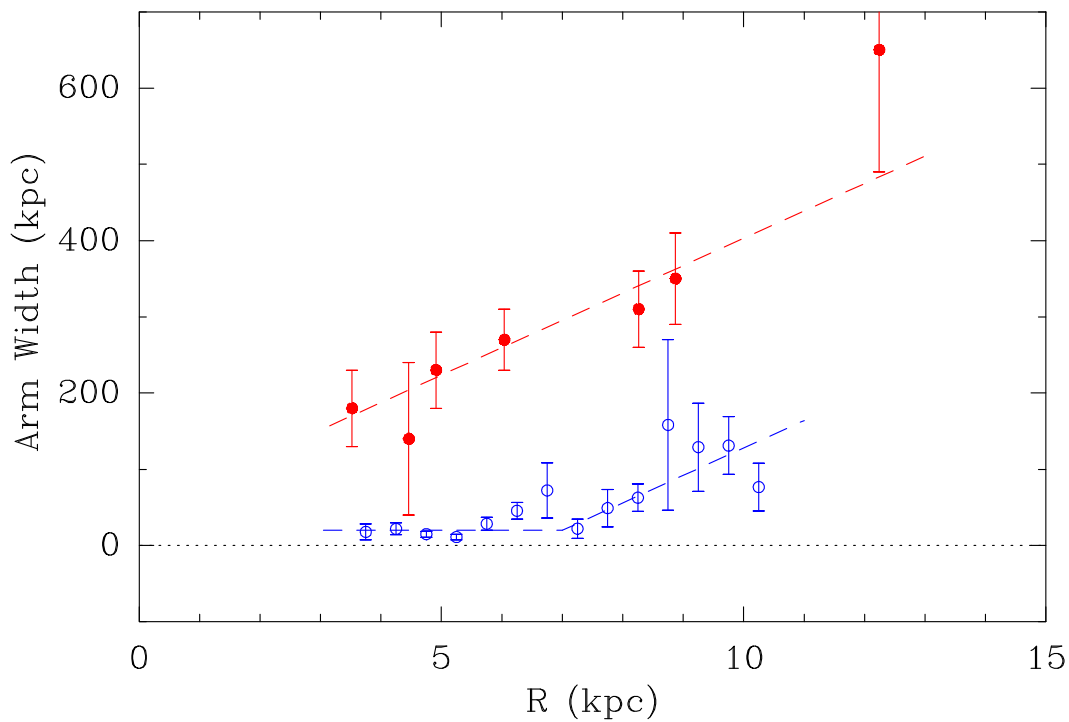}
\caption{\small Widths of spiral arms in the plane ({\it red dots}) and
perpendicular to the plane ({\it blue circles}) as a function
of Galactocentric radius.  The in-plane widths represent averages over the 3-kpc, 
Norma, Scutum, Sagittarius, Local, Perseus and Outer segments 
(in order of increasing average radius).  
The {\it dashed red line} is a best-fit straight line
with a width of 336 pc at 8.15 kpc radius and a slope of 36 pc kpc$^{-1}$.
The out-of-plane widths are rms values after removing the effects of warping.
The {\it dashed blue line} is a ``by-eye'' fit.
        }
\label{fig:armwid}
\end{figure}

The best fitting arm widths in the Galactic plane are plotted versus radius in 
Fig.~\ref{fig:armwid}, extending the radial range and updating the results of 
\citet{Reid:14}.  We find that spiral arms widen with Galactocentric radius as 
$w(R) = 336 + 36( R({\rm kpc}) - 8.15 )$ pc.

\subsection{Distributions Perpendicular to the Plane}

\begin{figure}[htp]
\epsscale{0.85} 
\plotone{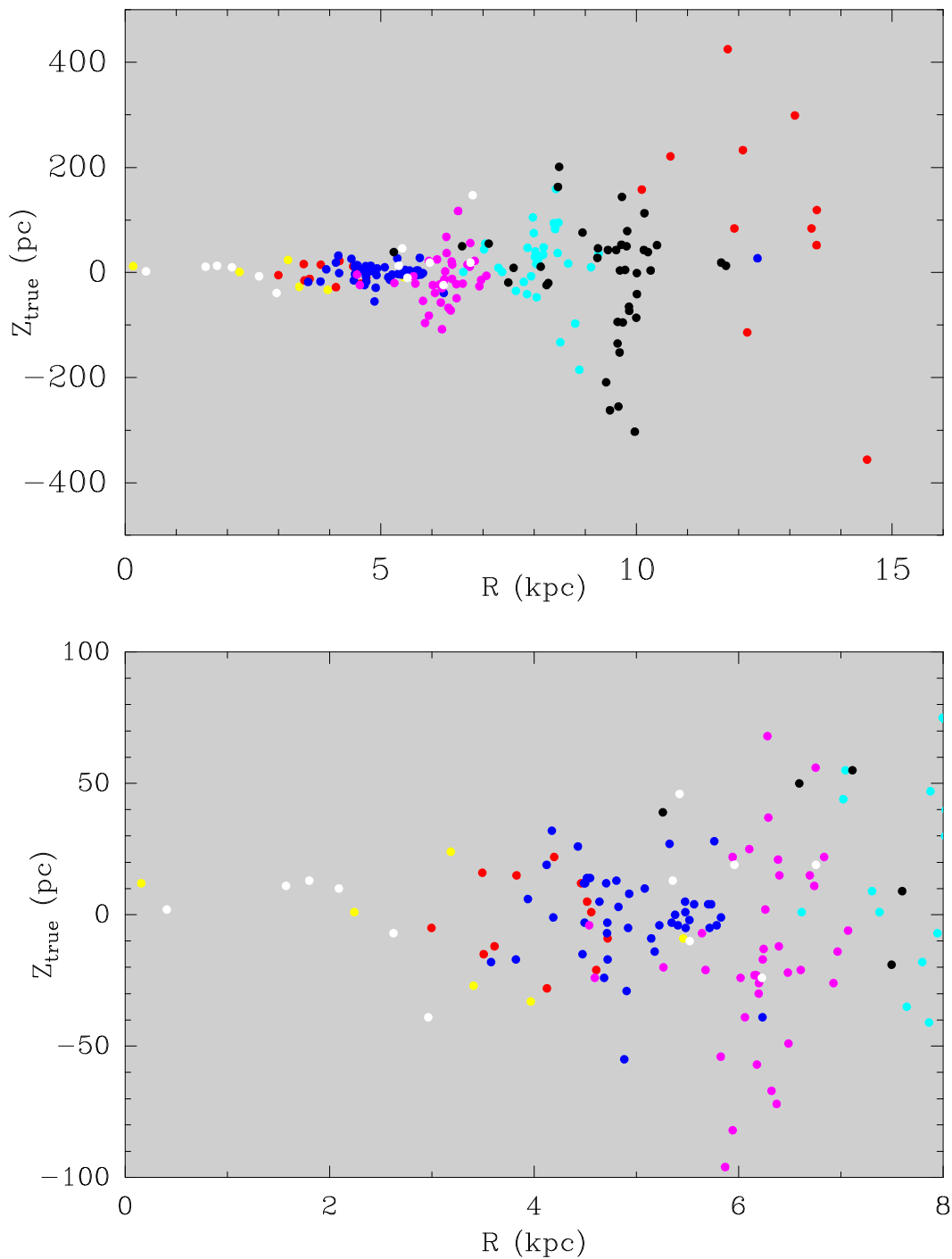}
\caption{\small 
The true $Z$-distribution of young high-mass stars perpendicular
to the Galactic plane as a function of Galactocentric radius.  Sources
are color-coded by spiral arm as in Fig. \ref{fig:parallaxes}, and $Z_{true}$
values are corrected for our view from 5.5 pc above the plane,
assuming the orientation of the IAU plane. 
{\it Top panel:} All stars in Table \ref{table:parallaxes}
with Galactocentric radii less than 16 kpc.
{\it Bottom panel:} A zoomed view of sources within $R=8$ kpc and 
$|Z|<100$ pc.
        }
\label{fig:Zs}
\end{figure}

The distribution perpendicular to the Galactic plane of the young high-mass 
stars with maser emission is shown in the top panel of Fig. \ref{fig:Zs}.
Plotted are true $Z$-heights, adjusted for the Sun's vantage point at 5.5 pc 
above the plane, but with no correction for the warping of the plane.
The increasing scatter in $Z$ beyond $\approx6-7$ kpc from the Galactic center
is a combination of intrinsic scatter and the effects of warping.  
The lower panel of Fig. \ref{fig:Zs} is a zoomed view of the inner 8 kpc.
It reveals a very flat distribution with no signs of corrugations with
amplitudes in excess of $\approx10$ pc.   This is in contrast to \HI\ 
observations toward tangencies in the inner Galaxy by \citet{Malhotra:95},
which suggest corrugations with semi-amplitudes of $\approx30$ pc. 

In order to estimate the widths of arms perpendicular to the Galactic plane,
one needs to deal with the effects of warping.  We did this by smoothing
the $z$-heights along each arm with a window of $\pm20\deg$ of Galactic azimuth
centered on each source.  Provided there were at least five sources within the window,
we then subtracted the variance-weighted mean-$z$.  In order to minimize the effects of 
outliers, we iterated this process, removing $6\sigma$, $5\sigma$, $4\sigma$ and finally $3\sigma$ 
outliers. Finally, with the effects of warping removed, the data from all arms were placed 
into 0.5 kpc bins in radius and rms values calculated.  These rms values are displayed in
Fig.~\ref{fig:armwid}.  These $z$-widths can be reasonably described by 
$\sigma_z(R) = 20 + 36( R({\rm kpc}) - 7.0 )$ pc, for $R>7.0$ kpc, and constant 
at 20 pc inside that radius.

\subsection{Updated Spiral Arm Model}\label{sect:armmodel}

Compared to \citet{Reid:14},
we now have nearly double the number of young, massive stars with maser parallaxes. 
Based on these data, the arms can be more clearly traced and some have been extended.   
Notable additions are spurs between the Local and Sagittarius arms \citep{Xu:16} 
and between the Sagittarius and Scutum arms \citep{Hu18}.  
Fig.~\ref{fig:parallaxes}, in addition to plotting the locations of stars
with measured parallaxes, shows traces of four spiral arms and some arm segments 
and spurs.  We now describe details of the observational constraints used to generate 
individual arms models.

\subsubsection{Norma-Outer Arm}\label{sect:Norma-Outer}

The Norma arm in the $1^{st}$ quadrant displays a quasi-linear, spur-like structure, 
as seen in Fig.~\ref{fig:parallaxes}, starting at $(X,Y) = (3,2)$ kpc near the end 
of the bar and extending to about $(2,5)$ kpc at Galactic azimuth of $\approx18\deg$.
This segment of the arm has a large pitch angle of $\approx20\deg$.  Proceeding counter 
clockwise from that azimuth, in order to pass through the observed tangency in the $4^{th}$ 
quadrant at $\ell\approx328\deg$, the pitch angle of this segment must be near zero.  
Most likely the Norma arm wraps around the far side of
the Galactic center and becomes the Outer arm as shown in Fig.~\ref{fig:parallaxes},
where we have adjusted their pitch angles to join them together.   
Were the Norma arm instead to connect to the Perseus arm, it would have to have a 
negative pitch angle over a large azimuthal range, starting at (or beyond) the Norma tangency 
in the $4^{th}$ quadrant.   
Alternatively, were it to connect to the Outer-Scutum-Centaurus arm, it would need a very 
large pitch angle starting at (or beyond) the tangency.  While neither of these 
possibilities seem likely, in order to rule them out parallaxes to some Norma-Outer 
arm sources beyond the Galactic center are needed.  Note, we removed one outlier source 
(G001.15$-$0.12) when fitting spiral segments.

The Outer arm appears to have a kink in the $2^{nd}$ quadrant near Galactocentric 
azimuth $18\deg$ with a near-zero pitch angle as it proceeds into the $3^{rd}$ quadrant. 
The motivation for such a kink is based in part on a revised distance for S269.  
A new and robust BeSSeL Survey distance to S269 of $4.15\pm0.22$ kpc, from a 16-epoch 
set of observations \citep{Nunez:19}, resolves the difference in parallax between 
VERA results by \citet{Honma:07} and \citet{Asaki14} in favor of the latter result.    
With this new information, it appears likely that two other sources 
(G160.14+03.15 and G211.59+01.05)
previously thought to lie between the Perseus and Outer arm, are likely part of an 
Outer arm segment which does not continue the $9\d4$ pitch angle from the 
$1^{st}$ and $2^{nd}$ quadrants into the $3^{rd}$ quadrant.  This revision of the 
structure of the Outer arm is shown in Fig.~\ref{fig:parallaxes}.

\subsubsection{Scutum-Centaurus-OSC Arm}\label{sect:OSC}

The Scutum arm may originate near $\beta\approx90\deg$ in the $1^{st}$ quadrant,
and then wind counter-clockwise into the $4^{th}$ quadrant as the Centaurus arm
with a tangency at $\approx306\deg$.  Over this large range of azimuth, one can
reasonably fit a spiral with a pitch angle of $\approx13\deg$, since fitted
pitch angles around a potential kink near $23\deg$ azimuth are statistically
consistent (see Table~\ref{table:pitchangles}).  The fits in the Table excluded six sources 
(G030.22$-$0.17, G030.41$-$0.23, G030.74$-$0.04, G030.81$-$0.05, G030.97$-$0.14, G031.41$+$0.30),
which have unusually large proper motions \citep{Immer18} and may be located in
a spur-like structure.

As discussed by \citet{Sanna17}, extending this model beyond the Galactic center, it 
passes near the parallax source G007.47+00.05 and back into the $1^{st}$ quadrant as the
``Outer Scutum Centaurus'' (OSC) arm \citep{Dame:11,Sun:15}.
The association of the Scutum-Centaurus and OSC arm segments is pivotal information
for a complete picture of the Milky Way, since it provides strong motivation for 
connecting the Norma and Outer arms, as both are directly interior to the 
Scutum-Centaurus-OSC arm.  Were we not to connect the Scutum-Centaurus and OSC arm 
segments, it would require adding a fifth spiral arm tightly packed in a region 
that is difficult to measure.  Occam's razor suggests that we adopt the 4-arm approach.

\subsubsection{Sagittarius-Carina Arm}

In Fig.~\ref{fig:parallaxes}, the Sagittarius arm has a 2-kpc long gap centered at 
azimuth $\beta\approx24\deg$ near $(X,Y) = (2,6)$ kpc.  When fitting
spiral segments, we removed two outliers (G032.74$-$0.07, G033.64$-$0.22). 
We found the best-fitting pitch angle for sources
with $\beta>24\deg$ to be nearly zero (\ie\ $\psi=1\d0\pm2\d1$). 
The arm segment for $\beta<24\deg$ in the $1^{st}$ quadrant to the $4^{th}$ quadrant 
tangency near $\beta=-33\deg$  appears to have a pitch angle of 
$\approx17\deg$, based on the locations of parallax sources as well as its 
$4^{th}$ quadrant tangency.
This results in a significant kink, which was first suggested in a prescient
paper by \citet{Burton:70} and has appeared in models of the Milky
Way, such as in the electron density model of \citet{Taylor:93}.  
Beyond the $4^{th}$ quadrant tangency, where we currently have no parallax
information, we decrease the pitch angle to $10^\deg$ in order to match the
CO $(l-v)$ trace of the Carina arm, assuming kinematic distances (see Fig. \ref{fig:LV}).
Invoking symmetry with the Norma arm, we extend the Sagittarius arm inward to the
Galactic bar near $(-3,-3)$ kpc.

\subsubsection{Perseus Arm}

The Perseus arm has been thought to be one of two dominant spiral arms 
(along with the Scutum-Centaurus arm) of the Milky Way \citep{Drimmel:00,Churchwell:09}.  
While the arm has a large number of massive star forming regions in the $2^{nd}$ quadrant, 
it appears to weaken and possibly die out in the $3^{rd}$ quadrant \citep{Koo:17}.  
Also, spiraling inward and through the $1^{st}$ quadrant, there is a clear decrease 
in active star formation \citep{Zhang:13,Zhang18} extending for about 8 kpc along 
the arm between longitudes of $90^\circ$ and $50^\circ$.  
Spiral pitch angles in each quadrant are similar and near $9\deg$.  
Extrapolating the arm with this pitch angle inward into the $4^{th}$ quadrant, 
it passes radially near but outside the bar and may originate near 
$(X,Y)=(-3.5,-0.5)$ kpc as depicted in Fig.~\ref{fig:parallaxes}.  
If this picture is correct, the Perseus arm is not a dominant arm as measured
by high-mass star formation activity over most of its length.  

\subsection{Other Arm Segments}

The Local arm has been mapped in detail by \citet{Xu:13,Xu:16}.  It appears
to be an isolated arm segment with very significant massive star formation.
However, it has not linked up with other segments to form a ``true'' arm,
like those listed above, which can wrap fully around the Galaxy.  Since we 
have only just begun to trace the 3-kpc and Connecting arms with parallax
measurements, their natures are not yet well defined.  However, these ``arms''
are associated with the Galactic bar and may not be true spiral arms.

%=========================================================================
\section{Modeling the Galaxy} \label{sect:modeling}

In order to estimate the distance to the Galactic center and rotation curve
parameters, we used the Bayesian MCMC approach described in \citet{Reid:14}.
We treat as data the three-dimensional components of velocity and model these
as arising from an axisymmetric Galactic rotation, with allowance for an
average streaming (non-circular) motion in the plane of the Galaxy.  
Previously, we adopted the ``Universal'' form for the rotation curve (URC) 
of \citet{Persic:96}, since it fitted the data as well or better than other 
rotation curves, and it well models the rotation of a large number of external 
spiral galaxies.  In our previous paper, we used a three-parameter formulation 
for the URC: $a1=V(R_{opt})$, $a2=R_{opt}/\Ro$, and $a3=1.5(L/L^*)^{0.2}$,
where $R_{opt}$ and $V(R_{opt})$ are the radius  enclosing 83\% of the optical light
and the circular velocity at that radius for a galaxy with luminosity $L$ relative 
to an $L^*$ galaxy with $M_b = -20.6$ mag.  However, Persic, Salucci and Stel, in their 
note added in proof, present a simplified two-parameter formulation with 
parameter $a1$ removed via scaling relations between optical radius, velocity 
and luminosity.  Since the two- and three-parameter formulations produce similar 
rotation curves over radial ranges of 0.5 to 2.0 $R_{opt}$, we now adopt the 
two-parameter version, with the rotation curve defined by only parameters $a2$ 
and $a3$.  (See Appendix B for a {\footnotesize FORTRAN} subroutine that
returns a circular rotation speed for a given Galactocentric radius.)

As in \citet{Reid:14}, given the current uncertainty in the value for the circular component 
(\V) of solar motion and the magnitude of the average peculiar motions of masers
associated with massive young stars, we present fits with four sets of priors:
\begin{itemize}
\item[A)]{} Adopting a loose prior for the \V\ component of solar motion,
$\U = 11.1\pm1.2$, $\V = 15\pm10$, $\W = 7.2\pm1.1$ \kms, and for the average 
peculiar motion for the masing stars of $\Usbar = 3\pm10$ and $\Vsbar =-3\pm10$ \kms.
\item[B)]{} Using no priors for the average peculiar motion of the stars, but a 
tighter prior for $\V = 12.2\pm2.1$ \kms\ from \citet{Schoenrich:10}.
\item[C)]{} Using no priors for the solar motion, but tighter priors on the 
average peculiar motion of the stars of $\Usbar = 3\pm5$ and $\Vsbar =-3\pm5$ \kms.
\item[D)]{} Using essentially no priors for either the solar motion or 
average peculiar motion of the stars, but bounding the \V\ and \Vsbar\ parameters
with equal probability within $\pm20$ \kms\ of the set-A values and zero
probability outside that range.
\end{itemize}

Since we expect large non-circular motions in the vicinity of the Galactic bar,
we removed the 19 sources that are within 4 kpc of the Galactic center.
Next, we removed 22 sources whose fractional parallax uncertainties exceeded 20\%,
in order to avoid significant complications arising from highly asymmetric
PDFs when inverting parallax to estimate distance \citep[\eg][]{Bailer-Jones:15}{}.
Finally, as discussed in \citet{Reid:14}, we expect some outliers in the 
motion data, owing for example to the effects of super-bubbles that can accelerate
gas which later forms the stars with masers which we observe.   
Therefore, we used the ``conservative formulation'' of \citet{Sivia:06}, 
which uses a ``Lorentzian-like'' PDF for motion uncertainties when fitting a 
preliminary model of Galactic rotation to the data.  The best-fitting parameters are 
listed in Table~\ref{table:fits} in column A1.  This fit is largely insensitive to 
outliers, and allows us to identify and remove them in a prescribed and objective 
manner. Defining an outlier as having greater than a $3\sigma$ residual in any
motion component, we removed 11 sources from further consideration\footnote{
G015.66$-$00.49, G029.86$-$00.04, G029.98$+$00.10, G030.19$-$00.16, G030.22$-$00.18,
G030.41$-$00.23, G030.81$-$00.05, G032.74$-$00.07, G078.12$+$03.63, G108.59$+$00.49,
G111.54$+$00.77}. 

With the resulting ``clean'' data set of 147 sources, we assumed a Gaussian
PDF for the data uncertainties (equivalent to least-squares fitting) and
the best fitting parameter values are given in Table \ref{table:fits}
in column A5.  Note that we present estimates of \To\ in Table \ref{table:fits},
calculated from the rotation curve, even though it was not a fitted parameter.
We also combine parameters to generate the marginalized PDF for the full circular 
rotation rate of the Sun in in linear, (\To+\V), and angular, (\To+\V)/\Ro, units.  
We adopt the A5 fit as the best model.  However, for completeness we follow 
\citet{Reid:14} and also present model fits using the different priors listed
above.  Table~\ref{table:fits} summarizes the best-fitting parameter estimates for
priors B, C and D in the last three columns, which yield parameter values similar to 
those for fit A5. Note that \citet{Nunez:17} have shown that fits of mock datasets  
demonstrate that parameter estimates should be unbiased.

The correlation between parameters \Ro\ and \To\ in fit A5 is modest: 
$r_{\Ro,\To}=0.45$.  This is similar to our previous value reported in \citet{Reid:14},
as well as found in simulations by \citet{Nunez:17},
since it depends on the range and distribution of parallax sources which has not
significantly changed.   Parallax measurements from a VLBI array in the southern 
hemisphere are needed to further reduce this correlation.  As we have previously
noted, \To, \V\ and \Vsbar\ can be highly correlated, hence the need for 
priors for some of these parameters.  For the set-A priors, we find the
following correlations: $r_{\To,\V}=-0.74$, $r_{\To,\Vsbar}=+0.74$, and
$r_{\V,\Vsbar}=-0.99$.

\section{Peculiar Motions}\label{sect:peculiar_motions}

\begin{figure}[htp]
\epsscale{0.85} 
\plotone{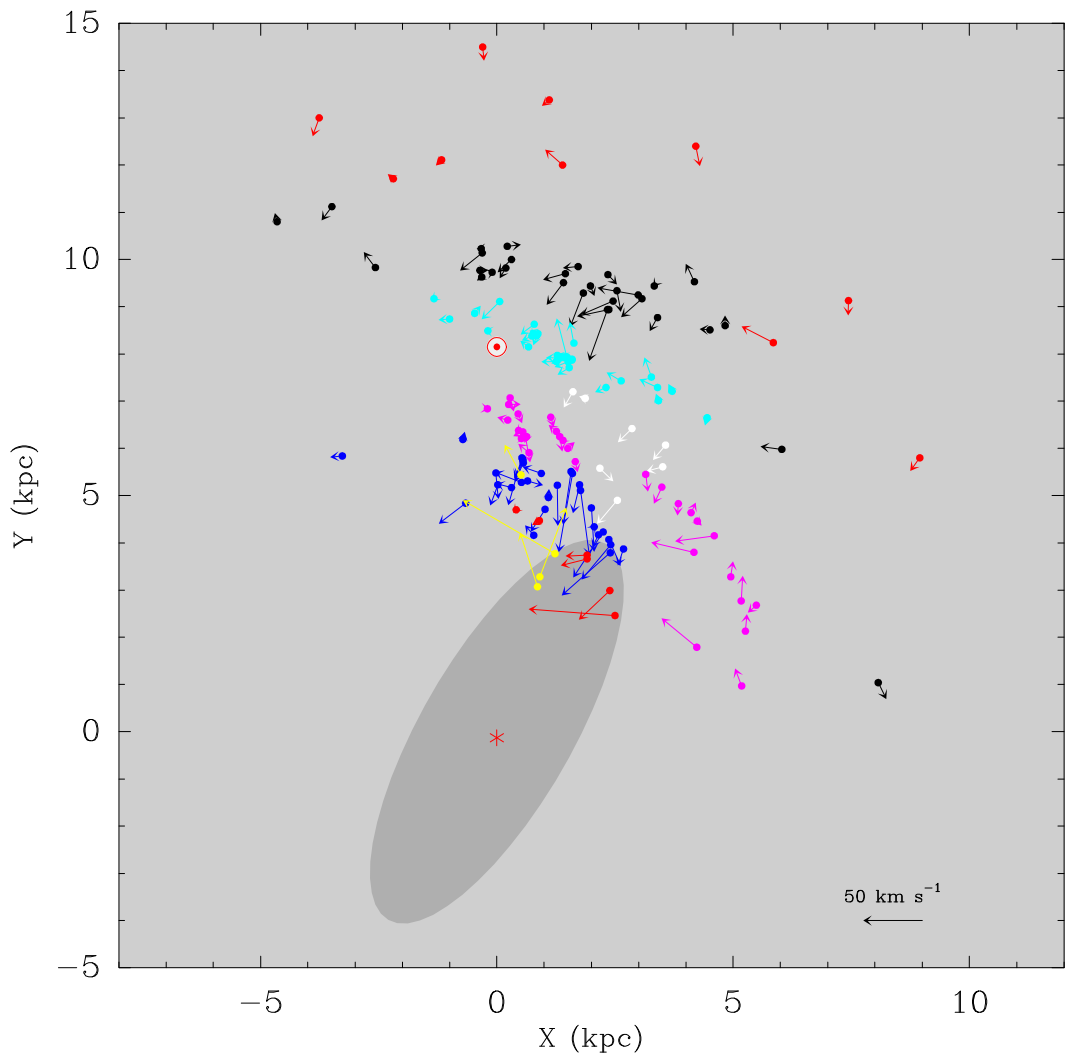}
\caption{\small 
Non-circular (peculiar) motions of massive young stars in the plane of the Galaxy, 
adopting fit-A5 values for Galactic rotation and solar motion, 
without removing average streaming motions (\ie\ setting $\Usbar=\Vsbar=0$).   
Only sources with vector uncertainties $<20$ \kms\ are shown.
A 50 \kms\ scale vector is shown in the lower-right corner.  Sources are color coded by
arm as in Fig. \ref{fig:parallaxes}.   A schematic ``long'' bar from \citet{Wegg:15}
is indicated with the shaded ellipse.  
        } 
\label{fig:peculiar_motions}
\end{figure}

Using the Galactic parameters and solar motion values from fit A5, we can transform
to a reference frame that rotates as a function of radius within the Galaxy.  The
resulting non-circular or peculiar motions in the plane of the Galaxy are shown 
in Fig. \ref{fig:peculiar_motions}.  We only plot sources whose motion uncertainties are 
$<20$ \kms.  While the vast majority of sources have moderate peculiar motions of 
$\lax10$ \kms, we can identify two regions in the Galaxy with significantly larger 
peculiar motions.  The first anomalous region is in the Perseus arm at Galactic longitudes 
between about $105\deg$ and $135\deg$.  This anomaly has been well documented in the
literature \citep[\eg][]{Humphreys:78,Xu:06,Sakai19}.  The second anomalous region 
is within a Galactocentric radius of $\approx5$ kpc.  Large peculiar motions are seen near 
the end of the long bar for sources in the 3-kpc, Norma, and Scutum arms.

The average peculiar motion of our sources for all fits in Table \ref{table:fits}
indicate small positive values for \Usbar\ (toward the Galactic center) 
and small negative values for \Vsbar\ (in the direction of Galactic rotation), 
possibly resulting from streaming motions of massive young stars
toward the Galactic center and counter to Galactic rotation.  
Adopting fit A5, which has loose priors for the Solar Motion component 
in the direction of Galactic rotation ($\Vo=15\pm10$ \kms) and for \Usbar\ and 
\Vsbar\ ($3\pm10$ and $-3\pm10$ \kms, respectively), we find $\Usbar = 6.0\pm1.4$ and 
$\Vsbar=-4.3\pm5.6$ \kms.  The $\Usbar$ value appears significant and qualitatively
consistent with theoretically expected values for the formation of stars from gas which
was shocked when entering a spiral arm of low pitch angle \citep[\eg][]{Roberts:69}.

%=======================================================================
\section{The Galactic Plane}\label{sect:GalacticPlane}

Given a source's distance and Galactic coordinates, we can calculate its 
3-dimensional location in the Galaxy.  Since one expects massive young stars to be
distributed very closely to the plane, this offers an opportunity to
refine the parameters of the IAU-defined Galactic plane, as recently investigated
by \citet{Anderson:19}.  Here we fit for the Sun's location perpendicular to the 
plane and a 2-dimensional tilt of the true plane with respect to the IAU-defined plane.  
Then, with improved parameters, we estimate the scale height of our sources.

\begin{figure}[ht]
\epsscale{1.00} 
\plottwo{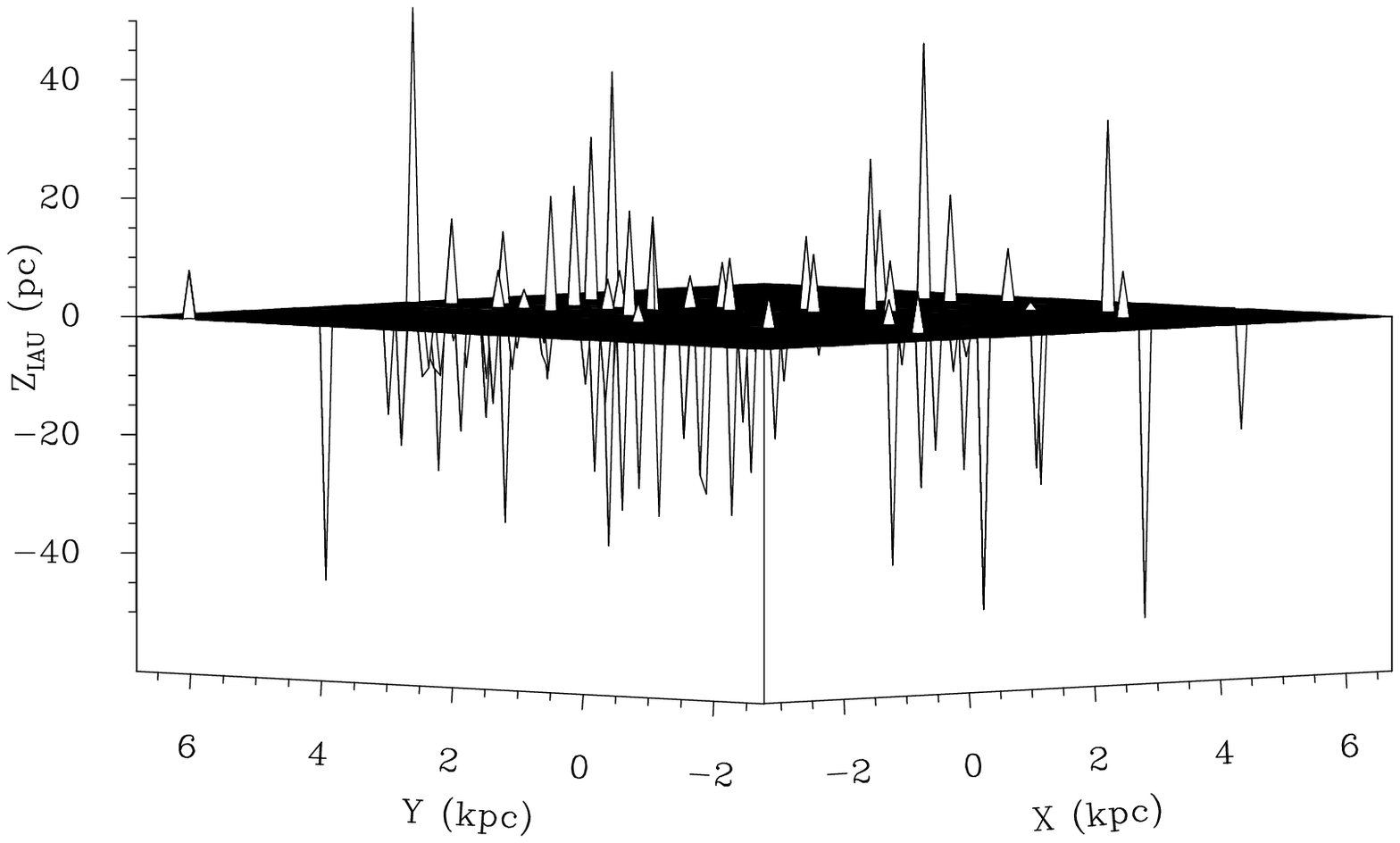}{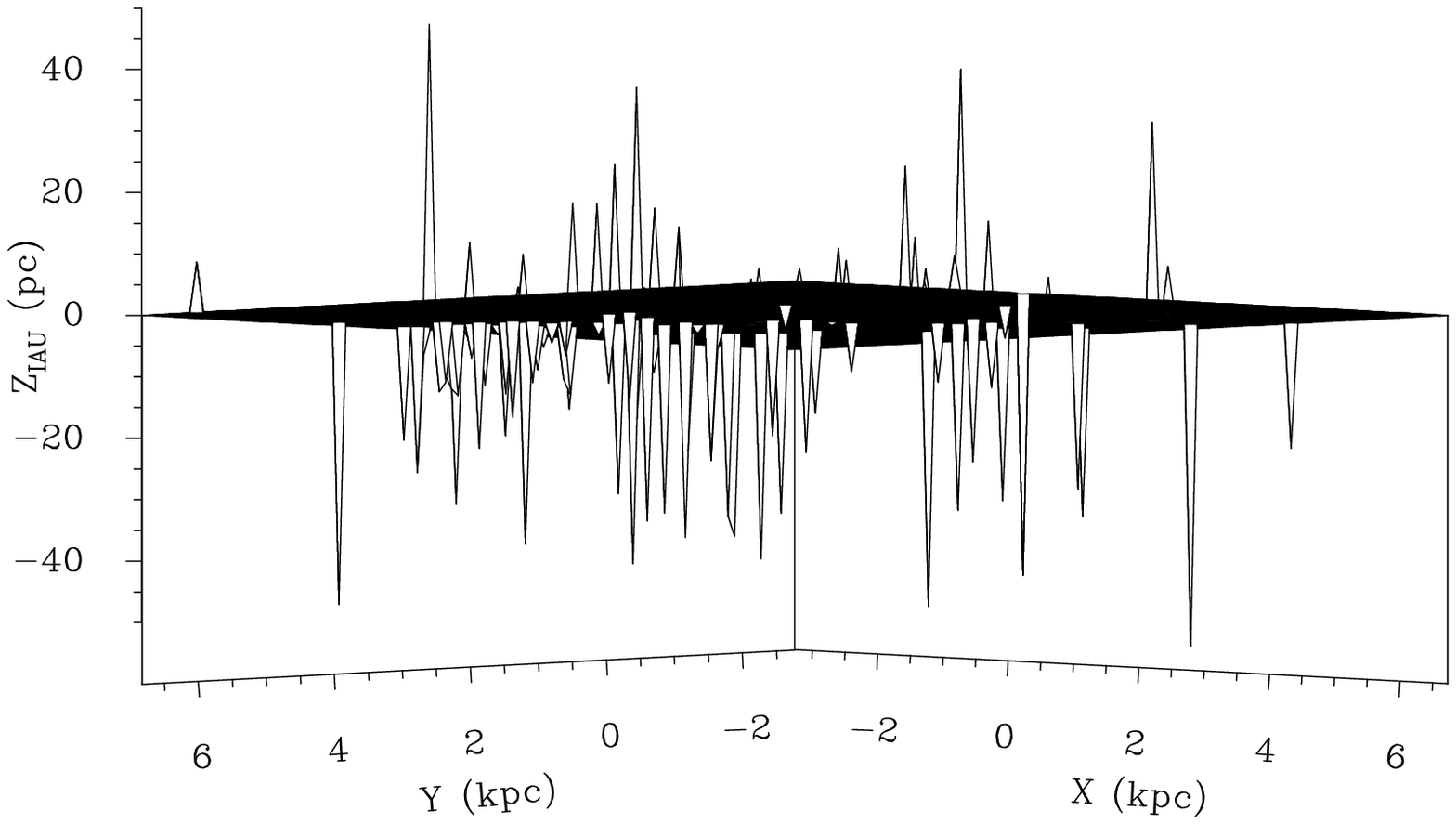}
\caption{\small Perspective plots of the 3-dimensional locations 
({\it cone tips}) of massive young stars with respect to the 
IAU-defined Galactic plane.  Only 96 sources within 7 kpc of the center 
are shown to avoid regions of Galactic warping.  
Note the \ZIAU\ scale is in pc, whereas the $X$ and $Y$ scales are kpc. 
{\it Left} is a view from 3 degrees above the plane and {\it right} is
from 3 degrees below the plane.  Two views are shown to better display the
slight asymmetry in \ZIAU\ favoring negative values, owing to our viewpoint
from the Sun which is above the plane.
        }
\label{fig:IAU_Zs}
\end{figure}

We define Cartesian Galactocentric coordinates $(X,Y,Z)$, where $X$ is 
distance perpendicular to the Sun-Galactic center line (positive in 
$1^{st}$ and $2^{nd}$ quadrants), $Y$ is distance from the Galactic center toward 
the Sun, and $Z$ is distance perpendicular to the plane (positive toward the north 
Galactic pole).  In order to avoid complications of Galactic warping, 
we follow \citet{Gum:60} and  \citet{Blaauw:60} and 
restrict the range of sources fitted to within a Galactocentric radius of 
7.0 kpc, giving a sample of 116 sources.  
Anticipating a scale height for massive young stars near 20 pc (see below), 
we also remove possible ``outlying'' sources more than 60 pc from the plane.  Some
outliers would be expected if gas compressed and accelerated by superbubbles 
forms high-mass stars.  
With these restrictions, we retained 96 massive young stars and plot their 
3-dimensional locations in the Galaxy relative to the IAU-defined plane in 
Fig.~\ref{fig:IAU_Zs}.  

These stars are very tightly distributed in a plane.  However, the distribution about  
the IAU-defined plane is slightly asymmetric, with two-thirds (65) of the sources 
lying below that plane.  The mean offset is $-7.3$ pc with a standard error 
of the mean of $\pm2.1$ pc.  In general, this can be explained by a combination
of the Sun being offset from the true Galactic plane and/or a tilt of the true 
plane from the IAU plane.  Hence, we fitted an adjusted plane to these data.
We treat the observed ($\ZIAU$) values as data and model them as follows:
$$\ZIAU = -\Zsun + \Dp~\cos l~\tan\Ytilt + \Dp~\sin l~\tan\Xtilt~~,$$
where \Zsun\ is the Z-offset of the Sun, \Dp\ is the distance
of a source from the Sun projected in the plane, $l$ is Galactic
longitude, and \Xtilt\ and \Ytilt\ are tilt angles of
the true plane with respect to the IAU-defined plane (referred to as ``roll'' 
and ``tilt,'' respectively, by \citet{Anderson:19}).  
The $\ZIAU$ data were fitted using an MCMC technique, accepting or
rejecting trials with the Metropolis-Hastings algorithm.  Data uncertainties 
were based on parallax uncertainties and assumed to be Gaussian.  
In order to minimize the effects of distance biases, 
owing to asymmetric distance PDFs when converting parallaxes to distances 
\citep[\eg][]{Bailer-Jones:15}, we used only the 80 sources with fractional parallax 
uncertainties less than 20\%.

\begin{figure}[htp]
\epsscale{0.85} 
\plotone{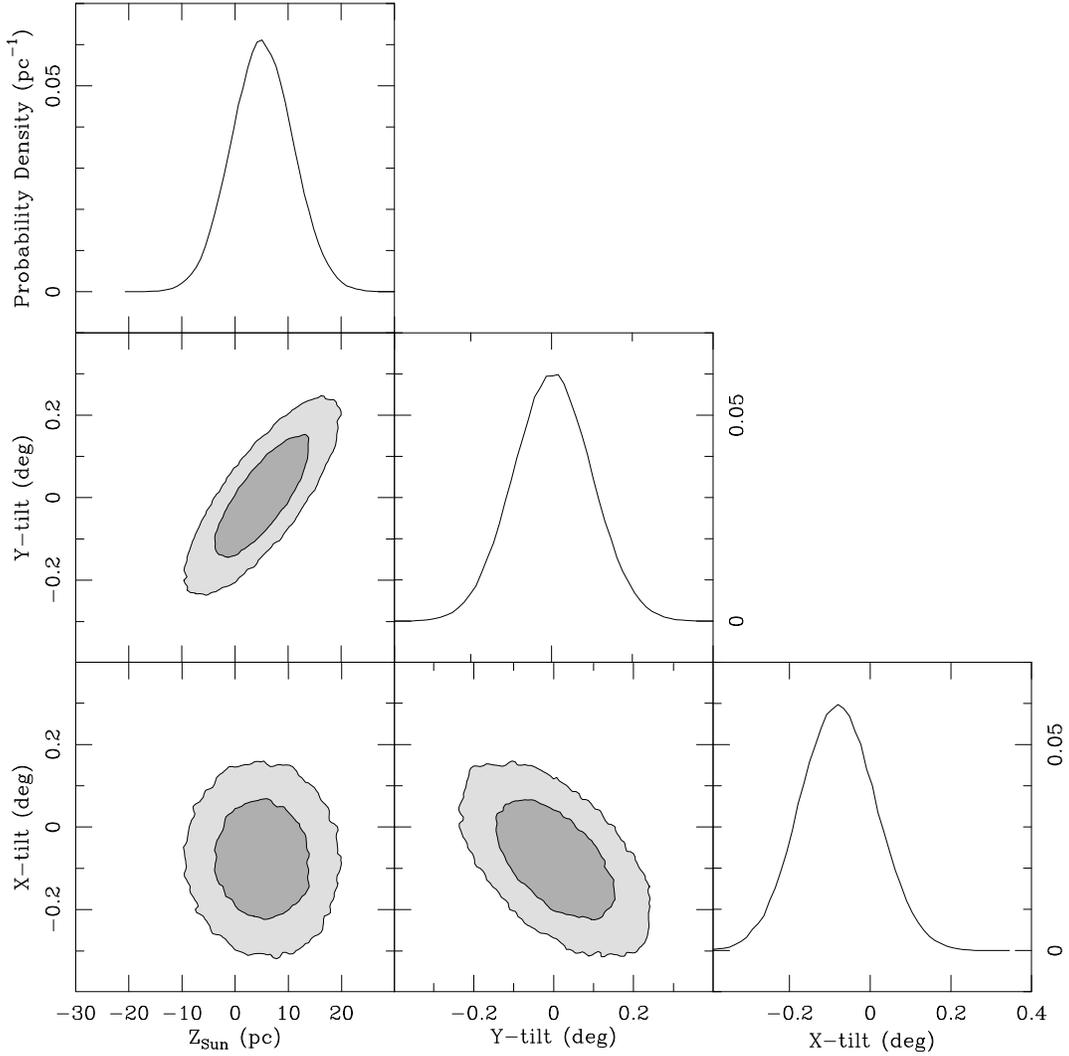}
\caption{\small
Results of fitting three parameters defining a tilted plane to the 
Galactic $Z$-heights of massive young stars within 7 kpc of the Galactic Center.  
Plotted are 2-dimensional 68\% and 95\% confidence contours based on the 
MCMC trial parameter values and their 1-dimensional marginalized
probability densities.  Z$_{\rm Sun}$ is the offset of the Sun from the
true plane toward the north Galactic pole; $\Xtilt$ and $\Ytilt$ are 
rotation angles of the true plane with respect to the IAU-defined plane
about lines toward the Galactic center and toward $90\deg$
Galactic longitude, respectively. 
        }
\label{fig:plane_pdfs}
\end{figure}

Fig. \ref{fig:plane_pdfs} displays the MCMC trials and marginalized PDFs for the 
three parameters.  The centers of the 68\% confidence ranges for the marginalized 
PDFs give the following parameter estimates:
$\Zsun = 5.5 \pm 5.8$ pc, $\Xtilt =-0\d08 \pm 0\d10$, and
$\Ytilt = 0\d00 \pm 0\d10$. Thus, we find very small changes in the 
orientation (``tilt'' parameters) of the true Galactic plane relative to the 
IAU-defined plane.  \citet{Anderson:19} find similarly small values of
$\Xtilt$ (between $-0\d04$ and +0\d15) and $\Ytilt$ (between $-0\d11$ and 
+0\d08) for different sub-samples.  Therefore, for simplicity, in the
following we assume the orientation of the plane follows the IAU definition,
but we correct for the Sun's $Z$-height above the plane.

\begin{figure}[htp]
\epsscale{0.65} 
\plotone{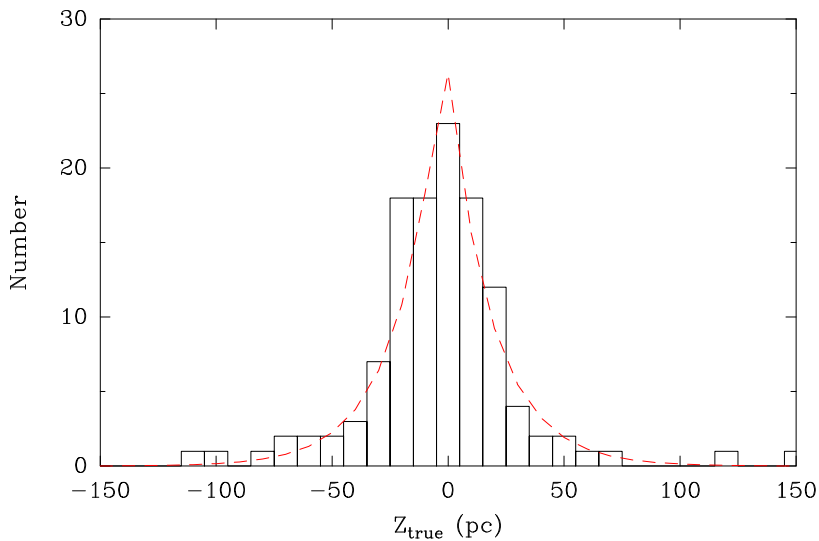}
\caption{\small Histogram of distances perpendicular to the ``true''
Galactic plane after correcting for the Sun's location of 5.5 pc
above the plane.  In order to avoid the effects of warping, only sources
within 7 kpc of the Galactic center are plotted.  The {\it dashed red line}  
is the best fitting exponential distribution with a scale height of 19 pc.
        }
\label{fig:z_histogram}
\end{figure}

Placing the Sun 5.5 pc above the IAU-defined plane, we calculate $Z$-offsets 
relative to the true (adjusted) Galactic plane, which we designate $Z_{true}$.  
Fig.~\ref{fig:z_histogram} presents a binned histogram of $Z_{true}$ values for 
our full sample of 120 massive young stars within 7.0 kpc of the
Galactic center and 200 pc of the plane, along with the best fitting exponential 
distribution.  The exponential distribution fits the data very well and gives a 
scale height of $19\pm2$ pc.

%===============================================================================
\section{Precession of the Plane} \label{sect:precession}

We now investigate the possibility that the Galaxy precesses, possibly
owing to torques from its triaxial halo and/or from Local Group galaxies.
A rigid-body precession of the plane could be inferred
from the z-component of velocity of our massive young stars, \Ws, following 
$$\Ws = W_1^x~(X/\Ro) + W_1^y~(Y/\Ro) - \W~~,~\eqno{(1)}$$
where $W_1^x$ and $W_1^y$ are the speeds of vertical motion at radius \Ro\
and location in the plane given by $X$ and $Y$.  We used the estimate of \citet{Schoenrich:10}
of the vertical motion of the Sun relative to the Solar Neighborhood of 
$\W=7.2\pm0.6$ \kms\ as a strong prior.  After removing \Ws\ measurements with 
uncertainties $>10$ \kms, we fitted the data using a Bayesian MCMC procedure similar 
to that described in Section \ref{sect:modeling}, first using the 
``conservative formulation'' of \citet{Sivia:06} to identify and remove 20 outliers 
and then refitting with Gaussian data uncertainties.   
Using sources at all Galactocentric radii, we find 
$W_1^x = 2.0\pm1.6$ \kms, $W_1^y = -0.1\pm0.8$ \kms, and $\W = 7.9\pm0.5$ \kms.  
Dividing the data into inner and outer regions,   
we find for $R<7$ kpc
$W_1^x = 1.2\pm2.1$ \kms, $W_1^y = -2.5\pm1.4$ \kms, and $\W = 7.4\pm0.6$ \kms; 
and for $R>7$ kpc 
$W_1^x = 3.1\pm2.2$ \kms, $W_1^y = +0.7\pm1.0$ \kms, and $\W = 7.6\pm0.6$ \kms.
The estimates of $W_1^y$ for the inner and outer Galaxy differ by
$3.2\pm1.7$ \kms\ with nearly $2\sigma$ significance, hinting at the possibility
of a radial dependence.

The Galaxy is known to exhibit significant warping, starting between about 7 and 8 kpc 
radius and reaching an amplitude of about 300 pc at a radius of 12 kpc \citep{Gum:60}.
Since the warping is likely dynamic in origin, one would expect vertical
motions leading to the warp to vary with radius.  Were the warping at 12 kpc radius
to develop over a time scale of $\sim10^8$ years (roughly one-third of an orbital period
at that radius), that would correspond to a characteristic vertical speed of
$\sim3$ \kms.  In order to investigate this possibility, we added second order terms 
to Eq.~(1) yielding
$$\Ws = W_1^x~(X/\Ro) + W_2^x~(X/\Ro)^2 + W_1^y~(Y/\Ro) + W_2^y~(Y/\Ro)^2 - \W~~.~\eqno{(2)}$$Re-fitting with no radial restriction, we find 
$W_1^x =-1.7\pm3.7$ \kms, $W_2^x =5.4\pm4.8$ \kms, $W_1^y = -0.9\pm1.2$ \kms, 
$W_2^y = 1.4\pm1.3$ \kms, and $\W = 7.8\pm0.5$ \kms.
Neither the first- nor second-order parameters differ significantly from zero, 
and for the $y$-terms reasonable upper limits for their magnitudes 
(and hence for vertical motions at $R=\Ro$) are $\approx4$ \kms.

%======================================================================
\section{Discussion }   \label{sect:discussion}

\subsection{An Updated View of the Milky Way}

Is the Milky Way a two- or four-arm spiral?  Based on the model described in Section 
\ref{sect:armmodel}, the answer is a four-arm spiral as traced by massive young stars.  
The four arms do not include sub-structures such as the 3-kpc arms, since they are 
likely a phenomenon associated with the bar, and the Local arm, which appears to be an 
isolated segment.  Were the Milky Way a two-arm spiral, its arms would have to wrap twice 
around the center in order to accommodate the parallax data.  This would require pitch 
angles to average near 5\deg.  However, based on the data in Table \ref{table:pitchangles} 
and weighting the pitch angles by their segment lengths, we find an average pitch angle 
of 10\deg\ for the major arms (Norma-Outer, Sct-Cen-OSC, Sgr-Car, and Perseus).

\begin{figure}[htp]
\epsscale{0.85} 
\plotone{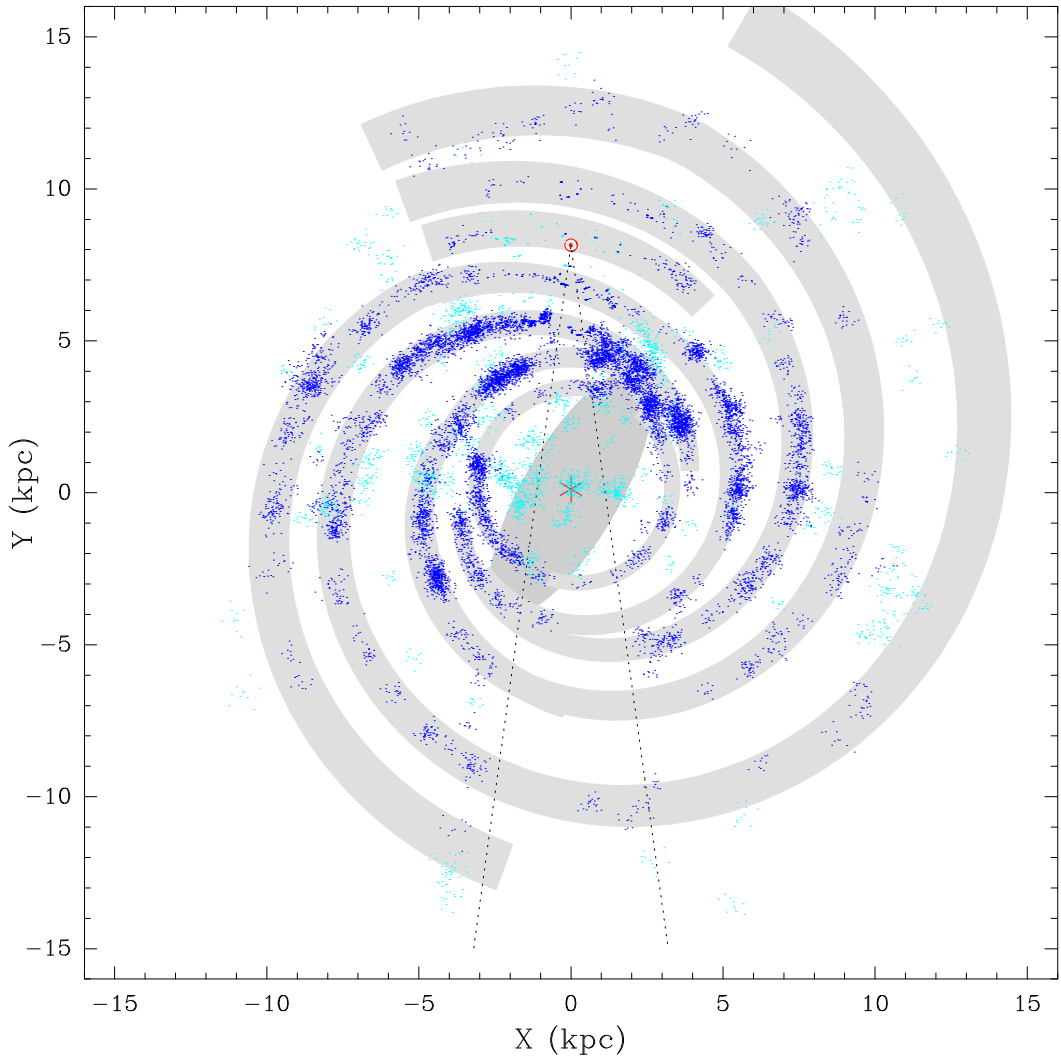}
\caption{\small 
Plan view of the Milky Way showing the locations of high-mass star forming regions 
estimated using version 2 of the parallax-based distance estimator of \citet{Reid:16}.  
Sources for which the distance estimator could reasonably assign to an arm are shown in blue,
otherwise they are in cyan.  The Galactic center ({\it red asterisk}) is at (0,0) and 
the Sun ({\it red Sun symbol}) is at (0,8.15) kpc.  The wedge-shaped region between 
{\it dashed lines} suffers from incompleteness as discussed in the text.
The background grey spirals are based on 
the parameters in Table~\ref{table:pitchangles}, with small adjustments to allow arm segments to 
connect smoothly beyond the Galactic center, and have widths of $1.65\sigma$, which would 
enclose 80\% of sources assuming a Gaussian distribution perpendicular to the arms.  
The ``long'' bar is indicated with the shaded ellipse.  
        } 
\label{fig:Bayesian_map}
\end{figure}

The form of spiral arm segments obtained in Section \ref{sect:armmodel} can be used to 
update the input model for the parallax-based distance estimator of \citet{Reid:16} 
(see Appendix A for details).
Coupling our revised model of spiral arm locations with line-of-sight velocities that trace
arms in CO or \HI\ emission in longitude-velocity plots, we arrive at a revised spatial-kinematic
model for arms.  Incorporating the revised arm model into our parallax-based distance estimator,
we can now take longitude-latitude-velocity values from surveys of spiral-arm tracers (\eg\ HII
regions, molecular clouds, star-forming masers) and estimate distances more accurately than 
from standard kinematic distances.  
Taking as input the catalogs of water masers from \citet{Valdettaro:01}, 
methanol masers from \citet{Pestalozzi:05} and \citet{Green:17},
\HII\ regions from \citet{Anderson:12}, and red {\it MSX} sources from \citet{Urquhart:14}, 
we now construct an improved visualization of the spiral structure of the Galaxy shown in 
Fig.~\ref{fig:Bayesian_map}.   

While the locations of spiral arms can be obtained from a sample of maser parallaxes that is
far from complete, the input catalogs for our visualization of the Milky Way have
better defined samples and are more complete.  For example, the red {\it MSX} catalog of
\citet{Urquhart:14} is estimated to be complete for star forming regions with bolometric
luminosities $>2\times10^4$ \Lsun\ to a distance of 18 kpc.  While the map in 
Fig.~\ref{fig:Bayesian_map} should be much more complete than that in 
Fig. \ref{fig:parallaxes}, we suspect that it under-represents distant sources,
owing to their intrinsic weakness as well as effects of confusion.  Therefore,
as a first attempt to correct for incompleteness, for sources more distant than 3 kpc 
we randomly ``sprinkle'' a number of points given by $(D{\rm (kpc)}/3)^2$ (capped at 10 points) 
from a Gaussian distribution whose width increases with Galactocentric radius
as found in Section \ref{sect:spiral_structure}.

The visualization of the pattern of sources associated with spiral arms in 
Fig.~\ref{fig:Bayesian_map} shows a dearth of sources toward the Galactic center within a cone 
of half angle $\approx12\deg$.  This is likely a result of survey bias against identifying 
distant, weak sources in the presence of numerous strong nearby sources.  Also, most sources
in the general direction of the Galactic center have LSR velocities close to zero, which 
can limit the information used to discriminate among arm assignments.

\subsection{Fundamental Galactic Parameters}

The distance to the Galactic center, \Ro, is a fundamental parameter affecting
a wide variety of astrophysical questions \citep[\eg][]{Reid:93}.  As such 
there are a large number of estimates for \Ro, and we restrict our discussion
to direct methods.  A trigonometric parallax measurement of water masers from
Sgr B2, a massive star forming region within $\approx100$ pc of the Galactic center,
indicates $\Ro = 7.9\pm0.8$ kpc \citep{Reid:09c}.  The most recent analyses of 
infrared observations tracing the orbits of stars around the supermassive black hole, 
\SgrA, provide geometric estimates of \Ro\ of $7.946\pm0.059$ kpc \citep{Do:19}  
and $8.178\pm0.026$ kpc \citep{Gravity:19}, where the improved accuracy of the latter 
measurement comes from infrared interferometric observations.  
Our result of $\Ro=8.15\pm0.15$ kpc is both independent of, and consistent with, 
these estimates.  

While there are moderate correlations among \Ro, \To, and \V, our parallax data 
strongly constrain the linear and angular speeds of the Sun in its Galactic orbit.  
Adopting the A5-fit results, we find $(\To+\V)=247\pm4$ \kms\ and 
$(\To+\V)/\Ro = 30.32\pm0.27$ \kms\ kpc$^{-1}$.  
This can be compared to an independent and direct estimate from the apparent proper 
motion of the supermassive black hole, \SgrA, assuming it is stationary at the
center of the Milky Way.  \citet{Reid:04} measure the apparent 
motion of \SgrA\ in Galactic longitude to be $-6.379\pm0.019$ \masy.  
This implies $(\To+\V)/\Ro = 30.24\pm0.12$ \kmsperkpc.  Thus, our 
estimate of the angular orbital speed of the Sun is in excellent agreement with
the {\it apparent} proper motion of \SgrA.

\begin{figure}[htp]
\epsscale{0.85} 
\plotone{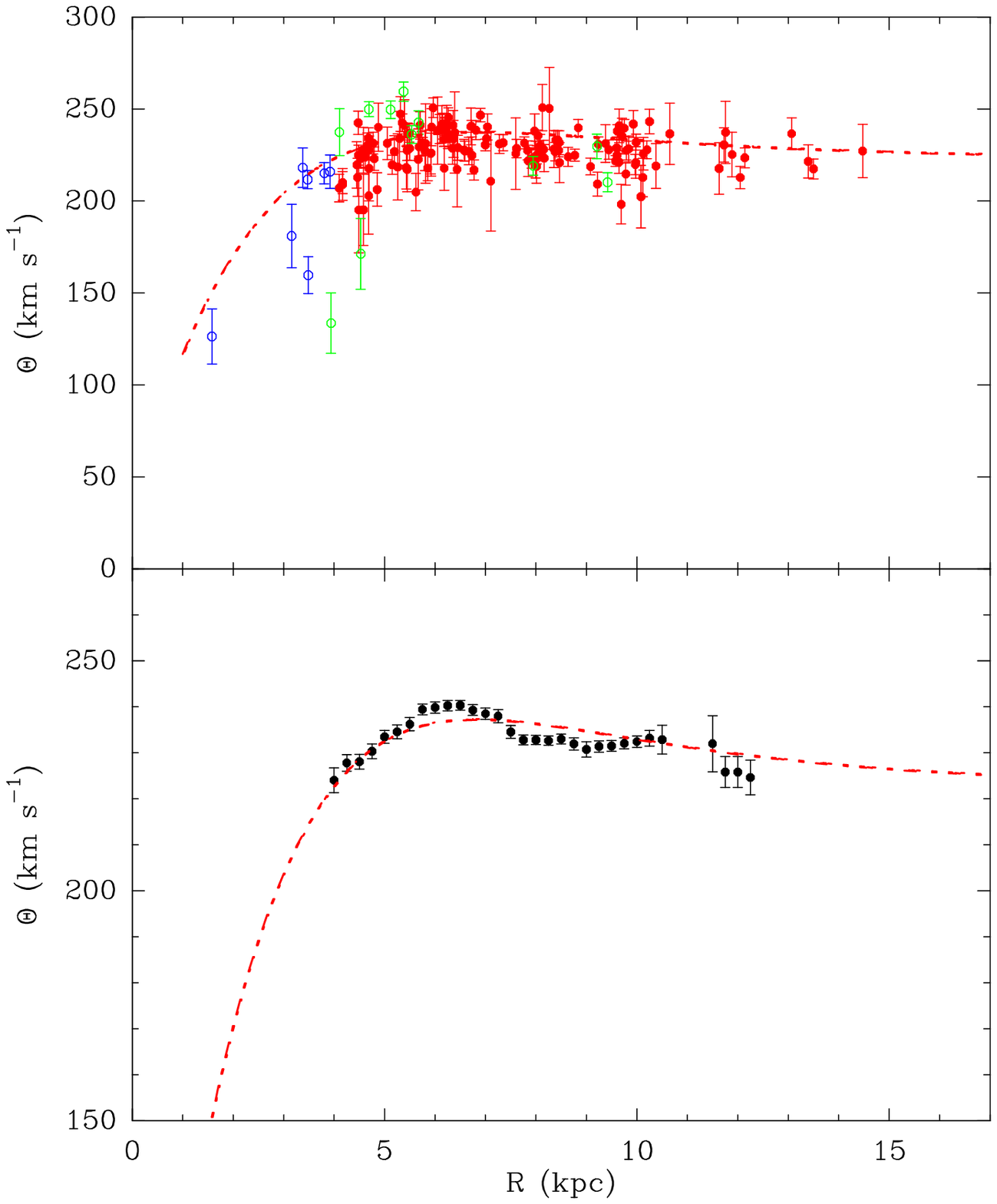}
\caption{\small
Rotation curve for masers associated with massive young stars based on measured parallaxes
and proper motions.  Plotted is the circular velocity component, $\Theta$, 
as a function of Galactocentric radius, $R$.  
{\it Top panel:} The transformation from heliocentric to Galactocentric frames uses the 
parameter values of fit A5, which is based on sources plotted with 
{\it filled red symbols}.
Sources not used in fit A5 with $R<4.0$ kpc are plotted with {\it open blue symbols}
and outliers (see text for details) are plotted with {\it open green symbols}.
The {\it dash-dot red line} is the fitted two-parameter ``Universal'' rotation curve
for spiral galaxies \citep{Persic:96}, with allowance for an average lag of 4.3
\kms\ for massive young stars.  {\it Bottom panel:} The {\it filled black} points are 
variance-weighted averages of the data used in the fitting (adjusted for the average lag) 
and smoothed with a boxcar window of 1.0 kpc full-width.
        } 
\label{fig:rotationcurve}
\end{figure}

Our best-fit model rotation curve for the Galaxy (fit A5) is shown in Fig.~\ref{fig:rotationcurve}
with the dot-dashed line.  This curve has the URC form and is specified by only two 
parameters ($a2=0.96$ and $a3=1.62$).  The slight bias of the curve above the data 
is a result of the tendency for massive young stars to lag Galactic rotation 
with $\Vsbar=-4.3$ km/s.  This rotation curve peaks at 237 \kms\ at a radius of 6.8 kpc 
and falls to 227 \kms\ at a radius of 14.1 kpc, corresponding to a slope of 1.4 \kmskpc\ or, 
alternatively, with a power-law index of $-0.059$ over that radial range. 

Plotted with black dots in the lower panel of Fig.~\ref{fig:rotationcurve} are 
variance-weighted averages of the data, after correcting for the lag, 
within a window of 1 kpc full-width in steps of 0.25 kpc 
(see Table \ref{table:rotationcurve}).  This model-independent rotation curve shows only 
very slight departures from the URC model.  The data could support 
a small decrease of $\approx5$ \kms\ between radii of 7 to 9 kpc, followed by a flattening 
out to 10 kpc.  However, there is no evidence for a $\approx20$ \kms\ dip in the rotation 
curve near $R=9$ kpc with a width of $\approx 2$ kpc, as has been suggested in the literature 
\citep[\eg][]{Sofue:09}.  Qualitatively, our rotation curve is similar to that
presented in the review by \citet{Bland-Hawthorn:16} and more recently in the paper
by \citet{Eilers:19} (allowing for their estimate of \To\ of 229 \kms, which is 
7 \kms\ below our value).

In the context of the URC formalism, $a2=R_{\rm opt}/\Ro$, implying $R_{\rm opt}=7.82$ kpc and 
$a3=1.5 (L/L^*)^{0.2}$ yielding $L = 1.47 L^*$.  \citet{Persic:96} give a halo mass of
$M_{\rm halo} = 1.6\times10^{12} (L/L^*)^{0.56}$ \Msun, which returns 
$M_{\rm halo} = 2.0\times10^{12}$ \Msun, and a disk scale length 
$R_{\rm disk} = R_{\rm opt}/3.2 =2.44$ kpc.

\citet{Reid:14} examined the impact of revised Galactic parameters on the Hulse-Taylor
binary-pulsar's orbital decay, owing to gravitational radiation as predicted by general 
relativity (GR).  This test of GR is sensitive to ($V^2/R$) accelerations from 
the Galactic orbits of the Sun and the pulsar.   Repeating that analysis with the Galactic
parameters of fit A5, we find that the binary's orbital decay rate is $0.99658\pm0.00035$
(nearly a $10\sigma$ deviation) of that expected from GR for the originally assumed pulsar 
distance of 9.9 kpc.  However, we note that, using our estimates of the values and 
uncertainties in \Ro\ and \To, a distance of $6.54\pm0.24$ kpc would remove 
the discrepancy from the GR prediction, providing a strong prediction that can be tested
with a direct pulsar parallax measurement.

\subsection{The Displacement of the Sun from the Galactic Plane}

In Section \ref{sect:GalacticPlane}, we modeled the distances perpendicular to the 
IAU-defined Galactic plane (\ZIAU) for our sample of massive young stars, allowing for 
the Sun's offset from the plane (\Zsun) and tilts of the true plane from the
IAU-plane in the X and Y directions.   We found no significant tilting
within uncertainties of $\pm 0\d1$.  
This finding is consistent with the \HI\ and radio continuum data used to define 
the (inner) plane with similar uncertainty \citep{Gum:60}.
Were we to assume no uncertainty in the orientation of the IAU-plane, 
we would obtain $\Zsun=7.3\pm2.1$ pc.
Allowing for the uncertainty in tilt of the true plane from the IAU-plane,
we find the Sun's displacement is $\Zsun=5.5\pm5.8$ pc toward the north Galactic pole. 

Our method to estimate \Zsun\ is direct and simple, essentially
averaging $Z$-values for extremely young and massive stars,
which should tightly trace the Galactic plane.  These are typically 
O-type stars which have recently formed or are still forming ($\lax$$10^5$ years).
Their very small scale height of $19$ pc attests to their being extreme population-I 
stars.  All distances are determined by trigonometric parallax, and we 
only used measurements with uncertainties less than 20\%, thus avoiding 
significant bias when converting parallax to distance.  
In contrast to optically based observations, which can be plagued 
by variable Galactic extinction, our radio observations are unaffected by 
extinction.  Finally, we have restricted our analysis to Galactocentric
radii $<7.0$ kpc in order to avoid serious complications from Galactic warping.
One shortcoming of our sample might be that we have almost no sources in the $4^{th}$
Galactic quadrant.  However, \HI\ observations have demonstrated little difference
in the $Z$ distribution between the $1^{st}$ and $4^{th}$ quadrants in the inner 
7 kpc of the Galaxy \citep{Gum:60}.

How does our estimate of \Zsun\ compare to previous values?
First, we note that it agrees with the results based on \HII\ regions of 
\citet{Anderson:19}, who provide a range of estimates for \Zsun\ between 3 and 6 pc, 
assuming no tilt of the true plane with respect to the IAU defined plane (see their Table 3),
but a larger range when a tilt is allowed (see their Table 4).
However, there is a very large scatter in estimates of \Zsun\ in the literature, with
most values ranging between about 5 and 40 pc \citep[see \eg][]{Bland-Hawthorn:16}.
Interestingly, dividing published results into three groups suggests systematic
differences: 1) radio and infrared studies typically find $\Zsun\approx10$ pc
\citep{Gum:60,Pandy:87,Cohen:95,Toller:90,Binney:97}; 
2) optical studies of OB and other young, massive stars give $\Zsun\approx20$ pc
\citep{Stothers:74,Pandy:88,Conti:90,Reed:97}; 
and 3) optical star counts yield $\Zsun\approx30$ pc
\citep{Stobie:87,Yamagata:92,Humphreys:95,Mendez:98,Chen:01,Maiz:01,Juric:08}.
The differences between group 1) and the other groups might be partially attributed to 
the effects of extinction, which complicates the optical measurements of
groups 2) and 3).  The differences between the two optical groups may be
more subtle, but generally studies that are restricted to the Solar Neighborhood
give larger \Zsun\ estimates than those with a larger Galactic reach. 
These differences suggest that a combination of extinction and Galactic warping 
may be affecting the optical results.  

Extinction is known to be irregular and could easily bias the ratio of star counts 
north and south of the Galactic plane, which is the basis for many estimates of \Zsun.
Changing the count ratio by 10\% would change the \Zsun\ estimates of
\citet{Humphreys:95} and \citet{Chen:01} by about 15 pc and that of
\citet{Yamagata:92} by about 30 pc.   
Additionally, differences between Solar Neighborhood and Galactic scale studies
might be attributed to the effects of warping of the Galactic plane.  
While the inner portion of the Galactic plane is known to be very flat,
starting between about 7 to 8 kpc from the center, the Galaxy warps upward
toward $l^{II}\approx50^\circ$ and downward toward $l^{II}\approx310^\circ$
\citep[see][noting their use of the old Galactic longitude system $l^I$]{Gum:60}.  
Since this warping can reach 100 pc at a radius of about 9.7 kpc, it would
not be surprising if star counts could be biased to yield \Zsun\ estimates
in error by $\sim10$ pc.  Thus, it is critical to reference \Zsun\ estimates
to the inner plane of the Milky Way as we have done.

Finally, we note that our best-fit value for \Zsun\ of 5.5 pc is very close to that 
needed to move the true Galactic center, defined by the location of
the supermassive black hole \SgrA, to its apparent latitude of 
$b=-0\d046$ (corresponding to $Z_{\rm \SgrA}=-6.5$ pc relative to the IAU 
plane at a distance of 8.15 kpc).

%\subsection{Hulse-Taylor Pulsar}

%With significantly improved fundamental Galactic parameters... 

\section{Concluding Remarks}\label{sect:future}

The original five-year phase of the BeSSeL Survey used about 3500 hours on the VLBA.
Results in this paper are nearly complete through the first four years of the project.
The final year's observations, which preferentially targeted distant sources and 
those especially important to better define spiral arms, are currently being 
analyzed.  

While Fig.~\ref{fig:Bayesian_map} presents the most complete picture of the spiral 
structure of the Milky Way to date, in regions lacking parallax data, distances
to arm segments are necessarily less accurate.  This model can be improved by adding 
parallaxes for sources in the $4^{th}$ quadrant.  In the near future, an extension of 
the BeSSeL Survey to the southern hemisphere is planned.  Using the University of 
Tasmania's AuScope array of four antennas spanning the Australian continent 
(Yarragadee in Western Australia, Katherine in the Northern Territory, 
Ceduna in South Astralia, and Hobart in Tasmania), agumented by an antenna in 
New Zealand (Warkworth), we anticipate starting observations in 2020.  
After several years observing, we hope to obtain parallaxes for 50 to 100 6.7-GHz 
masers in the Galactic longitude range $240\deg$ to $360\deg$, where currently only a 
few have been measured.  

In the long term, we would like to better explore the Milky Way well beyond its 
center, at distances of 10 to 20 kpc, in order to test and improve our current 
model of spiral structure.  While some progress can be made with existing VLBI arrays,
adding large numbers of parallaxes for these very distant regions may require the next generation 
of radio facilities, such as the SKA and ngVLA, with high sensitivity and long baselines.

\vskip 0.5truein 
This work was partially funded by the ERC Advanced Investigator Grant GLOSTAR (247078).
It was also sponsored by the MOST Grant No. 2017YFA0402701 and NSFC Grants 11933011,
11873019 and 11673066.
BZ and YWW are supported by the 100 Talents Project of the Chinese Academy of Sciences (CAS).
AB acknowledges support by the National Science Centre Poland through grant 
2016/21/B/ST9/01455.  KLJR acknowledges funding by the Advanced European Network of 
E-infrastructures for Astronomy with the SKA (AENEAS) project, supported by the European
Commission Framework Programme Horizon 2020 Research and Innovation action under
grant agreement No.~731016.

\vskip 0.5truein 
{\it Facilities:}  \facility{VLBA}, \facility{VERA}, \facility{EVN}, 
\facility{Australian LBA}

\newpage
\appendix
\appendixpage

\section{}

Version 2 of our parallax-based distance estimator \citep{Reid:16},
as well as the {\footnotesize FORTRAN} code and spiral arm data, 
can be found at {\tt http://bessel.vlbi-astrometry.org/}.
It incorporates the improved model for the locations of spiral arms in the Milky Way 
described in Section \ref{sect:armmodel}.  The previous model only included the 
locations of arm segments in the $1^{st}$, $2^{nd}$ and $3^{rd}$ Galactic quadrants.
We now provisionally extend the arm models to the $4^{th}$ quadrant by extrapolating 
arms interior to the solar circle from the $1^{st}$ quadrant guided by the directions 
of arm tangencies in the $4^{th}$ quadrant.  We also assume that arms do not cross each 
other on the far side of the Galactic center, and we make use of the parallax measurement 
of G007.47+00.05 \citep{Sanna17}, which ties the Scutum-Centaurus arm beyond the Galactic 
center with the ``Outer Scutum Centaurus'' arm.  It then follows that the Norma arm connects 
with the Outer arm beyond the Galactic center, and that the Perseus arm is interior to
the Norma arm, most likely originating near the far end of the Galactic bar.  
In addition, in Version 2 we include the following improvements:

\begin{itemize}

\item{}Source coordinates can now be entered either as J2000 (R.A.,Decl.) or Galactic ($l,b$). 
R.A. and Decl. must be in hhmmss.s and ddmmss.s format, and the program assumes values
smaller than 000360.0 and between $\pm$000090.0 are Galactic coordinates in degrees, 
since such small values for both R.A. and Decl. are far from the Galactic plane.

\item{}In Version 1, the spiral arm and Galactic latitude PDFs were plotted separately.
Since the arm-assignment probability terms, ${\rm Prob(arm|,}l,b,v,I{\rm )}$, 
are distance independent and depend on Galactic latitude, not $z$-height, alone they favor 
nearer arms over more distant arms.  While this bias was corrected for by considering the 
distance-dependent Galactic latitude PDF (which uses $z$-heights), for simplicity we now combine 
the sprial arm and Galactic latitude PDFs into a single PDF.

\item{}The uncertainty in \vlsr\ can now be entered and included in the probability densities.

\item{}Proper motions in both Galactic longitude and latitude and their uncertainties, 
if measured, can now be entered and used to estimate distances. 
Given a rotation curve and proper motion components, one can calculate a distance PDF
from the longitude motion in an analogous fashion to using Doppler velocities.  
This alternative type of ``kinematic distance'' \citep{Yamauchi16, Sanna17} has no distance 
ambiguity and is most effective toward the Galactic center and anticenter, the directions in 
which conventional kinematic distances based on radial velocities are least effective.  Also,
for a given velocity dispersion perpendicular to the Galactic plane, the latitude motion
should decrease with distance, and one can also obtain a distance PDF for this component of motion. 

\item{}Since sources near the Galactic bar can exhibit large peculiar motions, we now
inflate motion uncertainties for these sources in order to down-weight the impact of
their kinematic distance PDFs.  Starting inward from $R=6$ kpc, we add a peculiar motion
uncertainty in quadrature with measurement motion uncertainty.  This peculiar motion
uncertainty is zero at $R=6$ kpc and increases linearly to 25 \kms\ at $R=4$ kpc;  
inside of $R=4$ kpc we hold it constant at 25 \kms.

\item{}In addition to the Galactic bar region, large pecular motions are seen for
sources in the Perseus arm in the $2^{nd}$ quadrant within a region bounded
by $1.0 < X < 3.2$ kpc and $8.8 < Y < 10.0$ kpc.  For sources within this
reigon, we add 20 \kms\ in quadrature with measured uncertainties of each motion component.

\item{}For batch processing, the {\footnotesize FORTRAN} program now reads a source-information file
that contains the information for each source on a single ascii-text line.

\end{itemize}

\section{}
\footnotesize
\begin{lstlisting}

        subroutine Univ_RC_from_note ( r, a2, a3, Ro, 
     +                                 Tr ) 
c       Disk plus halo parameterization of rotation curve... 
c       see Persic, Salucci and Stel 1996 Note Added in Proof 

c       Input parameters:
c          r:   Galactocentric radius in kpc
c         a2:   Ropt/Ro  where Ropt=3.2*R_scalelength encloses 83% of light
c         a3:   1.5*(L/L*)^(1/5)   
c         Ro:   Ro in kpc
c       Output parameter:
c         Tr:   Circular rotation speed at r in km/s

        implicit real*8 (a-h,o-z)
        real*8   lambda, log_lam
           
        lambda = (a3/1.5d0)**5     ! L/L*
        Ropt   = a2 * Ro     
        rho    = r/Ropt

c       Calculate Tr...
        log_lam = log10( lambda )

        term1 = 200.d0*lambda**0.41d0
           
        top   = 0.75d0*exp(-0.4d0*lambda)
        bot   = 0.47d0 + 2.25d0*lambda**0.4d0
        term2 = sqrt( 0.80d0 + 0.49d0*log_lam + (top/bot) )

        top   = 1.97d0*rho**1.22d0
        bot   = (rho**2 + 0.61d0)**1.43d0
        term3 = (0.72d0 + 0.44*log_lam) * (top/bot)

        top   = rho**2
        bot   = rho**2 + 2.25*lambda**0.4d0
        term4 = 1.6d0*exp(-0.4d0*lambda) * (top/bot)

        Tr    = (term1/term2) * sqrt(term3 + term 4)            ! km/s 

        return
        end

\end{lstlisting}

\begin{deluxetable}{llrrrrrrll}
\tabletypesize{\tiny} 
%\tabletypesize{\scriptsize}           
%\rotate
\tablecolumns{10} \tablewidth{0pc}
\tablecaption{Parallaxes \& Proper Motions of High-mass Star Forming Regions}
\tablehead {
  \colhead{Source} & \colhead{Alias}& \colhead{R.A.} & \colhead{Dec.} &
  \colhead{Parallax} & \colhead{$\mu_x$} & \colhead{$\mu_y$} &
  \colhead{\vlsr}  & \colhead{Spiral}  & \colhead{Refs.}
\\
  \colhead{}      &\colhead{}      & \colhead{(hh:mm:ss)} & \colhead{(dd:mm:ss)} &
  \colhead{(mas)} & \colhead{(\masy)} & \colhead{(\masy)} &
  \colhead{(\kms)}& \colhead{Arm}   & \colhead{}      
          }
\startdata
G305.20$+$00.01 &               &13:11:16.8912 &$-$62:45:55.008 &$ 0.250\pm 0.050$ &$  -6.90\pm 0.33$ &$  -0.52\pm 0.33$ &$  -38\pm\phantom{1} 5$ &CtN &61           \\
G339.88$-$01.25 &               &16:52:04.6776 &$-$46:08:34.404 &$ 0.480\pm 0.080$ &$  -1.60\pm 0.52$ &$  -1.90\pm 0.52$ &$  -34\pm\phantom{1} 3$ &CtN &37           \\
G348.70$-$01.04 &               &17:20:04.0360 &$-$38:58:30.920 &$ 0.296\pm 0.026$ &$  -0.73\pm 0.31$ &$  -2.83\pm 0.59$ &$   -9\pm\phantom{1} 5$ &CtN &1            \\
G351.44$+$00.65 &NGC6334        &17:20:54.6010 &$-$35:45:08.620 &$ 0.752\pm 0.069$ &$   0.31\pm 0.58$ &$  -2.17\pm 0.90$ &$   -8\pm\phantom{1} 3$ &CrN &2,76         \\
G359.13$+$00.03 &               &17:43:25.6109 &$-$29:39:17.551 &$ 0.165\pm 0.031$ &$  -3.98\pm 0.36$ &$  -6.50\pm 0.79$ &$   -1\pm 10$ &GC  &52           \\
G359.61$-$00.24 &               &17:45:39.0697 &$-$29:23:30.265 &$ 0.375\pm 0.021$ &$   1.00\pm 0.40$ &$  -1.50\pm 0.50$ &$   21\pm\phantom{1} 5$ &CtN &59,52        \\
G359.93$-$00.14 &               &17:46:01.9183 &$-$29:03:58.674 &$ 0.181\pm 0.029$ &$  -1.33\pm 0.44$ &$  -1.80\pm 0.78$ &$  -10\pm 10$ &GC  &52           \\
G000.31$-$00.20 &               &17:47:09.1092 &$-$28:46:16.278 &$ 0.342\pm 0.042$ &$   0.21\pm 0.39$ &$  -1.76\pm 0.64$ &$   18\pm\phantom{1} 3$ &ScN &53           \\
G000.37$+$00.03 &               &17:46:21.4012 &$-$28:35:39.821 &$ 0.125\pm 0.047$ &$  -0.77\pm 0.28$ &$  -2.67\pm 0.40$ &$   37\pm 10$ &3kF &7            \\
G000.67$-$00.03 &SgrB2          &17:47:20.0000 &$-$28:22:40.000 &$ 0.129\pm 0.012$ &$  -0.78\pm 0.40$ &$  -4.26\pm 0.40$ &$   62\pm\phantom{1} 5$ &GC  &3            \\
G001.00$-$00.23 &               &17:48:55.2845 &$-$28:11:48.240 &$ 0.090\pm 0.057$ &$  -3.87\pm 0.28$ &$  -6.23\pm 0.56$ &$    2\pm\phantom{1} 5$ &??? &51           \\
G001.14$-$00.12 &               &17:48:48.5410 &$-$28:01:11.350 &$ 0.194\pm 0.161$ &$   0.82\pm 0.33$ &$  -3.59\pm 0.69$ &$  -16\pm\phantom{1} 3$ &Nor &51           \\
G002.70$+$00.04 &               &17:51:45.9766 &$-$26:35:57.070 &$ 0.101\pm 0.105$ &$  -3.13\pm 0.87$ &$  -9.36\pm 1.24$ &$   93\pm\phantom{1} 5$ &??? &51           \\
G005.88$-$00.39 &               &18:00:30.2801 &$-$24:04:04.576 &$ 0.334\pm 0.020$ &$   0.18\pm 0.34$ &$  -2.26\pm 0.34$ &$    9\pm\phantom{1} 5$ &ScN &4            \\
G006.79$-$00.25 &               &18:01:57.7525 &$-$23:12:34.245 &$ 0.288\pm 0.021$ &$  -0.30\pm 0.30$ &$  -1.30\pm 0.31$ &$   22\pm\phantom{1} 5$ &Nor &59           \\
G007.47$+$00.05 &               &18:02:13.1823 &$-$22:27:58.981 &$ 0.049\pm 0.006$ &$  -2.43\pm 0.10$ &$  -4.43\pm 0.16$ &$  -14\pm 10$ &OSC &62,63        \\
G008.34$-$01.00 &VXSgr          &18:08:04.0510 &$-$22:13:26.566 &$ 0.640\pm 0.040$ &$   0.36\pm 0.76$ &$  -2.92\pm 0.78$ &$    5\pm\phantom{1} 5$ &SgN &64           \\
G009.21$-$00.20 &               &18:06:52.8421 &$-$21:04:27.878 &$ 0.303\pm 0.096$ &$  -0.41\pm 0.45$ &$  -1.69\pm 0.50$ &$   43\pm\phantom{1} 5$ &ScN &53           \\
G009.62$+$00.19 &               &18:06:14.6600 &$-$20:31:31.700 &$ 0.194\pm 0.023$ &$  -0.58\pm 0.13$ &$  -2.49\pm 0.29$ &$    2\pm\phantom{1} 3$ &3kN &5            \\
G010.32$-$00.15 &               &18:09:01.4549 &$-$20:05:07.854 &$ 0.343\pm 0.035$ &$  -1.03\pm 0.36$ &$  -2.42\pm 0.50$ &$   10\pm\phantom{1} 5$ &ScN &54,52        \\
G010.47$+$00.02 &               &18:08:38.2290 &$-$19:51:50.262 &$ 0.117\pm 0.008$ &$  -3.86\pm 0.19$ &$  -6.40\pm 0.14$ &$   69\pm\phantom{1} 5$ &Con &7            \\
G010.62$-$00.33 &               &18:10:17.9849 &$-$19:54:04.646 &$ 0.362\pm 0.076$ &$  -0.36\pm 0.39$ &$  -1.86\pm 0.38$ &$   -6\pm\phantom{1} 5$ &3kN &52           \\
G010.62$-$00.38 &               &18:10:28.5628 &$-$19:55:48.738 &$ 0.202\pm 0.019$ &$  -0.37\pm 0.50$ &$  -0.60\pm 0.25$ &$   -3\pm\phantom{1} 5$ &3kN &7            \\
G011.10$-$00.11 &               &18:10:28.2470 &$-$19:22:30.216 &$ 0.246\pm 0.014$ &$  -0.23\pm 0.38$ &$  -2.01\pm 0.41$ &$   29\pm\phantom{1} 5$ &ScN &53           \\
G011.49$-$01.48 &               &18:16:22.1327 &$-$19:41:27.219 &$ 0.800\pm 0.033$ &$   1.42\pm 0.52$ &$  -0.60\pm 0.65$ &$   11\pm\phantom{1} 3$ &SgN &2            \\
G011.91$-$00.61 &               &18:13:58.1205 &$-$18:54:20.278 &$ 0.297\pm 0.050$ &$   0.66\pm 0.69$ &$  -1.36\pm 0.75$ &$   37\pm\phantom{1} 5$ &ScN &4            \\
G012.02$-$00.03 &               &18:12:01.8400 &$-$18:31:55.800 &$ 0.106\pm 0.008$ &$  -4.11\pm 0.11$ &$  -7.76\pm 0.28$ &$  108\pm\phantom{1} 5$ &3kN &7            \\
G012.68$-$00.18 &W33B           &18:13:54.7457 &$-$18:01:46.588 &$ 0.416\pm 0.028$ &$  -1.00\pm 0.95$ &$  -2.85\pm 0.95$ &$   58\pm 10$ &ScN &8            \\
G012.81$-$00.19 &W33M           &18:14:14.0600 &$-$17:55:11.300 &$ 0.343\pm 0.037$ &$  -0.40\pm 0.70$ &$  -0.25\pm 0.70$ &$   34\pm\phantom{1} 5$ &ScN &8            \\
G012.88$+$00.48 &              &18:11:51.4450 &$-$17:31:29.412 &$ 0.404\pm 0.044$ &$   0.15\pm 0.43$ &$  -2.30\pm 0.52$ &$   35\pm\phantom{1} 7$ &ScN &8,10         \\
G012.90$-$00.24 &W33A           &18:14:34.4366 &$-$17:51:51.891 &$ 0.408\pm 0.025$ &$   0.19\pm 0.40$ &$  -2.52\pm 0.40$ &$   36\pm\phantom{1} 5$ &ScN &8            \\
G012.90$-$00.26 &W33A           &18:14:39.5714 &$-$17:52:00.382 &$ 0.396\pm 0.032$ &$  -0.36\pm 0.40$ &$  -2.22\pm 0.40$ &$   37\pm\phantom{1} 5$ &ScN &8            \\
G013.71$-$00.08 &               &18:15:36.9814 &$-$17:04:32.108 &$ 0.264\pm 0.014$ &$  -0.29\pm 0.28$ &$  -2.16\pm 0.29$ &$   44\pm\phantom{1} 5$ &Nor &7            \\
G013.87$+$00.28 &               &18:14:35.8356 &$-$16:45:35.869 &$ 0.254\pm 0.024$ &$  -0.25\pm 2.00$ &$  -2.49\pm 2.00$ &$   48\pm 10$ &ScN &4            \\
G014.33$-$00.64 &               &18:18:54.6744 &$-$16:47:50.264 &$ 0.893\pm 0.101$ &$   0.95\pm 1.50$ &$  -2.40\pm 1.30$ &$   22\pm\phantom{1} 5$ &SgN &9            \\
G014.63$-$00.57 &               &18:19:15.5407 &$-$16:29:45.786 &$ 0.546\pm 0.022$ &$   0.22\pm 1.20$ &$  -2.07\pm 1.20$ &$   19\pm\phantom{1} 5$ &SgN &2            \\
G015.03$-$00.67 &               &18:20:24.8111 &$-$16:11:35.316 &$ 0.499\pm 0.026$ &$   0.68\pm 0.53$ &$  -1.42\pm 0.54$ &$   22\pm\phantom{1} 3$ &SgN &10,67        \\
G015.66$-$00.49 &               &18:20:59.7470 &$-$15:33:09.800 &$ 0.220\pm 0.029$ &$  -1.16\pm 0.24$ &$  -5.30\pm 0.30$ &$   -4\pm\phantom{1} 5$ &3kN &59           \\
G016.58$-$00.05 &               &18:21:09.0840 &$-$14:31:48.556 &$ 0.279\pm 0.023$ &$  -1.13\pm 0.34$ &$  -2.59\pm 0.35$ &$   60\pm\phantom{1} 5$ &ScN &4            \\
G016.86$-$02.15 &               &18:29:24.4085 &$-$15:16:04.141 &$ 0.426\pm 0.092$ &$   0.32\pm 0.49$ &$  -2.48\pm 0.56$ &$   17\pm\phantom{1} 3$ &SgN &54           \\
G017.02$-$02.40 &               &18:30:36.2931 &$-$15:14:28.384 &$ 0.531\pm 0.108$ &$   0.10\pm 0.61$ &$  -2.81\pm 0.76$ &$   22\pm\phantom{1} 3$ &SgN &54           \\
G017.55$-$00.12 &IRC-10414      &18:23:17.9084 &$-$13:42:47.146 &$ 0.497\pm 0.038$ &$  -0.04\pm 1.04$ &$  -2.01\pm 1.04$ &$   44\pm 10$ &SgN &59           \\
G017.63$+$00.15 &               &18:22:26.3821 &$-$13:30:11.951 &$ 0.669\pm 0.019$ &$  -0.26\pm 1.41$ &$  -1.53\pm 1.41$ &$   25\pm 10$ &SgN &54           \\
G018.34$+$01.76 &               &18:17:58.1254 &$-$12:07:24.893 &$ 0.500\pm 0.019$ &$  -0.32\pm 0.53$ &$  -1.99\pm 0.76$ &$   28\pm\phantom{1} 3$ &SgN &54           \\
G018.87$+$00.05 &               &18:25:11.3530 &$-$12:27:36.546 &$ 0.297\pm 0.049$ &$  -0.17\pm 0.32$ &$  -1.90\pm 0.37$ &$   39\pm\phantom{1} 3$ &ScN &53           \\
G019.00$-$00.02 &               &18:25:44.7778 &$-$12:22:45.886 &$ 0.247\pm 0.063$ &$  -1.77\pm 0.29$ &$  -4.00\pm 0.42$ &$   56\pm\phantom{1} 5$ &Nor &60           \\
G019.36$-$00.03 &               &18:26:25.7796 &$-$12:03:53.267 &$ 0.352\pm 0.069$ &$  -1.12\pm 0.38$ &$  -2.50\pm 0.46$ &$   27\pm\phantom{1} 3$ &ScN &53           \\
G019.49$+$00.11 &               &18:26:09.1691 &$-$11:52:51.354 &$ 0.326\pm 0.100$ &$  -3.60\pm 0.42$ &$  -8.20\pm 1.03$ &$  121\pm\phantom{1} 5$ &??? &60           \\
G022.35$+$00.06 &               &18:31:44.1199 &$-$09:22:12.336 &$ 0.231\pm 0.108$ &$  -1.70\pm 0.28$ &$  -3.90\pm 0.31$ &$   80\pm\phantom{1} 5$ &Nor &60           \\
G023.00$-$00.41 &               &18:34:40.2800 &$-$09:00:38.360 &$ 0.205\pm 0.015$ &$  -1.72\pm 0.22$ &$  -4.12\pm 0.37$ &$   79\pm\phantom{1} 5$ &Nor &11,68        \\
G023.20$-$00.37 &               &18:34:55.1794 &$-$08:49:15.206 &$ 0.239\pm 0.034$ &$  -1.45\pm 0.50$ &$  -3.40\pm 0.51$ &$   82\pm 10$ &Nor &68           \\
G023.25$-$00.24 &               &18:34:31.2397 &$-$08:42:47.306 &$ 0.169\pm 0.051$ &$  -1.72\pm 0.23$ &$  -4.09\pm 0.28$ &$   63\pm\phantom{1} 3$ &ScN &52           \\
G023.38$+$00.18 &               &18:33:14.3240 &$-$08:23:57.500 &$ 0.208\pm 0.025$ &$  -1.57\pm 0.23$ &$  -3.92\pm 0.26$ &$   75\pm\phantom{1} 3$ &Nor &60           \\
G023.43$-$00.18 &               &18:34:39.1870 &$-$08:31:25.405 &$ 0.170\pm 0.032$ &$  -1.93\pm 0.15$ &$  -4.11\pm 0.13$ &$   97\pm\phantom{1} 3$ &Nor &11           \\
G023.65$-$00.12 &               &18:34:51.5650 &$-$08:18:21.305 &$ 0.313\pm 0.039$ &$  -1.32\pm 0.20$ &$  -2.96\pm 0.20$ &$   81\pm\phantom{1} 3$ &ScN &12           \\
G023.70$-$00.19 &               &18:35:12.3610 &$-$08:17:39.530 &$ 0.161\pm 0.024$ &$  -3.17\pm 0.18$ &$  -6.38\pm 0.21$ &$   73\pm\phantom{1} 5$ &Nor &7            \\
G024.63$-$00.32 &               &18:37:22.7091 &$-$07:31:42.093 &$ 0.242\pm 0.045$ &$  -0.31\pm 0.27$ &$  -2.73\pm 0.30$ &$   43\pm\phantom{1} 5$ &ScN &53           \\
G024.78$+$00.08 &               &18:36:12.5614 &$-$07:12:10.840 &$ 0.150\pm 0.016$ &$  -2.62\pm 0.16$ &$  -5.08\pm 0.20$ &$  111\pm\phantom{1} 3$ &Nor &60           \\
G024.85$+$00.08 &               &18:36:18.3867 &$-$07:08:50.834 &$ 0.176\pm 0.016$ &$  -2.42\pm 0.19$ &$  -5.23\pm 0.26$ &$  111\pm\phantom{1} 5$ &Nor &53           \\
G025.70$+$00.04 &               &18:38:03.1450 &$-$06:24:15.190 &$ 0.098\pm 0.029$ &$  -2.89\pm 0.11$ &$  -6.20\pm 0.36$ &$   93\pm\phantom{1} 5$ &ScF &4            \\
G027.36$-$00.16 &               &18:41:51.0570 &$-$05:01:43.443 &$ 0.125\pm 0.042$ &$  -1.81\pm 0.11$ &$  -4.11\pm 0.27$ &$   92\pm\phantom{1} 3$ &ScF &10           \\
G028.14$-$00.00 &               &18:42:42.5896 &$-$04:15:35.128 &$ 0.158\pm 0.023$ &$  -2.11\pm 0.17$ &$  -4.85\pm 0.17$ &$  100\pm\phantom{1} 5$ &ScF &52           \\
G028.30$-$00.38 &               &18:44:21.9666 &$-$04:17:39.904 &$ 0.221\pm 0.022$ &$  -1.62\pm 0.23$ &$  -3.97\pm 0.23$ &$   87\pm\phantom{1} 5$ &ScN &53           \\
G028.39$+$00.08 &               &18:42:51.9822 &$-$03:59:54.494 &$ 0.231\pm 0.015$ &$  -1.40\pm 0.24$ &$  -3.26\pm 0.24$ &$   75\pm\phantom{1} 5$ &ScN &53           \\
G028.83$-$00.25 &               &18:44:51.0865 &$-$03:45:48.378 &$ 0.200\pm 0.040$ &$  -1.89\pm 0.22$ &$  -4.54\pm 0.22$ &$   87\pm\phantom{1} 5$ &ScN &53           \\
G028.86$+$00.06 &               &18:43:46.2233 &$-$03:35:29.846 &$ 0.201\pm 0.019$ &$  -3.19\pm 0.42$ &$  -5.72\pm 0.48$ &$  100\pm 10$ &ScF &52           \\
G029.86$-$00.04 &W43S           &18:45:59.5708 &$-$02:45:06.581 &$ 0.221\pm 0.030$ &$  -2.33\pm 0.26$ &$  -5.47\pm 0.31$ &$  101\pm\phantom{1} 3$ &ScF &6,52         \\
G029.95$-$00.01 &W43S           &18:46:03.7402 &$-$02:39:22.328 &$ 0.207\pm 0.019$ &$  -2.36\pm 0.20$ &$  -5.38\pm 0.20$ &$   99\pm\phantom{1} 5$ &ScF &6,52         \\
G029.98$+$00.10 &               &18:45:39.9622 &$-$02:34:32.581 &$ 0.156\pm 0.010$ &$  -0.94\pm 0.32$ &$  -3.49\pm 0.32$ &$  109\pm 10$ &ScF &52           \\
G030.19$-$00.16 &               &18:47:03.0698 &$-$02:30:36.268 &$ 0.212\pm 0.010$ &$  -2.08\pm 0.22$ &$  -3.80\pm 0.25$ &$  108\pm\phantom{1} 3$ &ScN &53           \\
G030.22$-$00.18 &               &18:47:08.2979 &$-$02:29:29.330 &$ 0.284\pm 0.032$ &$  -0.87\pm 0.31$ &$  -4.70\pm 0.34$ &$  113\pm\phantom{1} 3$ &ScN &52           \\
G030.41$-$00.23 &               &18:47:40.7589 &$-$02:20:30.907 &$ 0.253\pm 0.021$ &$  -1.98\pm 0.28$ &$  -4.07\pm 0.29$ &$  103\pm\phantom{1} 3$ &ScN &52           \\
G030.70$-$00.06 &               &18:47:36.7983 &$-$02:00:54.341 &$ 0.153\pm 0.020$ &$  -2.33\pm 0.18$ &$  -4.95\pm 0.18$ &$   89\pm\phantom{1} 3$ &ScF &52           \\
G030.74$-$00.04 &               &18:47:39.7248 &$-$01:57:24.974 &$ 0.326\pm 0.055$ &$  -2.38\pm 0.38$ &$  -4.46\pm 0.38$ &$   88\pm\phantom{1} 3$ &ScN &52           \\
G030.78$+$00.20 &               &18:46:48.0864 &$-$01:48:53.946 &$ 0.140\pm 0.032$ &$  -1.85\pm 0.16$ &$  -3.91\pm 0.19$ &$   82\pm\phantom{1} 5$ &ScN &52           \\
G030.81$-$00.05 &               &18:47:46.9751 &$-$01:54:26.416 &$ 0.321\pm 0.037$ &$  -2.34\pm 0.35$ &$  -5.66\pm 0.35$ &$  105\pm\phantom{1} 5$ &ScF &52           \\
G030.97$-$00.14 &               &18:48:22.0433 &$-$01:48:30.750 &$ 0.294\pm 0.022$ &$  -2.14\pm 0.31$ &$  -3.82\pm 0.34$ &$   77\pm\phantom{1} 3$ &ScN &52           \\
G031.24$-$00.11 &               &18:48:45.0808 &$-$01:33:13.200 &$ 0.076\pm 0.014$ &$  -2.80\pm 0.19$ &$  -5.54\pm 0.19$ &$   24\pm 10$ &Per &55           \\
G031.28$+$00.06 &W43M           &18:48:12.3760 &$-$01:26:28.778 &$ 0.210\pm 0.029$ &$  -1.99\pm 0.19$ &$  -4.58\pm 0.22$ &$  110\pm\phantom{1} 5$ &ScN &6,53       \\
G031.41$+$00.30 &               &18:47:34.3105 &$-$01:12:46.659 &$ 0.267\pm 0.029$ &$  -1.58\pm 0.70$ &$  -5.82\pm 0.70$ &$   97\pm 15$ &ScN &52           \\
G031.58$+$00.07 &W43M           &18:48:41.6740 &$-$01:09:59.774 &$ 0.183\pm 0.032$ &$  -1.97\pm 0.40$ &$  -4.43\pm 0.40$ &$   96\pm\phantom{1} 5$ &ScF &6            \\
G032.04$+$00.05 &               &18:49:36.5754 &$-$00:45:45.539 &$ 0.198\pm 0.015$ &$  -2.05\pm 0.21$ &$  -4.77\pm 0.21$ &$   98\pm\phantom{1} 5$ &ScN &4,52         \\
G032.74$-$00.07 &               &18:51:21.8624 &$-$00:12:06.220 &$ 0.126\pm 0.016$ &$  -3.15\pm 0.27$ &$  -6.10\pm 0.29$ &$   37\pm 10$ &SgF &55           \\
G032.79$+$00.19 &               &18:50:30.7330 &$-$00:01:59.280 &$ 0.103\pm 0.031$ &$  -2.94\pm 0.24$ &$  -6.07\pm 0.25$ &$   16\pm 10$ &Per &51           \\
G033.09$-$00.07 &               &18:51:59.4085 &$+$00:06:34.262 &$ 0.134\pm 0.016$ &$  -2.46\pm 0.15$ &$  -5.32\pm 0.15$ &$  100\pm\phantom{1} 5$ &ScF &52           \\
G033.39$+$00.00 &               &18:52:14.6412 &$+$00:24:54.374 &$ 0.113\pm 0.029$ &$  -2.48\pm 0.13$ &$  -5.36\pm 0.13$ &$  102\pm\phantom{1} 5$ &ScF &53           \\
G033.64$-$00.22 &               &18:53:32.5630 &$+$00:31:39.100 &$ 0.131\pm 0.020$ &$  -3.10\pm 0.15$ &$  -6.27\pm 0.14$ &$   60\pm\phantom{1} 3$ &SgF &6            \\
G034.41$+$00.23 &               &18:53:18.0319 &$+$01:25:25.500 &$ 0.340\pm 0.011$ &$   0.12\pm 0.72$ &$  -4.18\pm 0.72$ &$   60\pm 10$ &SgN &56           \\
G034.79$-$01.38 &               &18:59:45.9838 &$+$01:01:18.947 &$ 0.381\pm 0.020$ &$  -0.31\pm 0.62$ &$  -2.80\pm 0.72$ &$   45\pm\phantom{1} 5$ &SgN &54           \\
G035.02$+$00.34 &               &18:54:00.6576 &$+$02:01:19.217 &$ 0.430\pm 0.040$ &$  -0.92\pm 0.90$ &$  -3.61\pm 0.90$ &$   52\pm\phantom{1} 5$ &SgN &2            \\
G035.19$-$00.74 &               &18:58:13.0517 &$+$01:40:35.674 &$ 0.456\pm 0.045$ &$  -0.18\pm 0.50$ &$  -3.63\pm 0.50$ &$   30\pm\phantom{1} 7$ &SgN &14           \\
G035.20$-$01.73 &               &19:01:45.5420 &$+$01:13:32.690 &$ 0.412\pm 0.014$ &$  -0.68\pm 0.44$ &$  -3.60\pm 0.44$ &$   43\pm\phantom{1} 5$ &SgN &14,54        \\
G035.79$-$00.17 &               &18:57:16.8905 &$+$02:27:58.007 &$ 0.113\pm 0.013$ &$  -2.96\pm 0.12$ &$  -6.23\pm 0.14$ &$   61\pm\phantom{1} 5$ &SgF &55           \\
G036.11$+$00.55 &               &18:55:16.7927 &$+$03:05:05.392 &$ 0.246\pm 0.056$ &$  -1.41\pm 0.26$ &$  -3.76\pm 0.27$ &$   75\pm\phantom{1} 5$ &AqS &56           \\
G037.42$+$01.51 &               &18:54:14.3481 &$+$04:41:39.647 &$ 0.532\pm 0.021$ &$  -0.45\pm 0.35$ &$  -3.69\pm 0.39$ &$   41\pm\phantom{1} 3$ &SgN &2            \\
G037.47$-$00.10 &               &19:00:07.1430 &$+$03:59:52.975 &$ 0.088\pm 0.030$ &$  -2.63\pm 0.07$ &$  -6.19\pm 0.15$ &$   58\pm\phantom{1} 3$ &SgF &55           \\
G038.03$-$00.30 &               &19:01:50.4676 &$+$04:24:18.900 &$ 0.095\pm 0.022$ &$  -3.01\pm 0.06$ &$  -6.20\pm 0.11$ &$   60\pm\phantom{1} 3$ &SgF &55           \\
G038.11$-$00.22 &               &19:01:44.1513 &$+$04:30:37.400 &$ 0.242\pm 0.035$ &$  -2.79\pm 0.27$ &$  -5.25\pm 0.31$ &$   76\pm\phantom{1} 5$ &AqS &56           \\
G040.28$-$00.21 &               &19:05:41.2146 &$+$06:26:12.698 &$ 0.297\pm 0.019$ &$  -1.80\pm 0.31$ &$  -3.82\pm 0.33$ &$   74\pm\phantom{1} 5$ &AqS &56           \\
G040.42$+$00.70 &               &19:02:39.6192 &$+$06:59:09.052 &$ 0.078\pm 0.013$ &$  -2.95\pm 0.09$ &$  -5.48\pm 0.10$ &$   10\pm\phantom{1} 5$ &Per &56           \\
G040.62$-$00.13 &               &19:06:01.6288 &$+$06:46:36.140 &$ 0.080\pm 0.021$ &$  -2.69\pm 0.09$ &$  -5.60\pm 0.25$ &$   31\pm\phantom{1} 3$ &Per &55           \\
G041.15$-$00.20 &               &19:07:14.3676 &$+$07:13:18.025 &$ 0.125\pm 0.018$ &$  -2.79\pm 0.14$ &$  -5.85\pm 0.16$ &$   60\pm\phantom{1} 3$ &SgF &55           \\
G041.22$-$00.19 &               &19:07:21.3772 &$+$07:17:08.115 &$ 0.113\pm 0.022$ &$  -2.82\pm 0.13$ &$  -5.89\pm 0.16$ &$   59\pm\phantom{1} 5$ &SgF &55           \\
G042.03$+$00.19 &               &19:07:28.1834 &$+$08:10:53.433 &$ 0.071\pm 0.012$ &$  -2.40\pm 0.09$ &$  -5.64\pm 0.11$ &$   12\pm\phantom{1} 5$ &Per &56           \\
G043.03$-$00.45 &               &19:11:38.9819 &$+$08:46:30.665 &$ 0.130\pm 0.019$ &$  -3.03\pm 0.15$ &$  -6.56\pm 0.20$ &$   56\pm\phantom{1} 5$ &SgF &55           \\
G043.16$+$00.01 &W49N           &19:10:13.4100 &$+$09:06:12.800 &$ 0.090\pm 0.007$ &$  -2.88\pm 0.20$ &$  -5.41\pm 0.20$ &$   10\pm\phantom{1} 5$ &Per &15           \\
G043.79$-$00.12 &OH43.8-0.1     &19:11:53.9868 &$+$09:35:50.325 &$ 0.166\pm 0.010$ &$  -3.02\pm 0.36$ &$  -6.20\pm 0.36$ &$   44\pm 10$ &SgF &2            \\
G043.89$-$00.78 &               &19:14:26.3906 &$+$09:22:36.568 &$ 0.134\pm 0.013$ &$  -3.11\pm 0.15$ &$  -5.35\pm 0.21$ &$   52\pm\phantom{1} 3$ &SgF &2,55         \\
G045.07$+$00.13 &               &19:13:22.0427 &$+$10:50:53.336 &$ 0.129\pm 0.007$ &$  -3.21\pm 0.26$ &$  -6.11\pm 0.26$ &$   59\pm\phantom{1} 5$ &SgF &2            \\
G045.45$+$00.06 &               &19:14:21.2658 &$+$11:09:15.872 &$ 0.119\pm 0.017$ &$  -2.34\pm 0.38$ &$  -6.00\pm 0.54$ &$   55\pm\phantom{1} 7$ &SgF &2            \\
G045.49$+$00.12 &               &19:14:11.3553 &$+$11:13:06.370 &$ 0.144\pm 0.024$ &$  -2.62\pm 0.17$ &$  -5.61\pm 0.16$ &$   58\pm\phantom{1} 3$ &SgF &55           \\
G045.80$-$00.35 &               &19:16:31.0795 &$+$11:16:11.985 &$ 0.137\pm 0.023$ &$  -2.52\pm 0.17$ &$  -6.08\pm 0.27$ &$   64\pm\phantom{1} 5$ &SgF &55           \\
G048.60$+$00.02 &               &19:20:31.1760 &$+$13:55:25.210 &$ 0.093\pm 0.005$ &$  -2.89\pm 0.13$ &$  -5.50\pm 0.13$ &$   18\pm\phantom{1} 5$ &Per &15           \\
G048.99$-$00.29 &               &19:22:26.1348 &$+$14:06:39.133 &$ 0.178\pm 0.017$ &$  -2.20\pm 0.48$ &$  -5.84\pm 0.63$ &$   67\pm 10$ &SgF &69           \\
G049.04$-$01.07 &               &19:25:22.2504 &$+$13:47:19.525 &$ 0.164\pm 0.022$ &$  -3.23\pm 0.23$ &$  -6.42\pm 0.43$ &$   38\pm\phantom{1} 5$ &SgN &55           \\
G049.19$-$00.33 &               &19:22:57.7705 &$+$14:16:09.983 &$ 0.197\pm 0.008$ &$  -3.08\pm 0.40$ &$  -5.50\pm 0.40$ &$   67\pm\phantom{1} 5$ &SgF &2,69         \\
G049.26$+$00.31 &               &19:20:44.8571 &$+$14:38:26.864 &$ 0.113\pm 0.016$ &$  -2.73\pm 0.15$ &$  -5.85\pm 0.19$ &$    0\pm\phantom{1} 5$ &Per &51           \\
G049.34$+$00.41 &               &19:20:32.4472 &$+$14:45:45.390 &$ 0.241\pm 0.031$ &$  -2.36\pm 0.26$ &$  -5.59\pm 0.33$ &$   68\pm\phantom{1} 5$ &SgF &55           \\
G049.41$+$00.32 &               &19:20:59.2098 &$+$14:46:49.613 &$ 0.132\pm 0.031$ &$  -3.15\pm 0.17$ &$  -4.49\pm 0.66$ &$  -12\pm\phantom{1} 5$ &Per &51           \\
G049.48$-$00.36 &W51IRS2        &19:23:39.8240 &$+$14:31:04.950 &$ 0.195\pm 0.071$ &$  -2.49\pm 0.14$ &$  -5.51\pm 0.16$ &$   56\pm\phantom{1} 3$ &SgN &16           \\
G049.48$-$00.38 &W51M           &19:23:43.8700 &$+$14:30:29.500 &$ 0.185\pm 0.010$ &$  -2.64\pm 0.20$ &$  -5.11\pm 0.20$ &$   58\pm\phantom{1} 4$ &SgN &17           \\
G049.59$-$00.24 &               &19:23:26.6068 &$+$14:40:16.955 &$ 0.218\pm 0.009$ &$  -2.25\pm 0.24$ &$  -6.12\pm 0.25$ &$   63\pm\phantom{1} 5$ &SgF &55           \\
G052.10$+$01.04 &               &19:23:37.3212 &$+$17:29:10.437 &$ 0.165\pm 0.013$ &$  -2.77\pm 1.40$ &$  -5.85\pm 1.40$ &$   42\pm 40$ &SgF &18,55        \\
G054.10$-$00.08 &               &19:31:48.7978 &$+$18:42:57.096 &$ 0.231\pm 0.031$ &$  -3.13\pm 0.58$ &$  -5.57\pm 0.61$ &$   40\pm\phantom{1} 5$ &LoS &70           \\
G058.77$+$00.64 &               &19:38:49.1269 &$+$23:08:40.205 &$ 0.299\pm 0.040$ &$  -2.70\pm 0.34$ &$  -6.10\pm 0.38$ &$   33\pm\phantom{1} 3$ &LoS &70           \\
G059.47$-$00.18 &               &19:43:28.3504 &$+$23:20:42.522 &$ 0.535\pm 0.024$ &$  -1.83\pm 1.22$ &$  -6.60\pm 1.21$ &$   26\pm\phantom{1} 3$ &LoS &70           \\
G059.78$+$00.06 &               &19:43:11.2464 &$+$23:44:03.270 &$ 0.463\pm 0.020$ &$  -1.50\pm 0.53$ &$  -5.12\pm 0.51$ &$   23\pm\phantom{1} 5$ &LoS &16,56        \\
G059.83$+$00.67 &               &19:40:59.2938 &$+$24:04:44.177 &$ 0.242\pm 0.014$ &$  -2.91\pm 0.26$ &$  -6.12\pm 0.26$ &$   34\pm\phantom{1} 3$ &LoS &70           \\
G060.57$-$00.18 &               &19:45:52.4949 &$+$24:17:43.237 &$ 0.121\pm 0.015$ &$  -3.26\pm 0.15$ &$  -5.66\pm 0.15$ &$    4\pm\phantom{1} 5$ &Per &51           \\
G069.54$-$00.97 &ON1            &20:10:09.0700 &$+$31:31:35.950 &$ 0.406\pm 0.013$ &$  -3.19\pm 0.40$ &$  -5.22\pm 0.40$ &$   12\pm\phantom{1} 5$ &Loc &19,20,21     \\
G070.18$+$01.74 &               &20:00:54.4874 &$+$33:31:28.224 &$ 0.156\pm 0.016$ &$  -2.82\pm 0.16$ &$  -5.17\pm 0.19$ &$  -23\pm\phantom{1} 5$ &Per &58           \\
G071.31$+$00.82 &IRAS20056+3350 &20:07:31.2593 &$+$33:59:41.491 &$ 0.213\pm 0.028$ &$  -2.62\pm 0.70$ &$  -5.65\pm 0.60$ &$    9\pm\phantom{1} 5$ &Loc &49           \\
G071.52$-$00.38 &               &20:12:57.8943 &$+$33:30:27.083 &$ 0.277\pm 0.026$ &$  -2.48\pm 0.29$ &$  -4.97\pm 0.30$ &$   11\pm\phantom{1} 3$ &Loc &70           \\
G073.65$+$00.19 &               &20:16:21.9320 &$+$35:36:06.094 &$ 0.075\pm 0.020$ &$  -2.31\pm 0.17$ &$  -4.20\pm 0.23$ &$  -76\pm 10$ &Out &1            \\
G074.03$-$01.71 &IRAS20231+3440 &20:25:07.1053 &$+$34:49:57.593 &$ 0.629\pm 0.017$ &$  -3.79\pm 1.30$ &$  -4.88\pm 1.50$ &$    5\pm\phantom{1} 5$ &Loc &21           \\
G074.56$+$00.84 &IRAS20143+3634 &20:16:13.3617 &$+$36:43:33.920 &$ 0.367\pm 0.083$ &$  -3.00\pm 0.81$ &$  -4.37\pm 0.98$ &$   -1\pm\phantom{1} 5$ &Loc &74           \\
G075.29$+$01.32 &               &20:16:16.0120 &$+$37:35:45.810 &$ 0.108\pm 0.010$ &$  -2.37\pm 0.11$ &$  -4.48\pm 0.17$ &$  -58\pm\phantom{1} 5$ &Out &22           \\
G075.76$+$00.33 &               &20:21:41.0860 &$+$37:25:29.280 &$ 0.285\pm 0.022$ &$  -3.08\pm 0.60$ &$  -4.56\pm 0.60$ &$   -9\pm\phantom{1} 9$ &Loc &21           \\
G075.78$+$00.34 &ON2N           &20:21:44.0120 &$+$37:26:37.480 &$ 0.261\pm 0.030$ &$  -2.79\pm 0.55$ &$  -4.66\pm 0.55$ &$    1\pm\phantom{1} 5$ &Loc &23           \\
G076.38$-$00.61 &               &20:27:25.4752 &$+$37:22:48.460 &$ 0.770\pm 0.053$ &$  -3.73\pm 3.00$ &$  -3.84\pm 3.00$ &$   -2\pm\phantom{1} 5$ &Loc &21           \\
G078.12$+$03.63 &IRAS20126+4104 &20:14:26.0700 &$+$41:13:32.690 &$ 0.645\pm 0.030$ &$  -2.06\pm 0.84$ &$   0.98\pm 0.84$ &$   -6\pm\phantom{1} 5$ &Loc &24,71        \\
G078.88$+$00.70 &AFGL2591       &20:29:24.8230 &$+$40:11:19.590 &$ 0.300\pm 0.024$ &$  -1.20\pm 0.72$ &$  -4.80\pm 0.66$ &$   -6\pm\phantom{1} 7$ &Loc &25           \\
G079.73$+$00.99 &IRAS20290+4052 &20:30:50.6730 &$+$41:02:27.540 &$ 0.737\pm 0.062$ &$  -2.84\pm 0.79$ &$  -4.14\pm 1.06$ &$   -3\pm\phantom{1} 5$ &Loc &25           \\
G079.87$+$01.17 &IRAS20286+4105 &20:30:29.1440 &$+$41:15:53.640 &$ 0.620\pm 0.027$ &$  -3.23\pm 1.30$ &$  -5.19\pm 1.30$ &$   -5\pm 10$ &Loc &21           \\
G080.79$-$01.92 &NMLCyg         &20:46:25.5400 &$+$40:06:59.400 &$ 0.620\pm 0.047$ &$  -1.55\pm 0.57$ &$  -4.59\pm 0.57$ &$   -3\pm\phantom{1} 3$ &Loc &26           \\
G080.86$+$00.38 &DR20           &20:37:00.9600 &$+$41:34:55.700 &$ 0.687\pm 0.038$ &$  -3.29\pm 0.73$ &$  -4.83\pm 0.77$ &$   -3\pm\phantom{1} 5$ &Loc &25           \\
G081.75$+$00.59 &DR21A          &20:39:01.9930 &$+$42:24:59.290 &$ 0.666\pm 0.035$ &$  -2.84\pm 0.45$ &$  -3.80\pm 0.47$ &$   -3\pm\phantom{1} 3$ &Loc &25           \\
G081.87$+$00.78 &W75N           &20:38:36.4260 &$+$42:37:34.800 &$ 0.772\pm 0.042$ &$  -1.97\pm 0.50$ &$  -4.16\pm 0.51$ &$    7\pm\phantom{1} 3$ &Loc &25           \\
G090.21$+$02.32 &IRAS21007+4951 &21:02:22.7010 &$+$50:03:08.309 &$ 1.483\pm 0.038$ &$  -0.67\pm 1.56$ &$  -0.90\pm 1.67$ &$   -3\pm\phantom{1} 5$ &Loc &21           \\
G090.92$+$01.48 &               &21:09:12.9685 &$+$50:01:03.664 &$ 0.171\pm 0.031$ &$  -2.96\pm 0.20$ &$  -2.61\pm 0.28$ &$  -70\pm\phantom{1} 5$ &Out &1            \\
G092.67$+$03.07 &               &21:09:21.7320 &$+$52:22:37.080 &$ 0.613\pm 0.020$ &$  -0.69\pm 1.42$ &$  -2.25\pm 1.42$ &$   -5\pm 10$ &Loc &21           \\
G094.60$-$01.79 &AFGL2789       &21:39:58.2701 &$+$50:14:20.994 &$ 0.221\pm 0.013$ &$  -3.49\pm 0.23$ &$  -2.65\pm 0.23$ &$  -43\pm\phantom{1} 3$ &Per &18,28,58     \\
G095.29$-$00.93 &               &21:39:40.5089 &$+$51:20:32.808 &$ 0.206\pm 0.015$ &$  -2.74\pm 0.30$ &$  -2.85\pm 0.37$ &$  -38\pm\phantom{1} 5$ &Per &27           \\
G097.53$+$03.18 &               &21:32:12.4343 &$+$55:53:49.689 &$ 0.133\pm 0.017$ &$  -2.94\pm 0.29$ &$  -2.48\pm 0.29$ &$  -73\pm\phantom{1} 5$ &Out &27           \\
G098.03$+$01.44 &               &21:43:01.4400 &$+$54:56:17.800 &$ 0.366\pm 0.161$ &$  -3.28\pm 1.00$ &$  -0.63\pm 1.00$ &$  -61\pm\phantom{1} 5$ &Per &58           \\
G100.37$-$03.57 &               &22:16:10.3651 &$+$52:21:34.113 &$ 0.289\pm 0.016$ &$  -3.66\pm 0.60$ &$  -3.02\pm 0.60$ &$  -37\pm 10$ &Per &28           \\
G105.41$+$09.87 &               &21:43:06.4810 &$+$66:06:55.340 &$ 1.129\pm 0.063$ &$  -0.21\pm 1.20$ &$  -5.49\pm 1.20$ &$  -10\pm\phantom{1} 5$ &Loc &21           \\
G107.29$+$05.63 &IRAS22198+6336 &22:21:26.7280 &$+$63:51:37.920 &$ 1.288\pm 0.107$ &$  -2.47\pm 1.40$ &$   0.26\pm 1.40$ &$  -11\pm\phantom{1} 5$ &Loc &29           \\
G108.18$+$05.51 &               &22:28:51.4117 &$+$64:13:41.201 &$ 1.092\pm 0.018$ &$   0.26\pm 1.15$ &$  -2.26\pm 1.20$ &$  -11\pm\phantom{1} 3$ &Loc &19           \\
G108.20$+$00.58 &               &22:49:31.4775 &$+$59:55:42.006 &$ 0.227\pm 0.037$ &$  -2.25\pm 0.50$ &$  -1.00\pm 0.50$ &$  -49\pm 10$ &Per &28           \\
G108.42$+$00.89 &IRAS22480+6002 &22:49:58.8760 &$+$60:17:56.650 &$ 0.400\pm 0.050$ &$  -2.58\pm 0.60$ &$  -1.91\pm 0.50$ &$  -51\pm\phantom{1} 5$ &Per &75           \\
G108.47$-$02.81 &               &23:02:32.0800 &$+$56:57:51.400 &$ 0.309\pm 0.010$ &$  -3.13\pm 0.49$ &$  -2.79\pm 0.46$ &$  -54\pm\phantom{1} 5$ &Per &28           \\
G108.59$+$00.49 &               &22:52:38.3150 &$+$60:00:51.888 &$ 0.405\pm 0.033$ &$  -5.56\pm 0.40$ &$  -3.40\pm 0.40$ &$  -52\pm\phantom{1} 5$ &Per &28           \\
G109.87$+$02.11 &               &22:56:18.0559 &$+$62:01:49.562 &$ 1.228\pm 0.024$ &$  -1.03\pm 1.30$ &$  -2.62\pm 1.33$ &$   -7\pm\phantom{1} 5$ &Loc &30,70        \\
G110.19$+$02.47 &IRAS22555+6213 &22:57:29.8060 &$+$62:29:46.837 &$ 0.314\pm 0.070$ &$  -2.14\pm 0.68$ &$  -0.32\pm 0.77$ &$  -63\pm\phantom{1} 5$ &Per &76           \\
G111.23$-$01.23 &               &23:17:20.7888 &$+$59:28:46.970 &$ 0.300\pm 0.081$ &$  -4.37\pm 0.60$ &$  -2.38\pm 0.60$ &$  -53\pm 10$ &Per &28           \\
G111.25$-$00.76 &               &23:16:10.3420 &$+$59:55:28.596 &$ 0.280\pm 0.015$ &$  -2.69\pm 0.28$ &$  -1.75\pm 0.33$ &$  -40\pm\phantom{1} 3$ &Per &28           \\
G111.54$+$00.77 &NGC7538        &23:13:45.3600 &$+$61:28:10.550 &$ 0.378\pm 0.017$ &$  -2.45\pm 0.40$ &$  -2.44\pm 0.40$ &$  -57\pm\phantom{1} 5$ &Per &30           \\
G115.05$-$00.04 &PZCas          &23:44:03.2816 &$+$61:47:22.187 &$ 0.356\pm 0.040$ &$  -3.70\pm 0.78$ &$  -2.00\pm 0.81$ &$  -36\pm\phantom{1} 5$ &Per &77           \\
G121.29$+$00.65 &L1287          &00:36:47.3530 &$+$63:29:02.160 &$ 1.077\pm 0.039$ &$  -0.86\pm 1.19$ &$  -2.29\pm 1.23$ &$  -23\pm\phantom{1} 5$ &Loc &19           \\
G122.01$-$07.08 &IRAS00420+5530 &00:44:58.3970 &$+$55:46:47.600 &$ 0.460\pm 0.020$ &$  -3.70\pm 0.50$ &$  -1.25\pm 0.50$ &$  -50\pm\phantom{1} 5$ &Per &31           \\
G123.06$-$06.30 &NGC281         &00:52:24.7008 &$+$56:33:50.527 &$ 0.355\pm 0.030$ &$  -2.79\pm 0.62$ &$  -2.14\pm 0.70$ &$  -30\pm\phantom{1} 5$ &Per &32           \\
G123.06$-$06.30 &NGC281W        &00:52:24.1960 &$+$56:33:43.180 &$ 0.421\pm 0.022$ &$  -2.69\pm 0.31$ &$  -1.77\pm 0.29$ &$  -29\pm\phantom{1} 3$ &Per &19           \\
G133.94$+$01.06 &W3OH           &02:27:03.8190 &$+$61:52:25.230 &$ 0.512\pm 0.010$ &$  -1.20\pm 0.32$ &$  -0.15\pm 0.32$ &$  -47\pm\phantom{1} 3$ &Per &33,34        \\
G134.62$-$02.19 &SPer           &02:22:51.7106 &$+$58:35:11.444 &$ 0.413\pm 0.017$ &$  -0.49\pm 0.50$ &$  -1.19\pm 0.48$ &$  -39\pm\phantom{1} 5$ &Per &35           \\
G135.27$+$02.79 &               &02:43:28.5680 &$+$62:57:08.388 &$ 0.167\pm 0.011$ &$  -1.22\pm 0.30$ &$   0.46\pm 0.36$ &$  -72\pm\phantom{1} 3$ &Out &36           \\
G136.84$+$01.16 &               &02:49:33.6070 &$+$60:48:27.680 &$ 0.470\pm 0.115$ &$   0.42\pm 0.50$ &$  -0.43\pm 0.50$ &$  -42\pm\phantom{1} 5$ &Per &58           \\
G160.14$+$03.15 &               &05:01:40.2437 &$+$47:07:19.026 &$ 0.244\pm 0.006$ &$   0.87\pm 0.35$ &$  -1.32\pm 0.29$ &$  -18\pm\phantom{1} 5$ &Out &27           \\
G168.06$+$00.82 &IRAS05137+3919 &05:17:13.7436 &$+$39:22:19.915 &$ 0.187\pm 0.022$ &$   0.44\pm 0.29$ &$  -0.78\pm 0.35$ &$  -25\pm\phantom{1} 5$ &Out &38,27        \\
G170.65$-$00.24 &IRAS05168+3634 &05:20:22.0700 &$+$36:37:56.630 &$ 0.532\pm 0.075$ &$   0.23\pm 1.20$ &$  -3.14\pm 0.60$ &$  -19\pm\phantom{1} 5$ &Per &78           \\
G173.48$+$02.44 &               &05:39:13.0658 &$+$35:45:51.281 &$ 0.594\pm 0.014$ &$   0.62\pm 0.63$ &$  -2.34\pm 0.63$ &$  -13\pm\phantom{1} 5$ &Per &58           \\
G174.20$-$00.07 &AFGL5142       &05:30:48.0173 &$+$33:47:54.568 &$ 0.467\pm 0.020$ &$   0.32\pm 0.56$ &$  -0.22\pm 0.68$ &$   -2\pm 10$ &Per &50           \\
G176.51$+$00.20 &               &05:37:52.1353 &$+$32:00:03.928 &$ 1.038\pm 0.021$ &$   1.84\pm 1.00$ &$  -5.86\pm 1.00$ &$  -17\pm\phantom{1} 5$ &Loc &21           \\
G182.67$-$03.26 &               &05:39:28.4248 &$+$24:56:31.946 &$ 0.157\pm 0.042$ &$   0.44\pm 0.35$ &$  -0.45\pm 0.36$ &$   -7\pm 10$ &Out &27           \\
G183.72$-$03.66 &               &05:40:24.2276 &$+$23:50:54.728 &$ 0.629\pm 0.012$ &$   0.38\pm 1.30$ &$  -1.32\pm 1.30$ &$    3\pm\phantom{1} 5$ &Per &28           \\
G188.79$+$01.03 &IRAS06061-2151 &06:09:06.9750 &$+$21:50:41.400 &$ 0.496\pm 0.103$ &$  -0.10\pm 0.50$ &$  -3.91\pm 0.50$ &$   -5\pm\phantom{1} 5$ &Per &39           \\
G188.94$+$00.88 &S252           &06:08:53.3441 &$+$21:38:29.099 &$ 0.476\pm 0.006$ &$  -0.32\pm 0.49$ &$  -1.94\pm 0.49$ &$    8\pm\phantom{1} 5$ &Per &18,40        \\
G192.16$-$03.81 &               &05:58:13.5300 &$+$16:31:58.900 &$ 0.660\pm 0.040$ &$   0.70\pm 0.78$ &$  -1.80\pm 0.86$ &$    5\pm\phantom{1} 5$ &Per &41           \\
G192.60$-$00.04 &S255           &06:12:54.0100 &$+$17:59:23.200 &$ 0.601\pm 0.039$ &$  -0.18\pm 0.60$ &$  -0.43\pm 0.60$ &$    7\pm\phantom{1} 5$ &Per &19,50        \\
G196.45$-$01.67 &S269           &06:14:37.6410 &$+$13:49:36.693 &$ 0.242\pm 0.011$ &$   0.20\pm 0.27$ &$  -0.73\pm 0.27$ &$   19\pm\phantom{1} 5$ &Out &42,66,65     \\
G209.00$-$19.38 &Orion          &05:35:15.8000 &$-$05:23:14.100 &$ 2.421\pm 0.019$ &$   3.13\pm 3.06$ &$  -1.10\pm 3.71$ &$    3\pm\phantom{1} 5$ &Loc &43,44,45,57  \\
G211.59$+$01.05 &               &06:52:45.3210 &$+$01:40:23.072 &$ 0.239\pm 0.010$ &$  -0.79\pm 0.62$ &$   0.05\pm 0.66$ &$   45\pm\phantom{1} 5$ &Out &58           \\
G213.70$-$12.59 &               &06:07:47.8584 &$-$06:22:56.586 &$ 1.148\pm 0.030$ &$  -1.30\pm 1.20$ &$   2.20\pm 1.20$ &$   11\pm\phantom{1} 3$ &Loc &1,72         \\
G217.79$+$01.05 &V838Mon        &07:04:04.8217 &$-$03:50:50.636 &$ 0.163\pm 0.016$ &$  -0.54\pm 0.23$ &$   0.08\pm 0.17$ &$   49\pm\phantom{1} 5$ &Out &48           \\
G229.57$+$00.15 &               &07:23:01.7718 &$-$14:41:34.339 &$ 0.218\pm 0.012$ &$  -1.33\pm 0.70$ &$   0.77\pm 0.70$ &$   53\pm 10$ &Per &28           \\
G232.62$+$00.99 &               &07:32:09.7800 &$-$16:58:12.800 &$ 0.596\pm 0.035$ &$  -2.17\pm 0.38$ &$   2.09\pm 0.60$ &$   21\pm\phantom{1} 3$ &Loc &40           \\
G236.81$+$01.98 &               &07:44:28.2367 &$-$20:08:30.606 &$ 0.326\pm 0.026$ &$  -2.49\pm 0.43$ &$   2.67\pm 0.41$ &$   53\pm 10$ &Per &28           \\
G239.35$-$05.06 &VYCMa          &07:22:58.3260 &$-$25:46:03.060 &$ 0.855\pm 0.057$ &$  -2.80\pm 0.58$ &$   2.60\pm 0.58$ &$   20\pm\phantom{1} 3$ &Loc &46,47        \\
G240.31$+$00.07 &               &07:44:51.9676 &$-$24:07:41.372 &$ 0.187\pm 0.014$ &$  -1.55\pm 0.50$ &$   2.57\pm 0.30$ &$   68\pm\phantom{1} 5$ &Per &28,73        \\
\enddata
\tablecomments {
Columns 1 and 2 give the Galactic source name/coordinates and an alias, when appropriate.
Right Ascension and Declination (J2000) are listed in columns 3 and 4.
Columns 5 through 7 give the parallax and proper motion in the eastward 
($\mu_x=\ura \cos{\delta}$) and northward directions ($\mu_y=\udec$).  
Column 8 lists Local Standard of Rest velocity. Column 9 indicates the
spiral arm segment in which it resides, based mostly on association with structure seen
in $\ell-V$ plots of CO and H~I emission.
Starting near the Galactic Center: GC=Galactic Center region; Con=Connecting arm; 
3kN/F=3-kpc arm (near/far); Nor=Norma (a.k.a 4-kpc) arm; ScN/F=Scutum-Centaurus arm (near/far); 
SgN/F=Sagittarius arm (near/far); Loc=Local arm; Per=Perseus arm; and Out=Outer arm; 
OSC=Outer-Scutum-Centaurus arm.  
In addition we list two spurs: LoS=Local arm spur; AqS=Aquarius spur.  
Sources indicated with ``???'' could not be confidently assigned to an arm.
Some parameter values listed here were preliminary ones and may be slightly different
from final values appearing in published papers.  Motion components and their uncertainties 
are meant to reflect that of the central star that excites the masers, and may be larger
than formal measurement uncertainties quoted in some papers.  Parallax uncertainties
for sources with multiple ($N$) maser spots have been adjusted upwards by $\sqrt{N}$,
if not done so in the original publications. 
References are as follows: 
(1) BeSSeL Survey unpublished; 
(2) \citet{Wu:14}; 
(3) \citet{Reid:09c}; 
(4) \citet{Sato:14}; 
(5) \citet{Sanna:09}; 
(6) \citet{Zhang:13b}; 
(7) \citet{Sanna:14}; 
(8) \citet{Immer:13}; 
(9) \citet{Sato:10a}; 
(10) \citet{Xu:11}; 
(11) \citet{Brunthaler:09}; 
(12) \citet{Bartkiewicz:08}; 
(13) \citet{Kurayama:11}; 
(14) \citet{Zhang:09}; 
(15) \citet{Zhang:13}; 
(16) \citet{Xu:09}; 
(17) \citet{Sato:10b}; 
(18) \citet{Oh:10}; 
(19) \citet{Rygl:10}; 
(20) \citet{Nagayama:11}; 
(21) \citet{Xu:13}; 
(22) \citet{Sanna:12}; 
(23) \citet{Ando:11}; 
(24) \citet{Moscadelli:11}; 
(25) \citet{Rygl:12}; 
(26) \citet{Zhang:12b}; 
(27) \citet{Hachi:15}; 
(28) \citet{Choi:13}; 
(29) \citet{Hirota:08}; 
(30) \citet{Moscadelli:09}; 
(31) \citet{Moellenbrock:09}; 
(32) \citet{Sato:08}; 
(33) \citet{Xu:06}; 
(34) \citet{Hachi:06}; 
(35) \citet{Asaki:10}; 
(36) \citet{Hachi:09}; 
(37) \citet{Krishnan15}; 
(38) \citet{Honma:11}; 
(39) \citet{Niinuma:11}; 
(40) \citet{Reid:09a}; 
(41) \citet{Shiozaki:11}; 
(42) \citet{Honma:07}; 
(43) \citet{Sandstrom:07}; 
(44) \citet{Menten:07}; 
(45) \citet{Kim:08}; 
(46) \citet{Choi:08}; 
(47) \citet{Zhang:12a}; 
(48) \citet{Sparks:08}; 
(49) \citet{Burns14}; 
(50) \citet{Burns17}; 
(51) \citet{Zhang18}; 
(52) \citet{Immer18}; 
(53) \citet{Li18}; 
(54) \citet{Rygl18}; 
(55) \citet{Wu19}; 
(56) \citet{Hu18}; 
(57) \citet{Kounkel:17}; 
(58) \citet{Sakai19}; 
(59) \citet{Sanna19}; 
(60) \citet{Moscadelli18}; 
(61) \citet{Krishnan17}; 
(62) \citet{Yamauchi16}; 
(63) \citet{Sanna17}; 
(64) \citet{S.Xu:18}; 
(65) \citet{Nunez:19}; 
(66) \citet{Asaki14}; 
(67) \citet{Chibueze16}; 
(68) \citet{Anna18}; 
(69) \citet{Nagayama:14}; 
(70) \citet{Xu:16}; 
(71) \citet{Nagayama15}; 
(72) \citet{Dzib16}; 
(73) \citet{Sakai15}; 
(74) \citet{Burns14a}; 
(75) \citet{Imai:12}; 
(76) \citet{Chibueze:14}; 
(77) \citet{Kusuno:13}; 
(78) \citet{Sakai:12}
                }
\label{table:parallaxes}
\end{deluxetable}

\begin{deluxetable}{lrcrlrrrr}
\tablecolumns{9} \tablewidth{0pc} 
\tablecaption{Spiral Arm Characteristics}
%\tabletypesize{\tiny} 
%\tabletypesize{\scriptsize}           
\tabletypesize{\footnotesize}           
%\rotate
\tablehead {
 \colhead{Arm} &\colhead{N}&$\ell$ tangency &$\beta$ range    &\colhead{$\beta_{kink}$} &\colhead{$R_{kink}$} &\colhead{$\psi_<$} &\colhead{$\psi_>$}&\colhead{Width}\\
 \colhead{}    &\colhead{} &\colhead{(deg)}&\colhead{(deg)}    &\colhead{(deg)}          &\colhead{(kpc)}     &\colhead{(deg)}    &\colhead{(deg)}   &\colhead{(kpc)} 
           }
\startdata
3-kpc(N)       &3          &337.0          &$15\rightarrow\p18$&$15$                     &$3.52\pm0.26$       &$-4.2\pm3.8$       &$-4.2\pm3.8$   &$0.18\pm0.05$     \\
Norma          &11         &327.5          &$\p5\rightarrow\p54$&$18\pm4$                &$4.46\pm0.19$       &$-1.0\pm3.3$       &$19.5\pm5.1$   &$0.14\pm0.10$     \\
Sct-Cen        &36         &306.1          &$\p0\rightarrow104$&$23$                     &$4.91\pm0.09$       &$14.1\pm1.7$       &$12.1\pm2.4$   &$0.23\pm0.05$     \\
Sgr-Car        &35         &285.6          &$\p2\rightarrow\p97$&$24\pm2$                &$6.04\pm0.09$       &$17.1\pm1.6$       &$\p1.0\pm2.1$  &$0.27\pm0.04$     \\
Local          &28         &...            &$-8\rightarrow\p34$ &$\p9$                   &$8.26\pm0.05$       &$11.4\pm1.9$       &$11.4\pm1.9$   &$0.31\pm0.05$     \\
Perseus        &41         &...            &$-23\rightarrow115$&$40$                     &$8.87\pm0.13$       &$10.3\pm1.4$       &$\p8.7\pm2.7$  &$0.35\pm0.06$     \\
Outer          &11         &...            &$-16\rightarrow\p71$&$18$                    &$12.24\pm0.36$      &$\p3.0\pm4.4$      &$\p9.4\pm4.0$  &$0.65\pm0.16$     \\
\enddata
\tablecomments{\footnotesize
~Spiral parameters from fitting log-periodic spirals for arms listed in column 1, based
on data from Table \ref{table:parallaxes}.  The number of massive young star parallaxes 
assigned to individual arms is given in column 2.  
Priors for Galactic longitudes of arm tangencies in quadrant 4 from
\citet{Bronfman:00} (3-kpc(N): $337\deg$; Norma: $328\deg$; Sct-Cen: $308\deg$; Sgr-Car: $283\deg$) 
with uncertainties of $\pm2\deg$ were used to constrain fits in regions where few parallaxes have been 
measured; column 3 lists posteriori values from the fits.  Column 4 gives the range of Galactocentric
azimuth for the parallax data.   The spiral model allowed for a ``kink'' at azimuth $\beta_{kink}$ 
(column 5) and radius $R_{kink}$ (column 6), with pitch angles $\psi_<$ (column 7) and $\psi_>$
(column 8) for azimuths $\le$~and~$>\beta_{kink}$, respectively.   If $\beta_{kink}$ is given without
uncertainty, it was not solved for and assigned a value based primarily on a gap in sources.
If $\psi_< = \psi_>$, only a single pitch angle was solved for.
Column 9 is the intrinsic (Gaussian $1\sigma$) arm width at $R_{kink}$, 
adjusted with the other parameters in the MCMC trials, which resulted in a $\chi_\nu^2\approx1$.  
The models assume $\Ro=8.15$ kpc.
               }
\label{table:pitchangles}
\end{deluxetable}

\begin{deluxetable}{lccccc}
\tablecolumns{6} \tablewidth{0pc} 
\tablecaption{Bayesian Fitting Results}
\tabletypesize{\footnotesize}           
%\tiny
\tablehead {
  \colhead{} & \colhead{A1}   & \colhead{A5}  & \colhead{B}   & \colhead{C}   & \colhead{D}   
           }
\startdata
\cutinhead{Fitted Parameters}
\Ro~(kpc)    &$8.22\pm0.22$   &$8.15\pm0.15$  &$8.15\pm0.14$  &$8.04\pm0.17$  &~$8.04\pm0.19$ \\
\U~(\kms)    &$10.8\pm1.2$    &$10.6\pm1.2$   &$10.7\pm1.2$   &$\p8.1\pm2.5$  &~$8.2\pm2.8$   \\
\V~(\kms)    &$13.6\pm6.7$    &$10.7\pm6.0$   &$11.9\pm2.1$   &$10.7\pm4.8$   &~$6.5\pm8.4$  \\
\W~(\kms)    &\p$7.6\pm0.9$   &$\p7.6\pm0.7$  &$\p7.7\pm0.7$  &$\p7.9\pm0.8$  &$\p7.9\pm0.9$  \\
\Usbar~(\kms)&$\p6.1\pm1.9$   &$\p6.0\pm1.4$  &$\p6.2\pm1.5$  &$\p3.6\pm2.6$  &$\p3.7\pm3.0$  \\
\Vsbar~(\kms)&$-2.1\pm6.5$    &$-4.3\pm5.6$   &$-3.1\pm2.2$   &$-4.4\pm4.5$   &$-8.7\pm7.9$  \\
\atwo        &$0.96\pm0.08$   &$0.96\pm0.05$  &$0.95\pm0.05$  &$0.95\pm0.05$  &$0.96\pm0.05$  \\     
\athr        &$1.62\pm0.03$   &$1.62\pm0.02$  &$1.62\pm0.02$  &$1.61\pm0.02$  &$1.62\pm0.03$  \\     
\cutinhead{Calculated Values}
\To~(\kms)         &~$237\pm8\q$    &$236\pm7\q$    &$236\pm5\q$    &$233\pm6\q$    &$238\pm8\q$    \\
$(\To+\V)$ ~(\kms) &~$249\pm7\q$    &$247\pm4\q$    &$247\pm4\q$    &$244\pm5\q$    &$244\pm6\q$    \\
$(\To+\V)/\Ro$ ~(\kms kpc$^{-1}$)&$30.46\pm0.43$ &$30.32\pm0.27$ &$30.32\pm0.28$ &$30.39\pm0.28$ &$30.36\pm0.28$ \\
\cutinhead{Fit Statistics}
$\chisq$         & 809.7      & 425.0         & 427.3         & 426.2         & 425.7         \\
$N_{dof}$        & 465        & 432           & 432           & 432           & 432           \\
$N_{sources}$    & 158        & 147           & 147           & 147           & 147           \\
$r_{\Ro,\To}$    &0.57        &0.45           &0.80           &0.63           &0.36           \\
\enddata
\tablecomments{\scriptsize
~Fit A1 used the 158 sources in Table~\ref{table:parallaxes} with
Galactocentric radii $>4$ kpc and fractional parallax uncertainty $<20$\%, 
an ``outlier tolerant'' probability distribution function for data uncertainty 
(see Section~\ref{sect:modeling}),
and set-A priors: Gaussian solar motion priors of $\U=11.1\pm1.2$, $\V=15\pm10$, 
$\W=7.2\pm1.1$ \kms\ and average source peculiar motion priors of 
$\Usbar=3\pm10$ and $\Vsbar=-3\pm10$ \kms.
Fit A5 removed 12 sources found in Fit A1 to have a
motion component residual greater than $3\sigma$, used a Gaussian
probability distribution function for the residuals (ie, least-squares), 
and the same priors as A1.
Fits B, C and D were similar to A5, except for the priors: B used
the solar motion priors of \citet{Schoenrich:10}
$(\U=11.1\pm1.2, \V=12.2\pm2.1, \W=7.2\pm1.1)$ \kms  
and no priors for source peculiar motions; C used
no solar motion priors and source peculiar motion priors
of $\Usbar=3\pm5$ and $\Vsbar=-3\pm5$ \kms; and 
D used flat priors for all parameters, except
\Vo\ and \Vsbar\ which were given unity probability between $\pm20$ \kms\ 
of the set-A prior values and zero probability outside this range.
The fit statistics listed are chi-squared ($\chisq$), the
number of degrees of freedom ($N_{dof}$), the number of sources used ($N_{sources}$),
and the Pearson product-moment correlation coefficient for parameters 
\Ro\ and \To\ ($r_{\Ro,\To}$). 
               }
\label{table:fits}
\end{deluxetable}

\begin{deluxetable}{lccccc}
\tablecolumns{3} \tablewidth{0pc} 
\tablecaption{Rotation Curve}
\tabletypesize{\footnotesize}           
%\tiny
\tablehead {
  \colhead{R (kpc)}   & \colhead{$\Theta_{\rm URC}$}  & \colhead{$\Theta_{\rm masers}$} 
           }
\startdata
     0.250 &     50.73  &  ...          \\
     0.500 &     77.36  &  ...          \\
     0.750 &     98.72  &  ...          \\
     1.000 &    116.96  &  ...          \\
     1.250 &    132.93  &  ...          \\
     1.500 &    147.06  &  ...          \\
     1.750 &    159.61  &  ...          \\
     2.000 &    170.76  &  ...          \\
     2.250 &    180.64  &  ...          \\
     2.500 &    189.38  &  ...          \\
     2.750 &    197.07  &  ...          \\
     3.000 &    203.80  &  ...          \\
     3.250 &    209.67  &  ...          \\
     3.500 &    214.73  &  ...          \\
     3.750 &    219.09  &  ...          \\
     4.000 &    222.79  &   224.06 $\pm$   2.72 \\
     4.250 &    225.93  &   227.82 $\pm$   1.81 \\
     4.500 &    228.54  &   228.10 $\pm$   1.63 \\
     4.750 &    230.70  &   230.36 $\pm$   1.62 \\
     5.000 &    232.46  &   233.51 $\pm$   1.40 \\
     5.250 &    233.87  &   234.60 $\pm$   1.52 \\
     5.500 &    234.97  &   236.25 $\pm$   1.42 \\
     5.750 &    235.81  &   239.46 $\pm$   1.19 \\
     6.000 &    236.41  &   239.88 $\pm$   1.22 \\
     6.250 &    236.82  &   240.30 $\pm$   1.14 \\
     6.500 &    237.06  &   240.38 $\pm$   1.07 \\
     6.750 &    237.16  &   239.32 $\pm$   1.17 \\
     7.000 &    237.14  &   238.53 $\pm$   1.24 \\
     7.250 &    237.02  &   238.01 $\pm$   1.41 \\
     7.500 &    236.82  &   234.57 $\pm$   1.41 \\
     7.750 &    236.55  &   232.82 $\pm$   1.08 \\
     8.000 &    236.23  &   232.83 $\pm$   1.00 \\
     8.250 &    235.87  &   232.67 $\pm$   1.00 \\
     8.500 &    235.48  &   233.04 $\pm$   1.05 \\
     8.750 &    235.06  &   231.99 $\pm$   1.33 \\
     9.000 &    234.63  &   230.78 $\pm$   1.66 \\
     9.250 &    234.19  &   231.41 $\pm$   1.24 \\
     9.500 &    233.75  &   231.52 $\pm$   1.15 \\
     9.750 &    233.30  &   232.08 $\pm$   1.21 \\
    10.000 &    232.86  &   232.46 $\pm$   1.22 \\
    10.250 &    232.42  &   233.21 $\pm$   1.69 \\
    10.500 &    231.99  &   232.89 $\pm$   3.15 \\
    10.750 &    231.57  &  ...          \\
    11.000 &    231.16  &  ...          \\
    11.250 &    230.76  &  ...          \\
    11.500 &    230.38  &   232.01 $\pm$   6.11 \\
    11.750 &    230.01  &   225.85 $\pm$   3.38 \\
    12.000 &    229.65  &   225.85 $\pm$   3.38 \\
    12.250 &    229.31  &   224.66 $\pm$   3.76 \\
    12.500 &    228.98  &  ...          \\
    12.750 &    228.67  &  ...          \\
    13.000 &    228.37  &  ...          \\
    13.250 &    228.08  &  ...          \\
    13.500 &    227.81  &  ...          \\
    13.750 &    227.54  &  ...          \\
    14.000 &    227.30  &  ...          \\
    14.250 &    227.06  &  ...          \\
    14.500 &    226.83  &  ...          \\
    14.750 &    226.62  &  ...          \\
    15.000 &    226.41  &  ...          \\
    15.250 &    226.22  &  ...          \\
    15.500 &    226.04  &  ...          \\
    15.750 &    225.86  &  ...          \\
    16.000 &    225.70  &  ...          \\
    16.250 &    225.54  &  ...          \\
    16.500 &    225.39  &  ...          \\
    16.750 &    225.25  &  ...          \\
    17.000 &    225.11  &  ...          \\
    17.250 &    224.98  &  ...          \\
    17.500 &    224.86  &  ...          \\
    17.750 &    224.74  &  ...          \\
    18.000 &    224.63  &  ...          \\
    18.250 &    224.53  &  ...          \\
    18.500 &    224.43  &  ...          \\
    18.750 &    224.33  &  ...          \\
    19.000 &    224.24  &  ...          \\
    19.250 &    224.16  &  ...          \\
    19.500 &    224.07  &  ...          \\
    19.750 &    224.00  &  ...          \\
    20.000 &    223.92  &  ...          \\
\enddata
\tablecomments{\footnotesize {$R$ is Galactocentric radius, 
$\Theta_{\rm URC}$ is the Universal rotation curve from
fit A5, and $\Theta_{\rm masers}$ are variance-weighted, circular velocities 
for the masers within a radial window of 1 kpc; these values have also been 
adjusted upwards by 4.3 \kms\ to account for the average lag found in fit A5.}
               }
\label{table:rotationcurve}
\end{deluxetable}


\begin{thebibliography}{}
\bibitem[Anderson \etal(2012)]{Anderson:12}
         Anderson, L. D., Bania, T. M., Balser, D. S. \& Rood, T. 2012
         \apj, 754, 62
\bibitem[Anderson \etal(2019)]{Anderson:19}
         Anderson, L. D., Wegner, T. V., Armentrout, W. P., Balser, D. S \& Bania, T. M.
         2019, \apj, 871, 145
\bibitem[Ando \etal(2011)]{Ando:11} Ando, K., Nagayama, T., Omodaka, T., \etal\ 2011,
         \pasj, 63, 45
\bibitem[Asaki \etal(2010)]{Asaki:10}
         Asaki, Y., Deguchi, S., Imai, H., Hachisuka, K., Miyoshi, M. \& Honma, M. 2010,
         \apj, 721, 267
\bibitem[Asaki \etal(2014)]{Asaki14} 
         Asaki, Y., Imai, H., Sobolev, A. M. \& Parfenov, S. Yu 2014, \apj, 787, 54
\bibitem[Bailer-Jones (2015)]{Bailer-Jones:15}
         Bailer-Jones, C. A. L. 2015, \pasp, 127, 994
\bibitem[Bartkiewicz \etal(2008)]{Bartkiewicz:08} Bartkiewicz, A., 
         Brunthaler, A., Szymczak, M. van Langevelde, H. J \& Reid, M. J. 2008, \aap, 490, 787
\bibitem[A. Bartkiewicz et al. (2019, in preparation)]{Anna18}
         Bartkiewicz et al. (2019, in preparation)
\bibitem[Benjamin(2008)]{Benjamin:08} Benjamin, R. A. 2008, 
         in ``Massive Star Formation: Observations Confront Theory,''
         ASP Conference Series, Vol. 387, eds. H. Beuther, H. Linz \& Th. Henning, p. 375
\bibitem[Binney, Gerhard \& Spergel(1997)]{Binney:97}
         Binney, J., Gerhard, O. \& Spergel, D. 1997, \mnras, 288, 365
\bibitem[Blaauw \etal(1960)]{Blaauw:60}
         Blaauw, A., Gum, C. S., Pawsey, J. L. \& Westerhout, G. 1960, \mnras, 121, 123 
\bibitem[Bland-Hawthorn \& Gerhard(2016)]{Bland-Hawthorn:16}
         Bland-Hawthorn, J. \& Gerhard, O. 2016, \araa, 54,529
\bibitem[Blitz \& Spergel(1991)]{Blitz:91} 
         Blitz, L. \& Spergel,, D. N. 1991, \apj, 379, 631
%\bibitem[Boehle \etal(2016)]{Boehle:16}
%         Boehle, A., Ghez, A. M., Sch\"odel, R., Meyer, L., Yelda, S. \etal 
%        2016, \apj, 830, 17
\bibitem[Bronfman \etal(2000)]{Bronfman:00} 
         Bronfman, L., Casassus, S., May, J. \& Nyman, L.-A., 2000, \aap, 358, 521
\bibitem[Brunthaler \etal(2009)]{Brunthaler:09}  Brunthaler, A., Reid, M. J.,
         Menten, K. M., Zheng, X. W., Moscadelli, L. \& Xu, Y. 
         2009, \apj, 693, 424 
\bibitem[Burns \etal(2014)]{Burns14}
         Burns, R. A., Nagayama, T., Handa, T. et al. 2014, \apj, 797, 39
\bibitem[Burns \etal(2014a)]{Burns14a}
         Burns, R. A., Yamaguchi, Y., Handa, T. et al. 2014a, \pasj, 66, 102
\bibitem[Burns \etal(2017)]{Burns17}
         Burns, R. A., Handa, T., Imai, H. et al. 2017, \mnras, 467, 2367
\bibitem[Burton \& Shane(1970)]{Burton:70}
         Burton, W. B. \& Shane, W. W. 1970, in 
         ``The Spiral Structure of our Galaxy: Proceedings of IAU Symp. 38'',
         Eds. Wilhelm Becker and Georgios Ioannou Kontopoulos
         (Dordrecht, D. Reidel Pub. Co.), p. 397
\bibitem[Chen \etal(2001)]{Chen:01}
         Chen, B., Stoughton, C., Smith, J. A. \etal 2001, \apj, 553, 184
\bibitem[Chibueze \etal(2014)]{Chibueze:14}
         Chibueze, J. O., Sakanoue, H., Omodaka, T., Handa, T., Nagayama, T. et al. 2014, \pasj, 66, 104
\bibitem[Chibueze \etal(2016)]{Chibueze16}
         Chibueze, J. O., Kamezaki, T., Omodaka, T., Handa, T., Nagayama, T. et al. 2016, \mnras, 460, 1839
\bibitem[Choi \etal(2008)]{Choi:08} 
         Choi, Y. K. \etal, 2008, \pasj, 60, 1007
\bibitem[Choi \etal(2014)]{Choi:13} 
         Choi, Y. K., Hachisuka, K., Reid, M. J., Xu, Y., Brunthaler, A. \etal\ 2014, \apj, 790, 99
\bibitem[Churchwell \& Benjamin (2009)]{Churchwell:09}
         Churchwell, E. \& Benjamin, R. 2009, \pasp, 121, 213
\bibitem[Cohen(1995)]{Cohen:95}
         Cohen, M. 1995, \apj, 444,874
\bibitem[Conti \& Vacca(1990)]{Conti:90}
         Conti, P. S. \& Vacca, W. D. 1990, \aj, 100(2), 431
\bibitem[Dame \& Thaddeus(2011)]{Dame:11}
         Dame, T. M. \& Thaddeus, P. 2011, \apj, 734, L24
\bibitem[Damour \& Taylor(1991)]{Damour:91}
         Damour, T. \& Taylor, J. H. 1991, \apj, 366, 501
\bibitem[Do \etal (2019)]{Do:19}
         Do, T., Hess, A., Ghez, A. \etal\ 2019, {\it Sci}, 365, 664
\bibitem[D'Onghia, Vogelsberger \& Hernquist(2013)]{DOnghia:13}
         D'Onghia, E., Vogelsberger, M. \& Hernquist, L. 2013 \apj, 766,34 
\bibitem[Drimmel (2000)]{Drimmel:00}
         Drimmel, R. 2000, \aap, 358, L13
\bibitem[Dzib \etal(2016)]{Dzib16}
         Dzib, S. A., Ortiz-Leon, G. N., Loinard, L. et al. 2016, \apj, 826, 201
\bibitem[Eilers \etal (2019)]{Eilers:19}
         Eilers, A.-C., Hogg, D. W., Rix, H.-W., \& Ness, M. K. 2019, arXiv:2018.09466v2
\bibitem[Gravity Collaboration(2019)]{Gravity:19}
         Gravity Collaboration: Abuter, R., Amorim, A., Baub\"ock, M. \etal 2019, 
         arXiv1904.05721
\bibitem[Green \etal(2017)]{Green:17}
         Green, J. A., Breen, S. L., Fuller, G. A. \etal 2017, \mnras, 469, 1383
\bibitem[Gum, Kerr \& Westerhout(1960)]{Gum:60}
         Gum, C. S., Kerr, F. J. \& Westerhout, G. 1960, \mnras, 121, 132
\bibitem[Hachisuka \etal(2006)]{Hachi:06} 
         Hachisuka, K. \etal\ 2006, \apj, 645, 337
\bibitem[Hachisuka \etal(2009)]{Hachi:09} 
         Hachisuka, K. Brunthaler, A., Menten, K. M., \etal\ 2009, \apj, 696, 1981
\bibitem[Hachisuka \etal(2015)]{Hachi:15}
         Hachisuka, K., Choi, Y. K., Reid, M. J., Brunthaler, A., Menten, K. M., et al. 2015, \apj, 800, 2
\bibitem[Hammersley \etal(2000)]{Hammersley:00}
         Hammersley, P. L., Garz\'on, F., Mahoney, T. J., L\'opez-Corredoira, M., Torres, M. A. P. 
         2000, \mnras, 317, 45
\bibitem[Hirota \etal(2008)]{Hirota:08} 
         Hirota, T, Ando, K., Bushimata, T., \etal\ 2008, \pasj, 60, 961
\bibitem[Honig \& Reid(2015)]{Honig:15}
         Honig, Z. N. \& Reid, M. J. 2015, \apj,800, 53
\bibitem[Honma \etal(2007)]{Honma:07} 
         Honma, M., Bushimata, T., Choi, Y. K., \etal\ 2007, \pasj, 59, 889
\bibitem[Honma \etal(2011)]{Honma:11} 
         Honma, M., Hirota, T., Kanya, Y., \etal\ 2007, \pasj, 63, 17
\bibitem[Honma \etal(2012)]{Honma:12} 
         Honma,M. \etal 2012, \pasj, 64, 136
\bibitem[B. Hu et al. (2019, in preparation)]{Hu18}
         Hu et al. (2019, in preparation)
\bibitem[Humphreys(1978)]{Humphreys:78}
         Humphreys, R. M. 1978, \apjs, 38, 309
\bibitem[Humphreys \& Larsen(1995)]{Humphreys:95}
         Humphreys, R. M. \& Larsen, J. A. 1995, \aj, 110(5), 2183
\bibitem[Imai \etal(2012)]{Imai:12}
         Imai, H., Sakai, N., Nakanishi, H., Sakanoue, H., Honma, M. et al. 2012, \pasj, 64, 98
\bibitem[Immer \etal(2013)]{Immer:13} 
         Immer, K., Reid, M. J., Menten, K. M., Brunthaler, A., \& Dame, T. M. 2013, \aap, 553, 117
\bibitem[Juri\'c(2008)]{Juric:08}
         Juri\'c, M., Ivezi\'c, Z., Brooks, A. \etal 2008, \apj, 673, 864
\bibitem[K. Immer et al. (2019, in preparation)]{Immer18}
         Immer et al. (2019, in preparation) 
\bibitem[Kim \etal(2008)]{Kim:08}
         Kim, M. K., Hirota, T., Honma, M., \etal\ 2008, \pasj, 60, 991
\bibitem[Koo \etal(2017)]{Koo:17}
         Koo, B.-C., Park, G., Kim, W.-T., Lee, M. G., Balser, D. S. \etal. 2017, \pasp, 129, 979
\bibitem[Krishnan \etal(2015)]{Krishnan15}
         Krishnan, V., Ellingsen, S. P., Reid, M. J. et al. 2015, \apj, 805, 129
\bibitem[Krishnan \etal(2017)]{Krishnan17}
         Krishnan, V., Ellingsen, S. P., Reid, M. J. et al. 2017, \mnras, 465, 1095
\bibitem[Kurayama \etal(2011)]{Kurayama:11}
         Kurayama, T., Nakagawa, A., Sawada-Satoh, S, \etal\ 2011, \pasj, 63, 513
\bibitem[Kusuno \etal(2013)]{Kusuno:13}
         Kusuno, K, Asaki, Y., Imai, H. \& Oyama, T. 2013, \apj, 774, 107
\bibitem[Kounkel \etal(2017)]{Kounkel:17}
         Kounkel, M., Hartmann, L., Loinard, L., Ortiz-L\'eon, G. N., Mioduszewski, A. J. et al. 2017, 
         \apj, 834, 142
\bibitem[J. Li et al. (2019, in preparation)]{Li18}
         Li et al. (2019, in preparation)
\bibitem[Ma\'iz-Appel\'aniz(2001)]{Maiz:01}
         Ma\'iz-Appel\'aniz, J. 2001, \apj, 121, 2737
\bibitem[Malhotra (1995)]{Malhotra:95}
         Malhotra, S. 1995, \apj, 448, 138
\bibitem[M\'endez \& van Altena(1998)]{Mendez:98}
         M\'endez, R. A. \& van Altena, W. F. 1998, \aap, 330, 910
\bibitem[Menten \etal(2007)] {Menten:07} Menten, K. M., Reid, M. J., Forbrich J. 
         \& Brunthaler, A. 2007, \aap, 474, 515
\bibitem[Moellenbrock, Claussen \& Goss(2009)]{Moellenbrock:09} Moellenbrock, G. A.,
         Claussen, M. J. \& Goss, W. M. 2009, \apj, 694, 192
\bibitem[Moscadelli \etal(2009)]{Moscadelli:09} 
         Moscadelli, L., Reid, M. J., Menten, K. M., Brunthaler, A., Zheng, X. W. \&
         Xu, Y.  2009, \apj, 693, 406
\bibitem[Moscadelli \etal(2011)]{Moscadelli:11}
         Moscadelli, L., Cesaroni, R., Rioja, M. J., Dodson, R., Reid, M. J., 2011,
         \aap, 526, 66
\bibitem[Moscadell, Sanna \& Goddi (2011)]{Moscadelli:11b}
         Moscadelli, L., Sanna, A. \& Goddi, C., 2011, \aap, 536, 38
\bibitem[L. Moscadelli et al. (2019, in preparation)]{Moscadelli18}
         Moscadelli et al. (2019, in preparation)
\bibitem[Nagayama \etal(2011)]{Nagayama:11}
         Nagayama, T., Omodaka, T., Nakagawa, A, \etal\ 2011, \pasj, 63, 23
\bibitem[Nagayama \etal(2014)]{Nagayama:14}
         Nagayama, T., Kobayashi, H., Omodaka, T., Yasuhiro, M. Burns, R. A. et al 2014, \pasj, 67, 65
\bibitem[Nagayama \etal(2015)]{Nagayama15}
         Nagayama, T. Omodaka, T., Handa, T., Burns, R. A., Chibueze, J. O. et al. 2015, \pasj, 67, 66
\bibitem[Niinuma \etal(2011)]{Niinuma:11}
         Niinuma, K. Nagayama, T., Hirota, T., \etal\ 2011, \pasj, 63, 9
\bibitem[Quiroga-Nu\~nez \etal (2017)]{Nunez:17}
         Quiroga-Nu\~nez, van Langevelde, H. J., Reid, M. J. \& Green, J. A.
         2017, \aap, 604, A72
\bibitem[Quiroga-Nu\~nez \etal (2019)]{Nunez:19}
         Quiroga-Nu\~nez, L. H., Immer, K., van Langevelde, H. J., Reid, M. J. \& Burns, R. A.
         2019, \aap, 625, 70 
\bibitem[Oh \etal(2010)]{Oh:10} 
         Oh, C. S., Kobayashi, H., Honma, M., Hirota, T., Sato, K., Ueno, Y., 2010,
         \pasj, 62, 101
\bibitem[Pandy \& Mahra(1987)]{Pandy:87}
         Pandy, A. K. \& Mahra, H. S. 1987, \mnras,226, 635
\bibitem[Pandy, Bhatt \& Mahra(1988)]{Pandy:88}
         Pandy, A. K., Bhatt, B. C. \& Mahra, H. S. 1988, \aap, 189, 66
\bibitem[Persic, Salucci \& Stel(1996)]{Persic:96} 
         Persic, M., Salucci, P. \& Stel, F. 1996, \mnras, 281, 27
\bibitem[Pestalozzi, Minier \& Booth (2001)]{Pestalozzi:05}
         Pestalozzi, M. R., Minier, V., Booth, R. S. 2005, \aap, 432, 737
\bibitem[Reed(1997)]{Reed:97}
         Reid, B. C. 1990, \pasp, 109, 1145
\bibitem[Reid(1993)]{Reid:93}
         Reid, M. J. 1993, \araa, 31, 345
\bibitem[Reid \& Brunthaler(2004)]{Reid:04} 
         Reid, M. J. \& Brunthaler, A., 2004, \apj, 616, 872
\bibitem[Reid \etal(2009a)]{Reid:09a} 
         Reid, M. J., Menten, K. M., Brunthaler, A., Zheng, X. W., Moscadelli, L. \&
         Xu, Y.  2009, \apj, 693, 397 
\bibitem[Reid \etal(2009b)]{Reid:09b}
         Reid, M. J., Menten, K. M., Zheng, X. W. \etal\ 2009, \apj, 700, 137
\bibitem[Reid \etal(2009c)]{Reid:09c}
         Reid, M. J., Menten, K. M., Zheng, X. W., Brunthaler, A., \& Xu, Y. 2009,
         \apj, 705, 1548
\bibitem[Reid \etal(2014)]{Reid:14}
         Reid, M. J., Menten, K. M., Brunthaler, A., Zheng, X. W., Dame, T. M. \etal\ 2014, \apj, 783, 130
\bibitem[Reid \etal(2016)]{Reid:16}
         Reid, M. J., Dame, T. M., Menten, K. M., Brunthaler, A.  2016, \apj, 823, 77
\bibitem[Roberts (1969)]{Roberts:69}
         Roberts, W. W. 1969, \apj, 158, 123
\bibitem[Rygl \etal(2010)]{Rygl:10} Rygl, K. L. J., Brunthaler, A., Reid, M. J., 
         Menten, K. M., van Langevelde, H. J., Xu, Y., 2010, \aap, 511, 2
\bibitem[Rygl \etal(2012)]{Rygl:12} Rygl, K. L. J., Brunthaler, A., Sanna, A., \etal\ 2011,
         \aap, 539, 79
\bibitem[K. Rygl et al. (2019, in preparation)]{Rygl18}
         Rygl et al. (2019, in preparation)
\bibitem[Sakai \etal(2012)]{Sakai:12}
         Sakai, N., Honma, M., Nakanishi, H., Sakanoue, T., Shibata, K. M. et al. 2012, \pasj, 64, 108
\bibitem[Sakai \etal(2015)]{Sakai15}
         Sakai, N., Nakanishi, H., Matsuo, M. et al. 2015, \pasj, 67, 69
\bibitem[Sakai \etal(2019)]{Sakai19}
         Sakai, N. Reid, M. J., Menten, K. M. \& Brunthaler, A. \& Dame,, T. M. 
         2019, \apj, 876, 30
\bibitem[Sandstrom \etal(2007)]{Sandstrom:07}
         Sandstrom, K. M., Peek, J. E. G., Bower, G. C., Bolatto, A. D. \& Plambeck, Richard L., 
         2007, \apj, 667, 1161
\bibitem[Sanna \etal(2009)]{Sanna:09}
         Sanna, A., Reid, M. J., Moscadelli, L., \etal\ 2009, \apj, 706, 464
\bibitem[Sanna \etal(2010a)]{Sanna:10a}
         Sanna, A., Moscadelli, L., Tarchi, A. \etal\ 2010a, \aap, 517, 78
\bibitem[Sanna \etal(2010b)]{Sanna:10b}
         Sanna, A., Moscadelli, L., Cesaroni, R. \etal\ 2010b, \aap, 517, 71
\bibitem[Sanna \etal(2012)]{Sanna:12}
         Sanna, A., Reid, M. J., Dame, T., \etal\ 2012, \apj, 745, 82
\bibitem[Sanna \etal(2014)]{Sanna:14}    
         Sanna, A., Reid, M. J., Menten, K. M., \etal\ 2014, \apj, 781, 108
\bibitem[Sanna \etal(2017)]{Sanna17}
         Sanna, A., Reid, M. J., Dame, T. M., Menten, K. M. \& Brunthaler, A. 2017, {\it Science,} 358, 227
\bibitem[A. Sanna et al. (2019)]{Sanna19}
         Sanna, A. \etal (2019, in preparation)
\bibitem[Sato \etal(2008)]{Sato:08} 
         Sato, M. \etal  2008, \pasj, 60, 975
\bibitem[Sato \etal(2010a)]{Sato:10a} 
         Sato, M., Hirota, T., Reid, M., \etal\  2010a, \pasj, 62, 287
\bibitem[Sato \etal(2010b)]{Sato:10b} 
         Sato, M., Reid, M. J., Brunthaler, A. \& Menten, K. M. 2010b, \apj, 720, 1055
\bibitem[Sato \etal(2014)]{Sato:14} 
         Sato, M., Wu, Y. W., Immer, K., \etal\ 2014, \apj, 793, 72
\bibitem[Schoenrich, Binney \& Dehnen(2010)]{Schoenrich:10}
         Schoenrich, R., Binney, J. \& Dehnen, W. 2010, \mnras, 403, 1829
\bibitem[Shiozaki \etal(2011)]{Shiozaki:11}
         Shiozaki, S., Imai, H., Tafoya, D., \etal\ 2011, \pasj, 63, 1219
\bibitem[Sivia \& Skilling(2006)]{Sivia:06} 
         Sivia, D. \& Skilling, J. 2006,
         Data Analysis: A Bayesian Tutorial (2nd ed.; New York, Oxford Univ. Press), 168
\bibitem[Sofue, Honma \& Omodaka (2009)]{Sofue:09}
         Sofue, Y., Honma, M. \& Omodaka, T. 2009, \pasj, 61, 227
\bibitem[Sparks \etal(2008)]{Sparks:08}
         Sparks, W. B., Bond, H. E., Cracraft, M. et al. 2008, \aj, 135, 605
\bibitem[Stobie \& Ishida(1987)]{Stobie:87}
         Stobie, R. S. \& Ishida, K. 1987, \aj, 93, 624
\bibitem[Stothers \& Frogel(1974)]{Stothers:74}
         Stothers, R. \& Frogel, J. A. 1974, \aj, 79, 456
\bibitem[Sun \etal(2015)]{Sun:15}
         Sun, Y. Xu, Y., Yang, J., Li, F.-C., Di, X.-Y. \etal 2015, \apj, 798, L27
\bibitem[Taylor \& Cordes(1993)]{Taylor:93}
         Taylor, J. H. \& Cordes, J. M. 1993, \apj, 411, 674
\bibitem[Toller(1990)]{Toller:90}
         Toller, G. N. 1990, in IAU Symp. 139, The Galactic and Extragalactic Background Radiation, 
         eds. S. Boyer \& C. Leinert (Dordrecht, Reidel), 21
\bibitem[Urquhart \etal(2014)]{Urquhart:14}
         Urquhart, J. S., Figura, C. C., Moore, T. J. T., Hoare, M. G., 
         Lumsden, S. L. \etal\ 2014, \mnras, 437, 1791
\bibitem[Valdettaro \etal(2001)]{Valdettaro:01}
         Valdettaro, R., Palla, F., Brand, J., \etal\ 2001, \aap, 368, 845
\bibitem[Wegg, Gerhard \& Portail(2015)]{Wegg:15}
         Wegg, C., Gerhard, O. \& Portail, M. 2015, \mnras, 450, 4050 
\bibitem[Wu \etal(2014)]{Wu:14}
         Wu, Y. W., Sato, M., Reid, M. J., Moscadelli, L., Zhang, B. et al. 2014, \aap, 566, 17
\bibitem[Y. Wu et al. (2019)]{Wu19}
         Wu, Y. W., Reid, M. J., Sakai, N. \etal\ 2019, \apj, 874, 94
\bibitem[Xu \etal(2018)]{S.Xu:18}
         Xu, S., Zhang, B., Reid, M., Menten, K. M., Zheng, X. et al. 2018, \apj, 859, 14
\bibitem[Xu \etal(2006)]{Xu:06}  
         Xu, Y., Reid, M. J., Zheng, W. W. \& Menten, K. M. 2006, 
         {\it Science}, 311, 54
\bibitem[Xu \etal(2009)]{Xu:09} Xu, Y., Reid, M. J., Menten, K. M., Brunthaler, A.,
         Zheng, X. W. \& Moscadelli, L.  2009, \apj, 693, 413
\bibitem[Xu \etal(2011)]{Xu:11}
         Xu, Y., Moscadelli, L., Reid, M. J., \etal\ 2011, \apj, 733, 25
\bibitem[Xu \etal(2013)]{Xu:13}
         Xu, Y., Li, J. J., Reid, M. J., \etal\ 2013, \apj, 769,15
\bibitem[Xu \etal(2016)]{Xu:16}
         Xu, Y., Reid, M., Dame, T., Menten, K., Sakai, N. et al. 2016, {\it SciA}, 2, 878
\bibitem[Yamagata \& Yoshii(1992)]{Yamagata:92}
         Yamagata, T. \& Yoshii, 1992, \aj, 103(1), 117
\bibitem[Yamauchi \etal(2016)]{Yamauchi16}
         Yamauchi, A., Yamashita, K., Honma, M., Sunada, K. \& Nakagawa, A. \etal 2016, \pasj, 68, 60
\bibitem[Zhang \etal(2009)]{Zhang:09}  
         Zhang, B., Zheng, X. W., Reid, M. J., \etal\  2009, \apj, 693, 419
\bibitem[Zhang \etal(2012a)]{Zhang:12a}
         Zhang, B., Reid, M. J., Menten, K. M., \& Zheng, X. W., 2012a,
         \apj, 744, 23
\bibitem[Zhang \etal(2012b)]{Zhang:12b}
         Zhang, B., Reid, M. J., Menten, K. M., Zheng, X. W., Brunthaler, A., 2012b,
         \aap, 544, 42
\bibitem[Zhang \etal(2013)]{Zhang:13}
         Zhang, B., Reid, M. J., Menten, K. M., \etal\ 2013a, \apj, 775, 79
\bibitem[Zhang \etal(2014)]{Zhang:13b}
         Zhang, B., Moscadelli, L., Sato, M., Reid, M. J., Menten, K. M. \etal\ 2014, \apj, 781, 89
\bibitem[Zhang et al.(2019)]{Zhang18}
         Zhang, B., Reid, M. J., Zhang, L., Wu, Y., Hu, B. \etal 2019, \aj, 157, 200
\end{thebibliography}
\end{document}